\documentclass[10pt,prd,longbibliography,superscriptaddress]{revtex4-2}

\usepackage[T1]{fontenc}
\usepackage[utf8]{inputenc}
\usepackage[english]{babel}

\usepackage{amsfonts,amsbsy,amssymb,amsmath,graphicx,float} 
\usepackage{color}

\graphicspath{{figures/}}

\usepackage[linktoc=all,colorlinks=true,citecolor=blue,linkcolor=blue]{hyperref}

\begin{document}

\title{Visualizing the Phase Space of the HeI$_2$ \\ van der Waals Complex using Lagrangian Descriptors}

\author{Makrina Agaoglou}
\email{makrina.agaoglou@bristol.ac.uk}
\affiliation{School of Mathematics, University of Bristol, \\ Fry Building, Woodland Road, Bristol, BS8 1UG, United Kingdom.}

\author{V\'ictor J. Garc\'ia-Garrido}
\email{vjose.garcia@uah.es}
\affiliation{Departamento de F\'isica y Matem\'aticas, Universidad de Alcal\'a, \\ Alcal\'a de Henares, 28871, Spain.}

\author{Matthaios Katsanikas}
\email{matthaios.katsanikas@bristol.ac.uk}
\affiliation{School of Mathematics, University of Bristol, \\ Fry Building, Woodland Road, Bristol, BS8 1UG, United Kingdom.}

\author{Stephen Wiggins}
\email{s.wiggins@bristol.ac.uk}
\affiliation{School of Mathematics, University of Bristol, \\ Fry Building, Woodland Road, Bristol, BS8 1UG, United Kingdom.}

\begin{abstract}
In this paper we demonstrate the capability of the method of Lagrangian descriptors to unveil the phase space structures that characterize transport in high-dimensional symplectic maps. In order to illustrate its use, we apply it to a four-dimensional symplectic map model that is used in che\-mistry to explore the nonlinear dynamics of van der Waals complexes. The advantage of this technique is that it allows us to easily and effectively extract the invariant manifolds that determine the dynamics of the system under study by means of examining the intersections of the underlying phase space structures with low-dimensional slices. With this approach, one can perform a full computational \textit{phase space tomography} from which three-dimensional representations of the higher-dimensional phase space can be systematically reconstructed. This analysis may be of much help for the visualization and understanding of the nonlinear dynamical mechanisms that take place in high-dimensional systems. In this context, we demonstrate how this tool can be used to detect whether the stable and unstable manifolds of the system intersect forming turnstile lobes that enclose a certain phase space volume, and the nature of their intersection.      
\end{abstract}

\maketitle

\noindent\textbf{Keywords:} Symplectic maps, Phase space structure, Lagrangian descriptors, Chemical reaction dynamics.

\section{Introduction}
\label{sec:intro}

Visualizing the phase space of a Hamiltonian system is of significant importance since it helps to better analyze and understand the dynamics of the system. For a two dimensional (2D) phase space this is a standard procedure. However, for higher dimensions it is not a straightforward process. This underlines the importance of developing tools that help to identify high dimensional phase space structures which play a crucial role in the study of transport mechanisms in the phase space of both discrete and continuous systems \cite{mackay1984}. This is concerned with the evolution from one phase space region to another or the escape from a region. Typically, the phase space of a system displays a mixture of chaotic and regular behavior, and this coexistence strongly influences the transport in the system, since chaotic orbits stick to the vicinity of regular structures, which is an effect known as trapping. In 2D maps, that correspond to Hamiltonian systems with 2 two degrees-of-freedom (DoF), this is well understood in terms of partial barriers and the turnstile mechanism \cite{meiss1992, meiss2015, backer2020}. In systems with more than two degrees of freedom, transport in higher dimensional systems is a relevant topic of current research, and in order to understand the transport mechanisms, it is of significant importance to understand the organisation of the phase space, which is more complicated for higher dimensional maps \cite{huebner,firmbach}.  

In a two DoF system, the phase space is 4D and since energy is conserved,  motion will be constrained to a 3D energy surface. It is possible to reduce the dynamics to a 2D phase space by using Poincar\'e surfaces of section, and therefore visualization becomes much easier because it is a 2D surface. On the other hand, for a three DoF Hamiltonian System the phase space is 6D. Due to the conservation of energy the dynamics is restricted to a 5D manifold and by introducing a Poincar\'e section we obtain a 4D symplectic map. For a deeper understanding of the underlying dynamics 4D maps, it is advantageous to visualize the geometry of the phase space structures of the map. Many different methods have been developed to help in the visualization of higher dimensional phase space structures, and usually they consist of reducing the order of complexity in the phase space by reducing the dimension in order to understand the dynamics better. For example, one can use 2D projections of the phase space of a 4D symplectic map, or the 4D Poincar\'e map of an autonomous 3 DoF Hamiltonian system \cite{froeschle1970, udry1988, pfenniger1985,pfenniger1985b, skokos1997,martinet1981,magnenat1982,vrahatis1996,paskauskas}  
or 3D projections \cite{martinet1981,magnenat1982,pavskauskas2009bottlenecks,vrahatis1996}, 
also including color to indicate the projected coordinate \cite{patsis1994,katsanikas2011a, katsanikas2011b,katsanikas2011c,katsanikas2013,zachilas2013structure,patsis2014phase}, frequency analysis \cite{martens1987,laskar1993,chandre2003,sethi2012}, or 2D plots of multi sections \cite{froeschle1972,froeschle2000}, and action-space plots \cite{bazzani1998}. In this work we emphasize the contribution of the method of Lagrangian descriptors (LDs) for the visualization of the phase space structures of a 4D symplectic map. It is a trajectory-based scalar diagnostic tool that was first introduced in \cite{madrid2009,Mancho1,mancho2013lagrangian,lopesino2017}. For an extended tutorial on LDs, with detailed explanations on its implementation and examples, refer to \cite{ldbook2020}. This technique focuses on integrating a positive scalar function along the trajectory of an initial condition of the system. The method of LDs has been widely applied in the literature to problems that arise in many areas such as chemical reaction dynamics
\cite{craven2015lagrangian,naik2019a,GG2019a}. In this paper we use Discrete Lagrangian Descriptors (DLDs) for maps that has been introduced in \cite{carlos,GG2019b}. We highlight the contribution of the LD method to reveal all the phase space structures. The method of LDs is efficient and easy to apply providing a very detailed visualization of the phase space structures, even in the case where relevant fixed points are  at infinity, as we will demonstrate in this paper. Using this method we can obtain information about the stable and unstable manifolds without the knowledge of the position of the fixed points. Moreover this tool can easily detect whether the geometrical characteristics of the  intersection of manifolds is such that they can give rise to lobes having the property that they bound 4D regions of the phase space \cite{wiggins90, ezra, beigie1995, beigie1995b, toda1995}. As a comparison, we will also apply the method introduced in \cite{richter2014} to visualize the KAM tori of the map.

In this work we apply DLDs to analyze the phase space of a 4D symplectic map model that is used in chemistry to explore the nonlinear dynamics of van der Waals complexes. This model system was originally introduced in \cite{gaspard1989}. The rare gas-diatomic halogen van der Waals (vdW) complexes, such as HeI$_{2}$ in Fig. \ref{fig:hei2_molecule} have been proven to be ideal model systems for fundamental investigations of molecular dynamics \cite{toda1995,ezra,montoya2020b}, because of their relative simplicity. The authors in \cite{gaspard1989} produced a model that is a three DoF system, and then, by imposing a periodic kick they converted it into a Poincar\'e map, which is the 4D symplectic map that we will analyze in this paper. In this work, we want to study the phase space structures in this model that  correspond to the intramolecular bonding, the dissociation and predissociation for the complex HeI$_2$. 

The contents of this paper are outlined as follows. In Sec \ref{sec:sec1} we introduce the 4D symplectic map that is used in chemical reaction dynamics to model the dissociation process of the HeI$_2$ van der Waals complex. Section \ref{sec:sec2} is devoted to describing the results of applying the method of Discrete Lagrangian descriptors to unveil and visualize the geometrical template of high-dimensional phase space structures governing the dynamical evolution of this physical model. Finally, in Sec. \ref{sec:conc} we discuss the main contributions of this paper and highlight some potential research lines and interesting applications where the use of this methodology could provide relevant insights to explore the intricate nonlinear dynamics and transport mechanisms that take place in a high-dimensional phase space. The appendix includes information about the equilibrium points of the system, and also a brief introduction to the method of Lagrangian descriptors and its implementation details for this work.

\section{The 4D Symplectic Map Model}
\label{sec:sec1}

In this work we study the phase space structure of the 4D symplectic map introduced in \cite{gaspard1989,ezra} to analyze the dynamics of the planar HeI$_2$ van der Waals complex. In this section we describe the Hamiltonian model that is the main focus of this work.

Consider the following Hamiltonian model:
\begin{equation}
H = \dfrac{p_{s}^{2}}{2\mu} + \dfrac{p_{r}^{2}}{2m} + \dfrac{p_{\gamma}^{2}}{2I}+\frac{L^{2}}{2mr^{2}} + V(s,r,\gamma)
\label{eq:ham_HeI2}
\end{equation}
where $\mu$ represents the reduced mass of I$_2$, $m$ is that of the HeI$_2$ complex, $I$ is the moment of inertia of the I$_{2}$ molecule, $L$ is its corresponding angular momentum, $r$ is the distance from the He atom to the center of mass of I$_{2}$, $s$ is the bond length for I$_{2}$, $\gamma$ is the angle between the I$_{2}$ diatom axis and the He to I$_{2}$ vector, and $p_{s},p_{r},p_{\gamma}$ are the conjugate momenta respectively. The description of the dynamical behavior that we carry out here assumes that the orbital angular momentum associated to rotation in the plane to be constant, and for simplicity we set $L = 0$. Therefore the Hamiltonian has the form:
\begin{equation}
   H = \dfrac{p_{s}^{2}}{2\mu} + \dfrac{p_{r}^{2}}{2m} + \dfrac{p_{\gamma}^{2}}{2I}+ V(s,r,\gamma)
\end{equation}
Moreover we assume that the bond length $s$ is fixed at its equilibrium configuration, i.e. $s = s_e$ and $p_s = 0$. A geometrical representation of the HeI$_2$ complex and the relevant DoF in the model is shown below in Fig. \ref{fig:hei2_molecule}.

\begin{figure}[htbp]
	\begin{center}
		\includegraphics[scale=0.45]{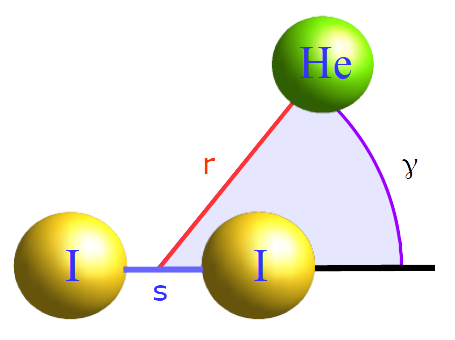} 
	\end{center}
	\caption{Degrees of freedom $(r,\gamma)$ for the HeI$_2$ van der Waals complex. The angle is measured in the counter-clockwise direction from the reference line in black. The variable $s$ represents the bond length between the atoms of the iodine molecule.}
	\label{fig:hei2_molecule}
\end{figure}

\begin{figure}[htbp]
	\begin{center}
		A)\includegraphics[scale=0.3]{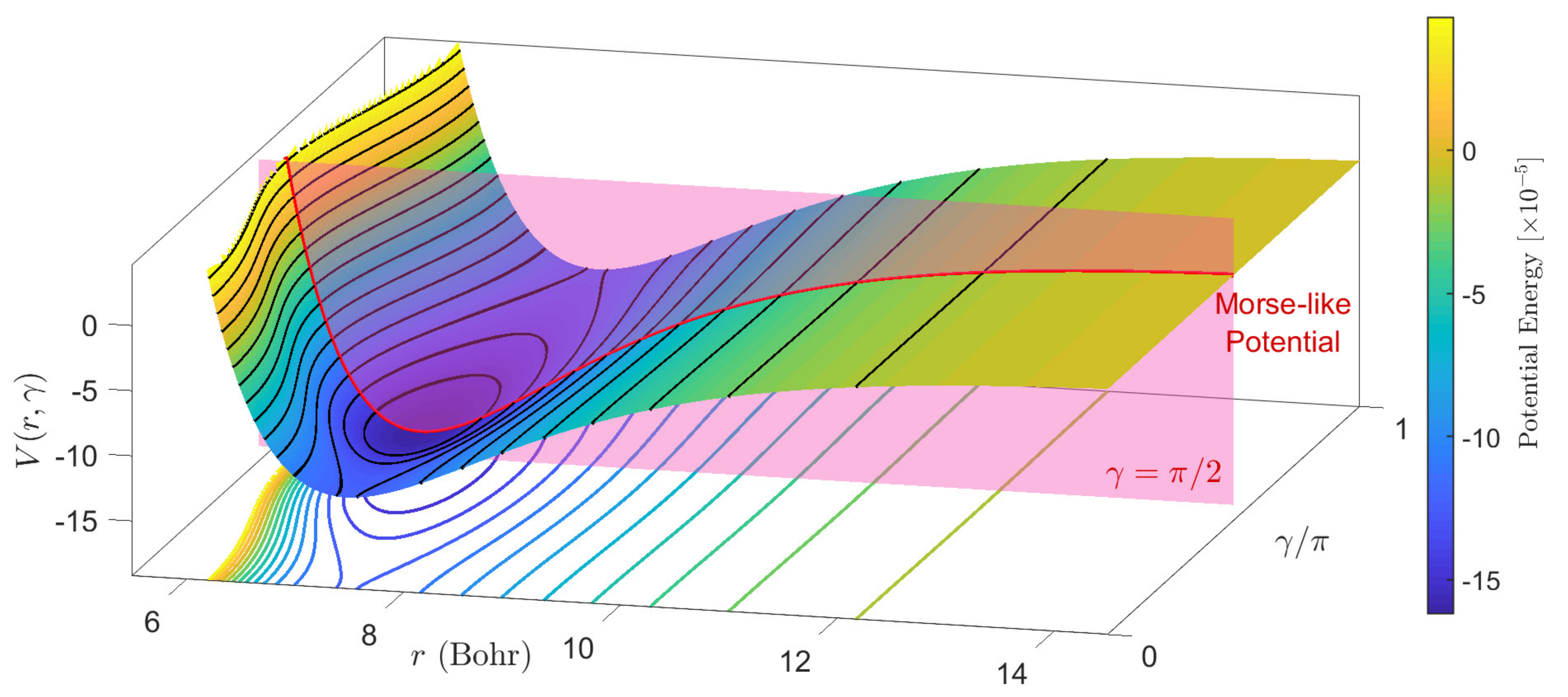} 
		B)\includegraphics[scale=0.31]{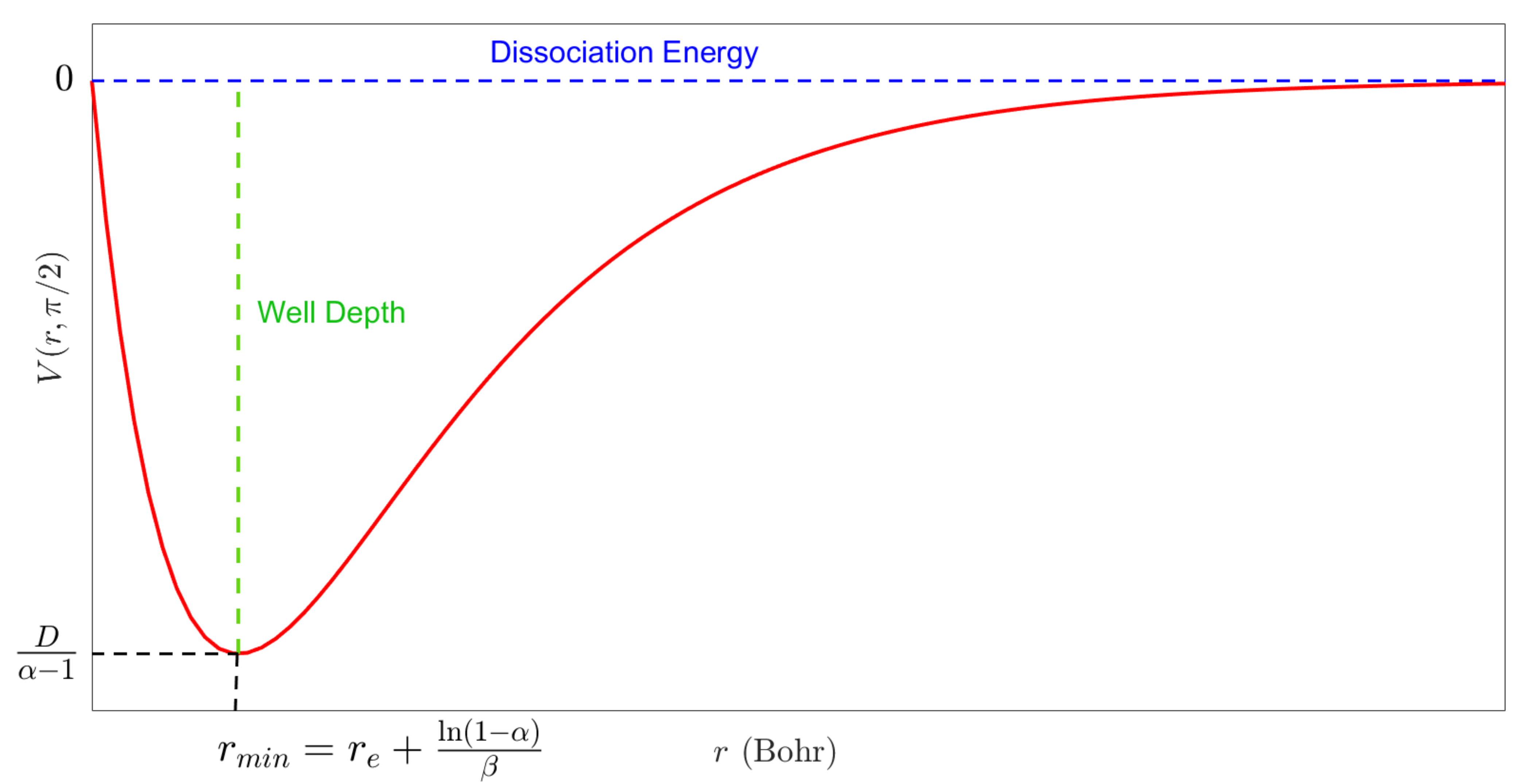}
	\end{center}
	\caption{A) van der Waals potential energy surface described in Eq. \eqref{eq:vdwpot}. B) Restricted to the plane $\gamma = \pi/2$, the resulting one-dimensional potential energy function resembles that of a classical Morse oscillator.}
	\label{fig:VdW_PES}
\end{figure}

The resulting Hamiltonian with 2 DoF has the form:
\begin{equation}
H = \dfrac{p_{r}^{2}}{2m} + \dfrac{p_{\gamma}^{2}}{2I} + V(s_e,r,\gamma)
 \label{eq:ham_HeI2_simp}
\end{equation}
and potential energy surface (PES) is known in the literature as the van der Waals (vdW) potential, and is described by a function that couples rotational and vibrational motion of the form:
\begin{equation}
V(r,\gamma) \equiv V(s_e,r,\gamma) = D \left(\left(1+\alpha\cos(2\gamma)\right) e^{-2\beta(r-r_{e})} - 2 \, e^{-\beta(r-r_{e})} \right)
\label{eq:vdwpot}
\end{equation}
where $D$ is the potential well depth (the dissociation energy), $\beta$ is the Morse parameter, $\gamma$ represents the DoF associated to the bending motion, $r$ is the stretching DoF, $r_{e}$ is the equilibrium internuclear distance (bond length), and $0 \leq \alpha < 1$ is the coupling parameter between the DoF of the system. Notice that if the DoF are uncoupled ($\alpha = 0$), the PES reduces to that of a Morse oscillator describing the intramolecular force between He and I$_2$ in the radial direction, as if it were a diatomic molecule:
\begin{equation}
V(r) = D \left(e^{-2\beta(r-r_{e})} - 2 \, e^{-\beta(r-r_{e})} \right) = D\left(1-e^{-\beta (r-r_{e})}\right)^{2} - D 
\label{morsepot}
\end{equation}
The physical parameters of the vdW potential are given by:
\begin{equation} 
\alpha = \dfrac{B/W}{2-(B/W)} \quad,\quad D = W(1-\alpha) \quad,\quad r_{e} = r_{min} - \dfrac{\ln(1-\alpha)}{\beta}
\label{model_params}
\end{equation}
and for the HeI$_2$ complex we have used the physical values in Table \ref{tab:tab1}. In Fig. \ref{fig:VdW_PES} A we show the energy landscape of the PES given in Eq. \eqref{eq:vdwpot}, and Fig. \ref{fig:VdW_PES} B presents the one-dimensional Morse-like potential energy function that results from the intersection with the plane $\gamma = \pi/2$.

\begin{table}[htbp]	
	\begin{tabular}{| c | c |}
		\hline
		\hspace{.3cm} Physical Quantity \hspace{.3cm} & \hspace{.3cm} \text{Description} \hspace{.3cm} \\
		\hline\hline
		$W = 1.62\times 10^{-4}$ & van der Waals Well Depth \\
		\hline
		$B = 4\times 10^{-5}$ & Height of Barrier to Internal Rotation \\
		\hline
		$r_{min} = 7$ & He-I Bond Distance at Potential Minimum \\
		\hline
		$\beta = 0.6033$ & Morse Parameter \\		
		\hline
		$m = 7069.4$ & Reduced Mass of HeI$_2$ \\		
		\hline
		$I = 3782220.8$ & Moment of Inertia of I$_2$ \\	
		\hline
	\end{tabular} 
	\caption{Physical parameters for the HeI$_{2}$ van der Waals complex, measured in atomic units.} 
	\label{tab:tab1} 
\end{table}

The dynamics of the HeI$_2$ complex is governed by Hamilton's equations:
\begin{equation}
\begin{cases}
\dot{r} = \dfrac{\partial H}{\partial p_r} = \dfrac{p_r}{m} \\[.4cm]
\dot{\gamma} = \dfrac{\partial H}{\partial p_\gamma} = \dfrac{p_\gamma}{I} \\[.4cm]
\dot{p}_{r} = -\dfrac{\partial H}{\partial r} = -\dfrac{\partial V}{\partial r} = -2 \beta D \left[1 - \left(1 + \alpha \cos (2\gamma )\right) e^{-\beta (r-r_e)}\right] e^{-\beta (r-r_e)}  \\[.4cm]
\dot{p}_{\gamma} = -\dfrac{\partial H}{\partial \gamma} = -\dfrac{\partial V}{\partial \gamma} = 2 \alpha D \sin (2\gamma) \, e^{-2\beta (r-r_e)} 
\end{cases}
\label{eq:hameq_2dof}
\end{equation}
so that the dynamics of the system takes place in a 4D phase space, and due to energy conservation, motion is restricted to a three-dimensional energy hypersurface. By examining the equations of motion it is easy to check that the following isoenergetic subspaces are invariant:
\begin{equation}
\begin{split}
\Sigma_1(H_0) & = \left\{(r,\gamma,p_r,p_\gamma) \in \mathbb{R}^{+} \times \mathbb{T} \times \mathbb{R}^2 \;\big|\; \gamma = 0 \, ,\, p_{\gamma} = 0 \, , \, H = H_0\right\} \\[.3cm]
\Sigma_2(H_0) & = \left\{(r,p_r,\gamma,p_\gamma) \in \mathbb{R}^{+} \times \mathbb{T} \times \mathbb{R}^2 \;\big|\; \gamma = \pi/2 \, ,\, p_{\gamma} = 0 \, , \, H = H_0\right\}
\end{split}
\label{eq:inv_subs}
\end{equation}
where $\mathbb{R}^{+}$ represents the non-negative real numbers and $\mathbb{T} = \mathbb{R} / \left(2\pi\mathbb{Z}\right)$ is a one-dimensional torus (or circle). The invariance condition means that if we take initial conditions on any of these subspaces and evolve them in time, they will remain in that subspace forever along their trajectory.

At this point, we are ready to derive the 4D symplectic map model that is the aim of this work from the 2 DoF Hamiltonian in Eq. \eqref{eq:ham_HeI2_simp}. To do so, we consider that the HeI$_2$ complex is periodically kicked with a period $T$. For small values of the kicking period the Hamiltonian mapping is a symplectic trajectory integrator. An interpretation of a symplectic 2N- dimensional map can be a Poincar\'e return map for a periodically driven N DoF. It is important to notice that the kicking period $T$ should be equal to the vibrational period of the $I_{2}$ molecule in its initial vibrational state. In this work we consider different values of the kicking period, between 1000 to 8000, for the analysis of the dynamics. This gives the time-dependent Hamiltonian:
\begin{equation}
H(r,\gamma,p_r,p_{\gamma}) = \dfrac{p_{r}^{2}}{2m} + \dfrac{p_{\gamma}^{2}}{2I} + TV(r,\gamma) \sum_{n \in \mathbb{Z}} \delta (t-nT)
\label{eq:ham_kick}
\end{equation}
Between kicks, the system is governed by Hamilton's equations:
\begin{equation}
\begin{cases}
\dot{r} = \dfrac{p_r}{m} \\[.4cm]
\dot{\gamma} = \dfrac{p_\gamma}{I}  \\[.3cm]
\dot{p}_{r} = - T \dfrac{\partial V}{\partial r} \displaystyle{\sum_{n \in \mathbb{Z}} \delta (t-nT)}  \\[.5cm]
\dot{p}_{\gamma} = - T \dfrac{\partial V}{\partial \gamma}\displaystyle{\sum_{n \in \mathbb{Z}} \delta (t-nT)} 
\end{cases}
\;\; \Leftrightarrow \quad 
\begin{cases}
dr = \dfrac{p_r}{m} \, dt \\[.4cm]
d\gamma = \dfrac{p_\gamma}{I} \, dt \\[.3cm]
d p_r = - T \dfrac{\partial V}{\partial r} \displaystyle{\sum_{n \in \mathbb{Z}} \delta (t-nT)} \, dt \\[.5cm]
d p_{\gamma} = - T \dfrac{\partial V}{\partial \gamma}\displaystyle{\sum_{n \in \mathbb{Z}} \delta (t-nT)} \, dt
\end{cases}
\label{eq:hameq_kick}
\end{equation}
By integrating in time over one period $T$ we obtain the following 4D symplectic map:
\begin{equation}
\begin{cases}
r_{n+1} = r_{n} + \dfrac{T}{m} \, p_{r,n+1} = r_{n} + \dfrac{T}{m} \, p_{r,n} - \dfrac{T^2}{m} \, \dfrac{\partial V}{\partial r} \left(r_{n},\gamma_n\right) \\[.4cm]
\gamma_{n+1} = \gamma_{n} + \dfrac{T}{I} \, p_{\gamma,n+1} = \gamma_{n} + \dfrac{T}{I} \, p_{\gamma,n} - \dfrac{T^2}{I} \, \dfrac{\partial V}{\partial \gamma} \left(r_{n},\gamma_n\right) \mod (2\pi) \\[.4cm]
p_{r,n+1} = p_{r,n} - T \, \dfrac{\partial V}{\partial r} \left(r_{n},\gamma_n\right) \\[.4cm]
p_{\gamma,n+1} = p_{\gamma,n} - T \, \dfrac{\partial V}{\partial \gamma} \left(r_{n},\gamma_n\right) \mod (2\pi)
\end{cases}
\; , \quad n \in \mathbb{N} \cup \lbrace0\rbrace
\label{eq:map4D}
\end{equation}
Note that the phase space $(r,\gamma,p_{r},p_{\gamma})$ is periodic in $p_{\gamma}$ in addition to the periodicity in the angle $\gamma$ that we have mentioned previously and therefore the subspace $(\gamma,p_{\gamma})$ is a 2 torus. 
It is important to highlight that this 4D map also has the same invariant subspaces as those we found for the 2 DoF continuous Hamiltonian system, which are described in Eq. \eqref{eq:inv_subs}. 
To finish this section, notice that the 4D map in Eq. \eqref{eq:map4D} is invertible and that the inverse map is given by:
\begin{equation}
\begin{cases}
r_{n} = r_{n+1} - \dfrac{T}{m} \, p_{r,n+1} \\[.4cm]
\gamma_{n} = \gamma_{n+1} - \dfrac{T}{I} \, p_{\gamma,n+1}  \mod (2\pi) \\[.4cm]
p_{r,n} = p_{r,n+1} + T \, \dfrac{\partial V}{\partial r} \left(r_{n},\gamma_n\right) \\[.4cm]
p_{\gamma,n} = p_{\gamma,n+1} + T \, \dfrac{\partial V}{\partial \gamma} \left(r_{n},\gamma_n\right) \mod (2\pi)
\end{cases}
\; , \quad n \in \mathbb{Z}^{-}
\label{eq:map4D_inv}
\end{equation}
This is important because we need both, the 4D map and its inverse, in order to calculate Lagrangian descriptors for the analysis of the resulting high-dimensional phase space of the system. 

\section{Results}
\label{sec:sec2}
In this section  we will study the ability of DLDs to reveal phase space structures  of a 4D symplectic map. Furthermore, we want to study the phase space  structures that  correspond to  intramolecular bonding,  dissociation and  predissociation.

In the case of intramolecular bonding the solutions of our model correspond to a  bounded distance between the atom of $He$ and the center of  mass of the two atoms of $I_2$. This is the reason that for this case we are looking for structures that are invariant and for the points stay there forever. On the contrary, in the case of  dissociation the solutions of our model correspond to an  unbounded distance (goes to infinity) between the atom of $He$ and the center mass of the two atoms of $I_2$. Finally, in the case of  predissociation, the complex $HeI_2$ typically climbs into a state above the dissociation threshold for a Van der Waals complex. Nevertheless, it does not fly apart immediately, since some time is needed to redistribute the energy "intramolecularly" (see \cite{gentry1984vibrationally,janda1985predissociation}). In this case,  the solutions of our model correspond to a  distance of  the atom of $He$ from the center mass of the two atoms of $I_2$   that is  bounded  for long discrete time (many iterations) until it goes to infinity. For this reason,  we are looking for structures  which give rise to this behavior.

In the first subsection, we study the phase space structures in  a special case of the 4D mapping model (a 2D mapping model). In the second subsection, we extend our analysis in the full  4D mapping model.  

\subsection{Analysis of the 2D Area-Preserving Map}

We begin this section by analyzing the phase space of the one DoF Hamiltonian system that results from restricting motion to the invariant subspace $\Sigma_{2}$ in Eq. \eqref{eq:inv_subs} where the potential energy function reduces to:
\begin{equation}
V(r) \equiv V(r,\pi/2) = D \left(\left(1-\alpha\right) e^{-2\beta(r-r_{e})}-2 \, e^{-\beta(r-r_{e})}\right)
\end{equation}
The integrable Hamiltonian is:
\begin{equation}
	\widetilde{H}(r,p_r) \equiv H(r,p_r,\pi/2,0) = \dfrac{p^{2}_{r}}{2m} + V(r) 
\label{ham_1D}
\end{equation}
and Hamilton's equations of motion are given by:
\begin{equation}
\begin{cases}
\dot{r} = \dfrac{p_{r}}{m} \\[.3cm]
\dot{p}_{r} = -2 \beta D \left[1 - \left(1 - \alpha\right) e^{-\beta (r-r_e)}\right] e^{-\beta (r-r_e)}
\end{cases}
\label{hameq_1D}
\end{equation}
Recall that this two-dimensional dynamical system has two equilibrium points located at:
\begin{equation}
(r^{\ast},p_{r}^{\ast}) = \left(r_{e} + \dfrac{\ln(1-\alpha)}{\beta},0\right) \;\; , \;\; (\infty,0)
\end{equation}
and that the equilibrium point at infinity is parabolic, and the other one (near the bottom of the potential well) is a center (stable fixed point). The separatrix of this system, which is a homoclinic trajectory comprised of the stable and unstable manifolds of the parabolic equilibrium point at infinity, is given by the curve with the same energy as that of the parabolic point:
\begin{equation}\label{sep}
\widetilde{H}(\infty,0) = 0 = \dfrac{p^{2}_{r}}{2m} + V(r)
\end{equation}

\begin{figure}[htbp]
	\begin{center}
		A)\includegraphics[scale=0.3]{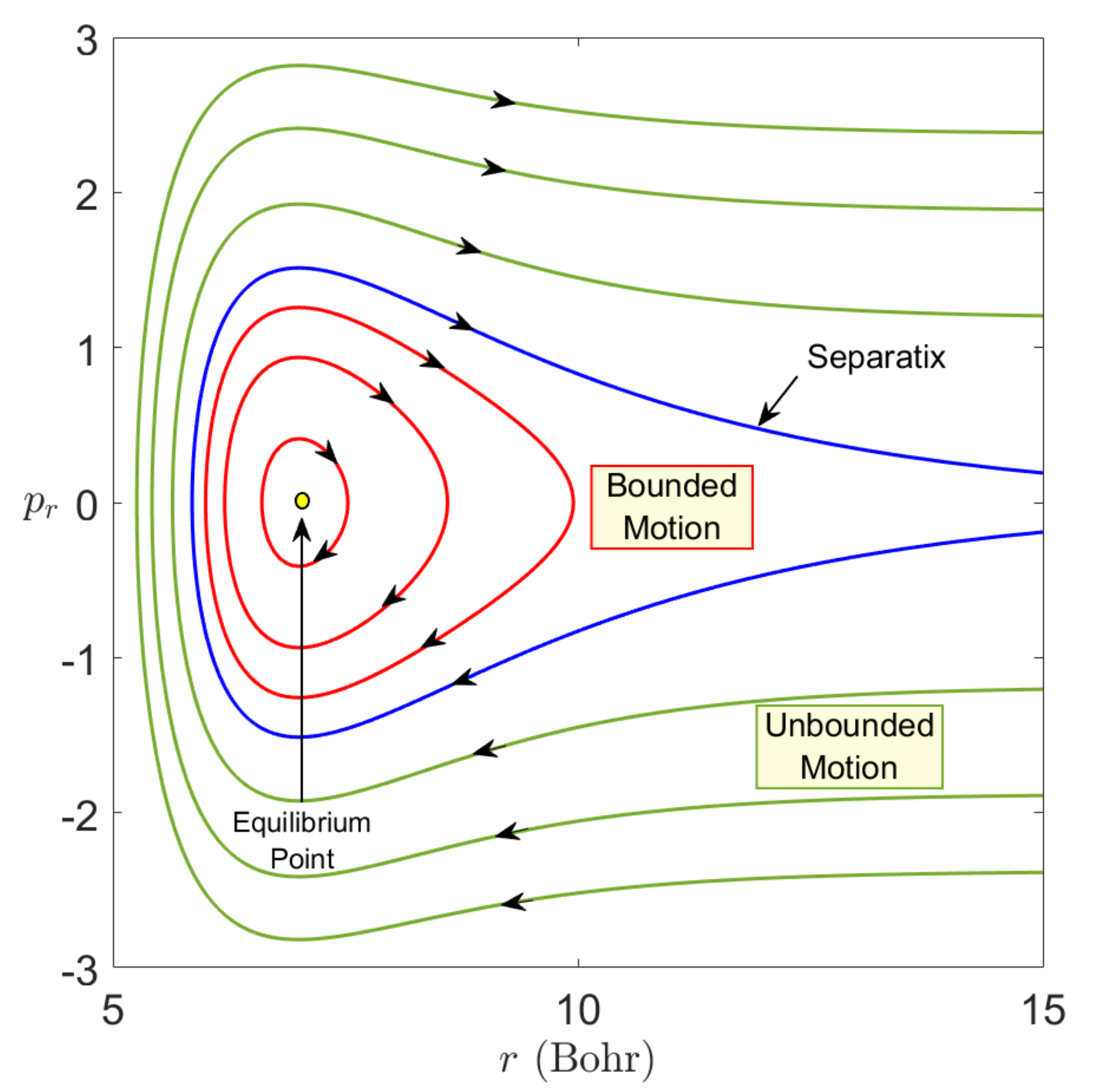} 
		B)\includegraphics[scale=0.3]{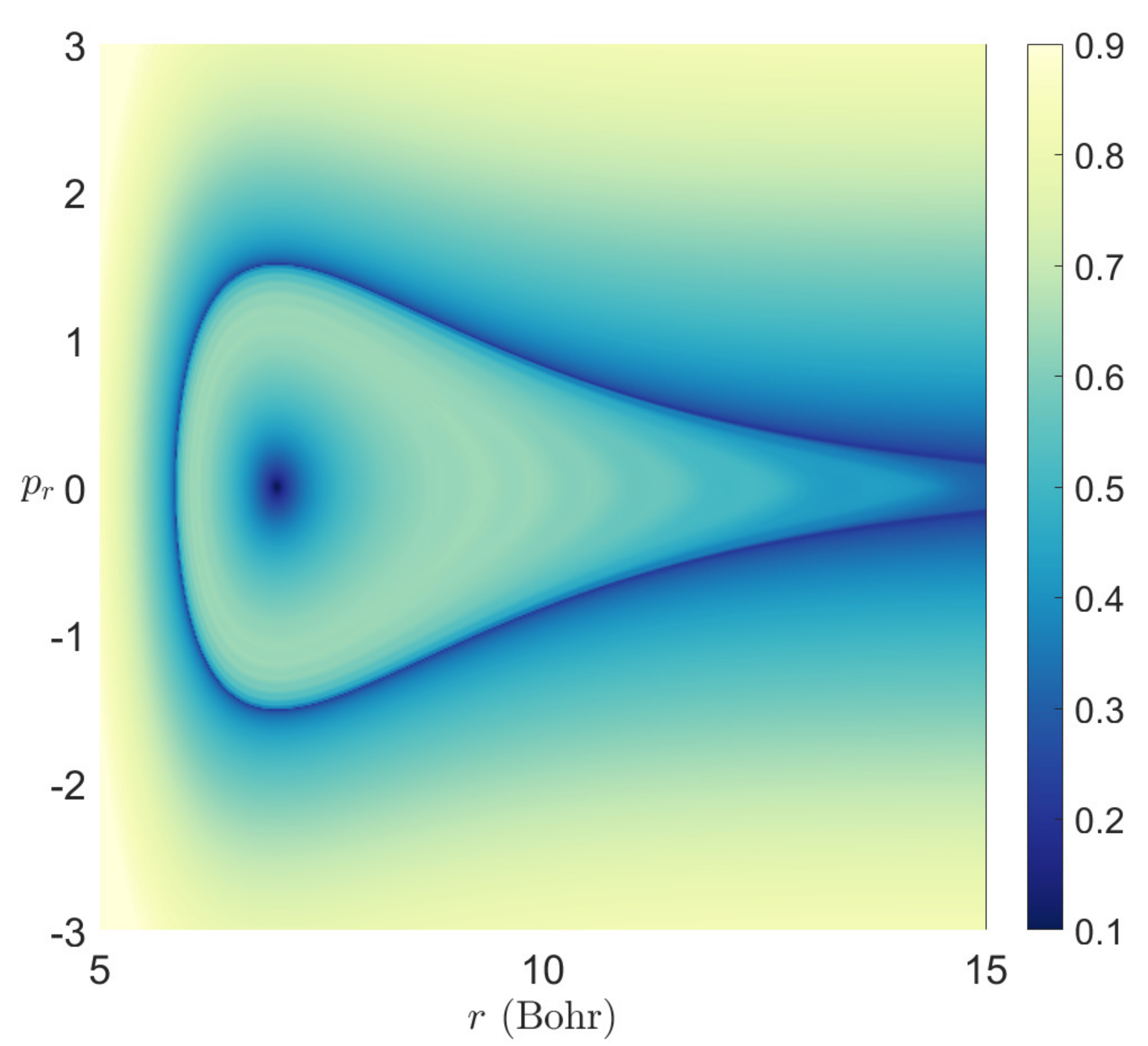}
	\end{center}
	\caption{A) Phase portrait for the dynamical system described in Eq. \eqref{hameq_1D}. B) Phase space structures revealed by applying Lagrangian descriptors with $p = 1$ and using an integration time $\tau = 500000$.}
	\label{fig:phase_port}
\end{figure}

In panel A of Fig. \ref{fig:phase_port} we show the phase space for the van der Waals potential for $\gamma = \pi/2$ where the separatrix (\ref{sep}) is depicted in blue. In the interior of the separatrix are the periodic orbits (bounded orbits, bounded motion of the complex) whereas in the exterior we can distinguish the scattering orbits (unbounded orbits, dissociation or break up of the complex) in green. In panel B of Fig. \ref{fig:phase_port} we have calculated LDs. The method reveals the phase space structures of the integrable system. The separatrix Eq. (\ref{sep}) in blue encloses all the bounded trajectories. Usually the method of LDs reveals the phase portrait of a system for a small integration time $\tau$. In this case the trajectories integrate very slowly and this is the reason that the value of $\tau$ here is very large.

Restricted to any of the invariant subspaces in Eq. \eqref{eq:inv_subs}, the 4D map reduces to an area-preserving 2D map in the $r$-$p_r$ subspace with the form:
\begin{equation}
\begin{cases}
r_{n+1} = r_{n} + \dfrac{T}{m} \, p_{r,n+1} \\[.3cm]
p_{r,n+1} = p_{r,n} - T \, \dfrac{\partial V}{\partial r}(r_n,\gamma^{\ast})
\end{cases}
\; , \quad n \in \mathbb{N} \cup \lbrace 0\rbrace
\label{2D_map}
\end{equation}
and the inverse mapping is:
\begin{equation}
\begin{cases}
r_{n} = r_{n+1} - \dfrac{T}{m} \, p_{r,n+1} \\[.3cm]
p_{r,n} = p_{r,n+1} + T \, \dfrac{\partial V}{\partial r}(r_n,\gamma^{\ast}) 
\end{cases}
\; , \quad n \in \mathbb{Z^{-}}
\label{inv_2D_map}
\end{equation}
where $\gamma^{\ast} = 0 \, , \, \pi/2$ depending on the subspace. The fixed points of this map in the plane $\gamma = \pi/2$ are:
\begin{equation}
\left(r_{e} + \dfrac{\ln(1-\alpha)}{\beta},0\right) \quad,\quad (\infty,0) 
\end{equation}

\begin{figure}[htbp]
	\begin{center}
		A)\includegraphics[scale=0.3]{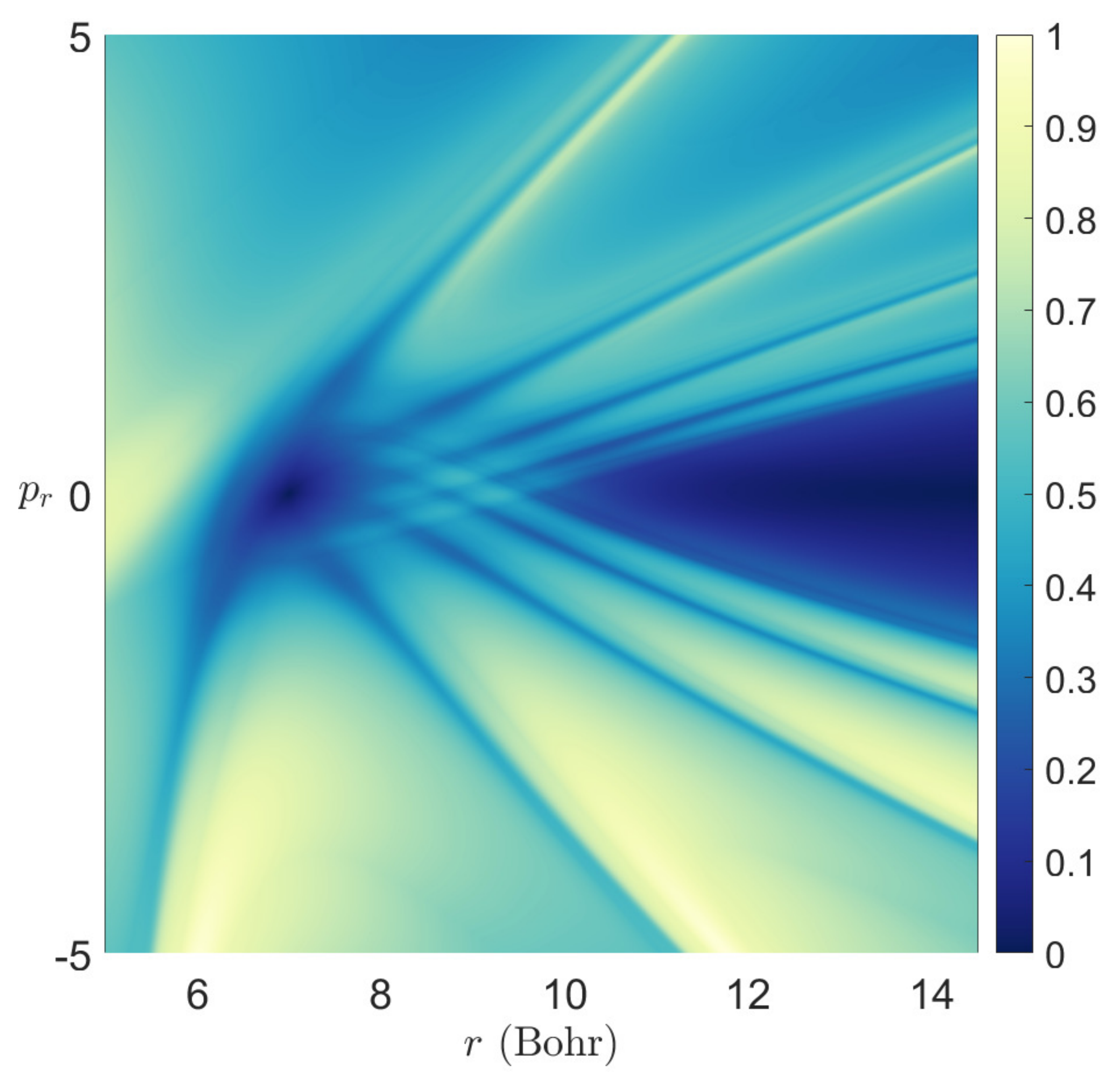} 
		B)\includegraphics[scale=0.3]{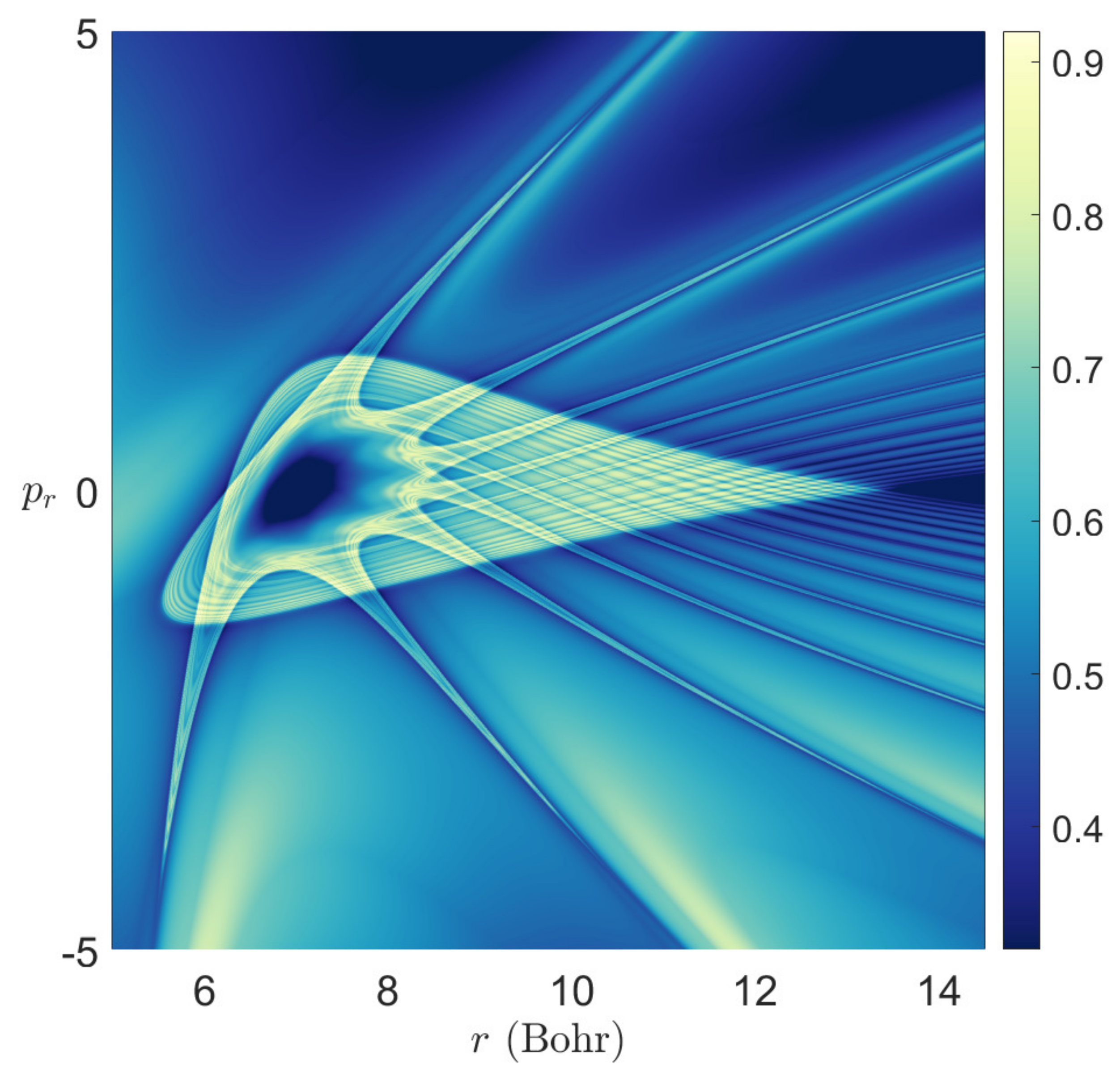}
	\end{center}
	\caption{Phase space revealed by applying Discrete Lagrangian descriptors ($p = 1$)  to the 2D map described in Eq. \eqref{2D_map} with a kicking period $T = 8000$ on the invariant subspace $\gamma = \pi / 2$ and $p_{\gamma} = 0$. A) $N = 5$ iterations; B) $N = 15$ iterations.}
	\label{fig:iter_complex}
\end{figure}

Now we will describe the dynamics in this model using the method of  Discrete Lagrangian Descriptors (DLD) as has been presented in \cite{carlos,GG2019b,GG2020a}. The method of DLD can reveal the geometry of the phase space with increasing complexity as the number of the iterations is increased. We illustrate this property in Fig. \ref{fig:iter_complex} where the period of the kick has been chosen as $T = 8000$. In Fig. \ref{fig:iter_complex} A we show the computation for $N = 5$ iterations, while panel B corresponds to $N = 15$ iterations. We notice that since $T \neq 0$ the homoclinic orbit breaks and therefore now the orbits that start in the phase space region of bounded motion can enter the region of the phase space corresponding to unbounded motion.

\begin{figure}[htbp]
	\begin{center}
		A)\includegraphics[scale=0.28]{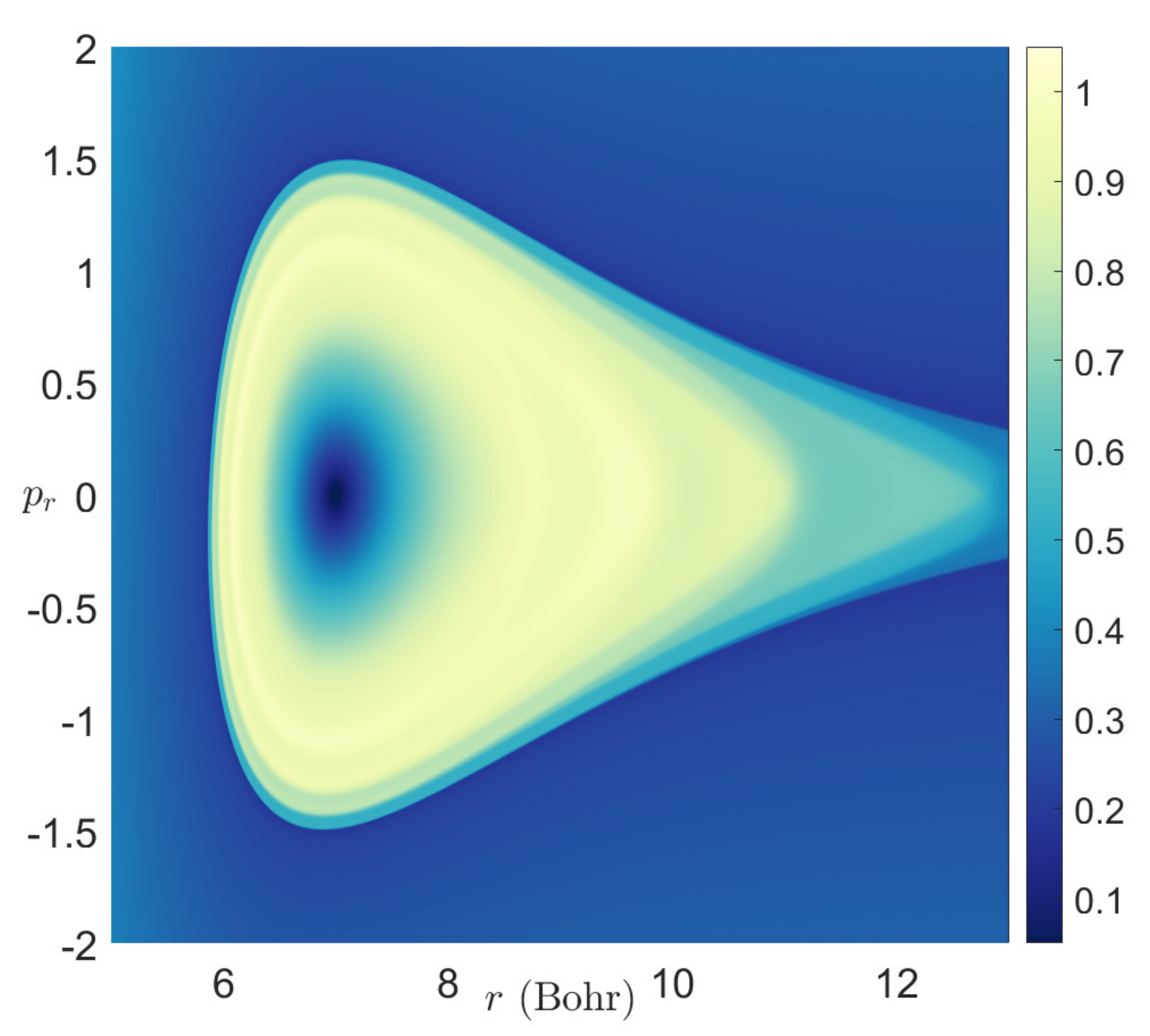} 
		B)\includegraphics[scale=0.275]{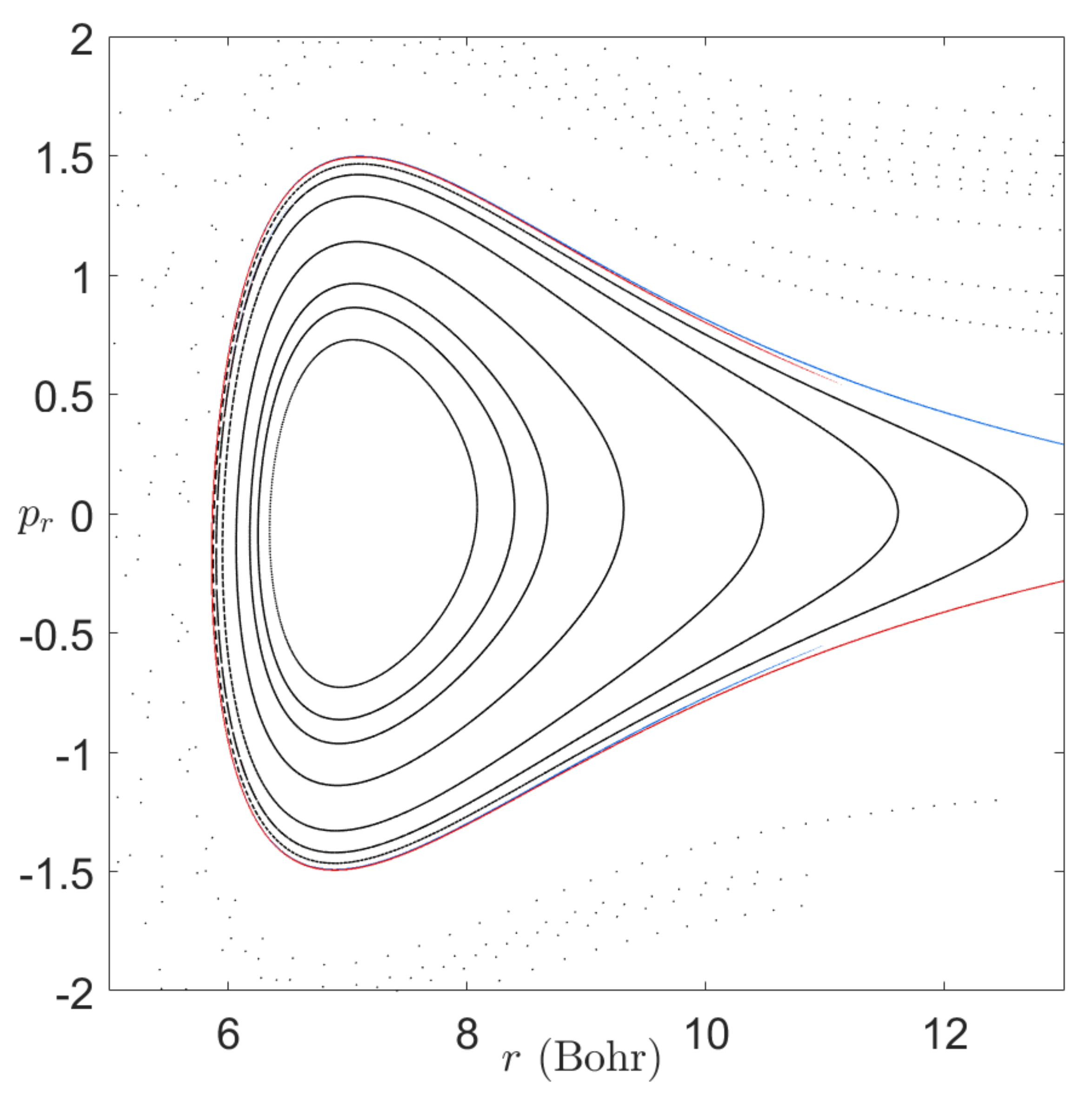}
		C)\includegraphics[scale=0.28]{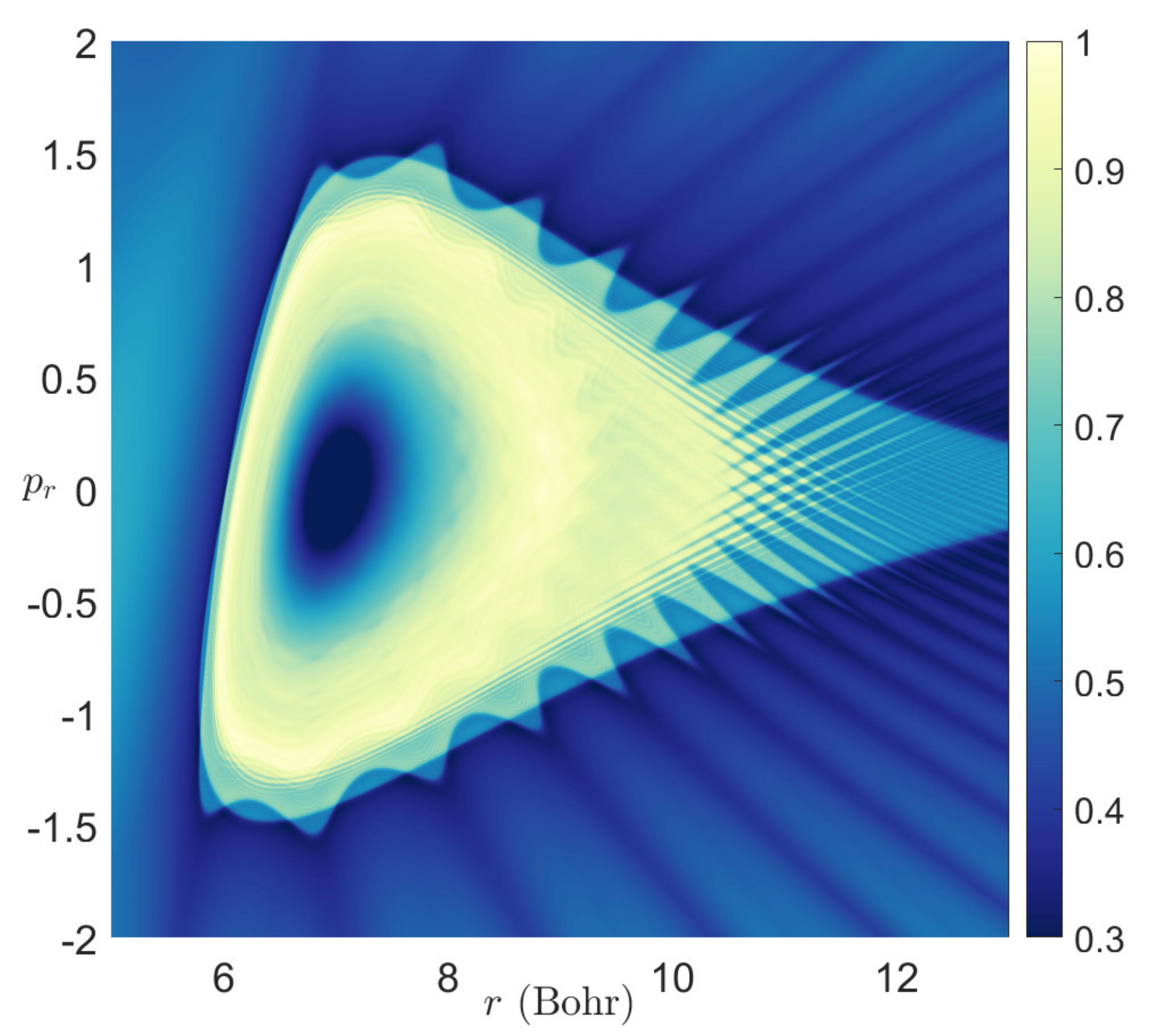}
		D)\includegraphics[scale=0.275]{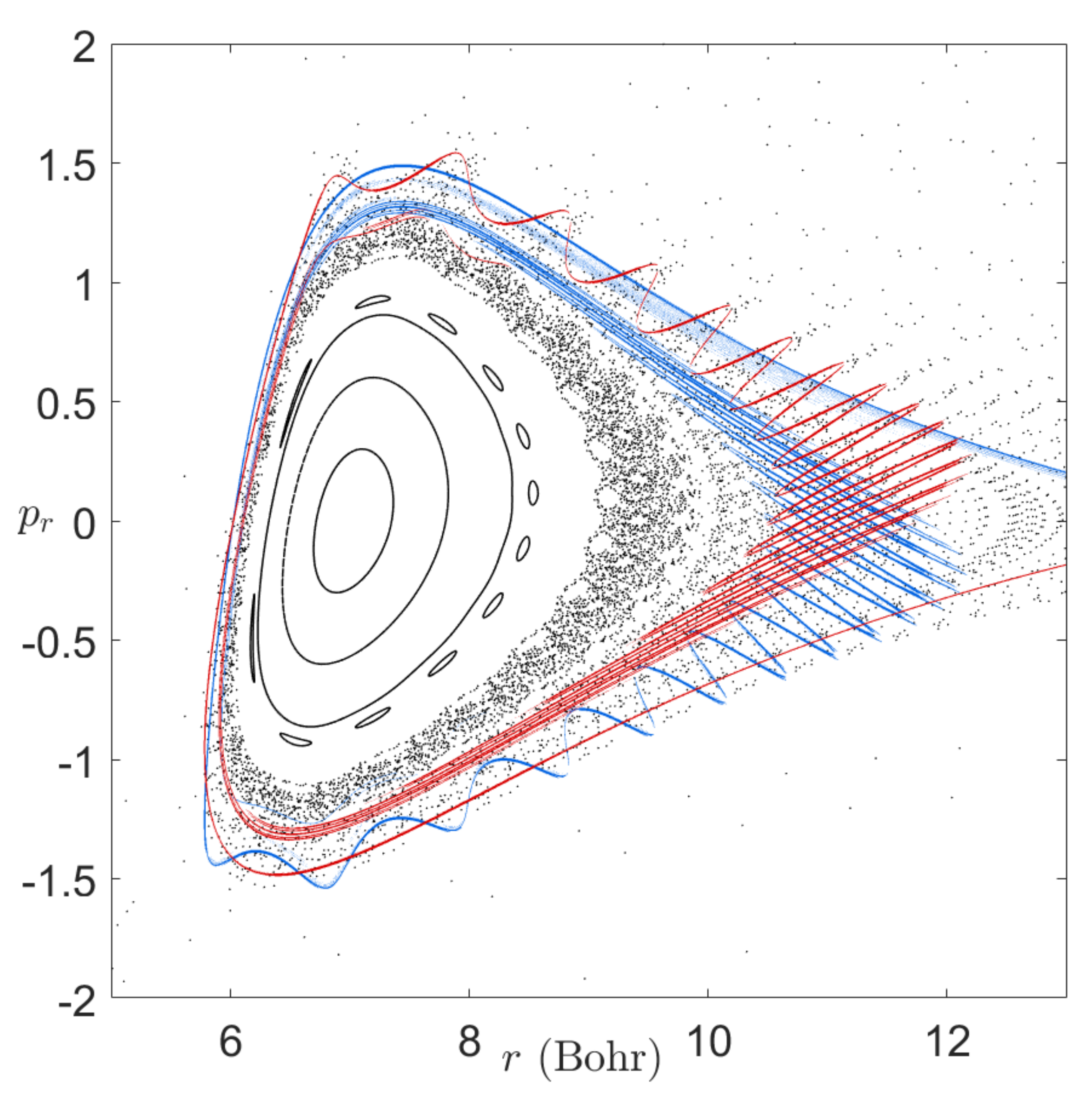}
		E)\includegraphics[scale=0.28]{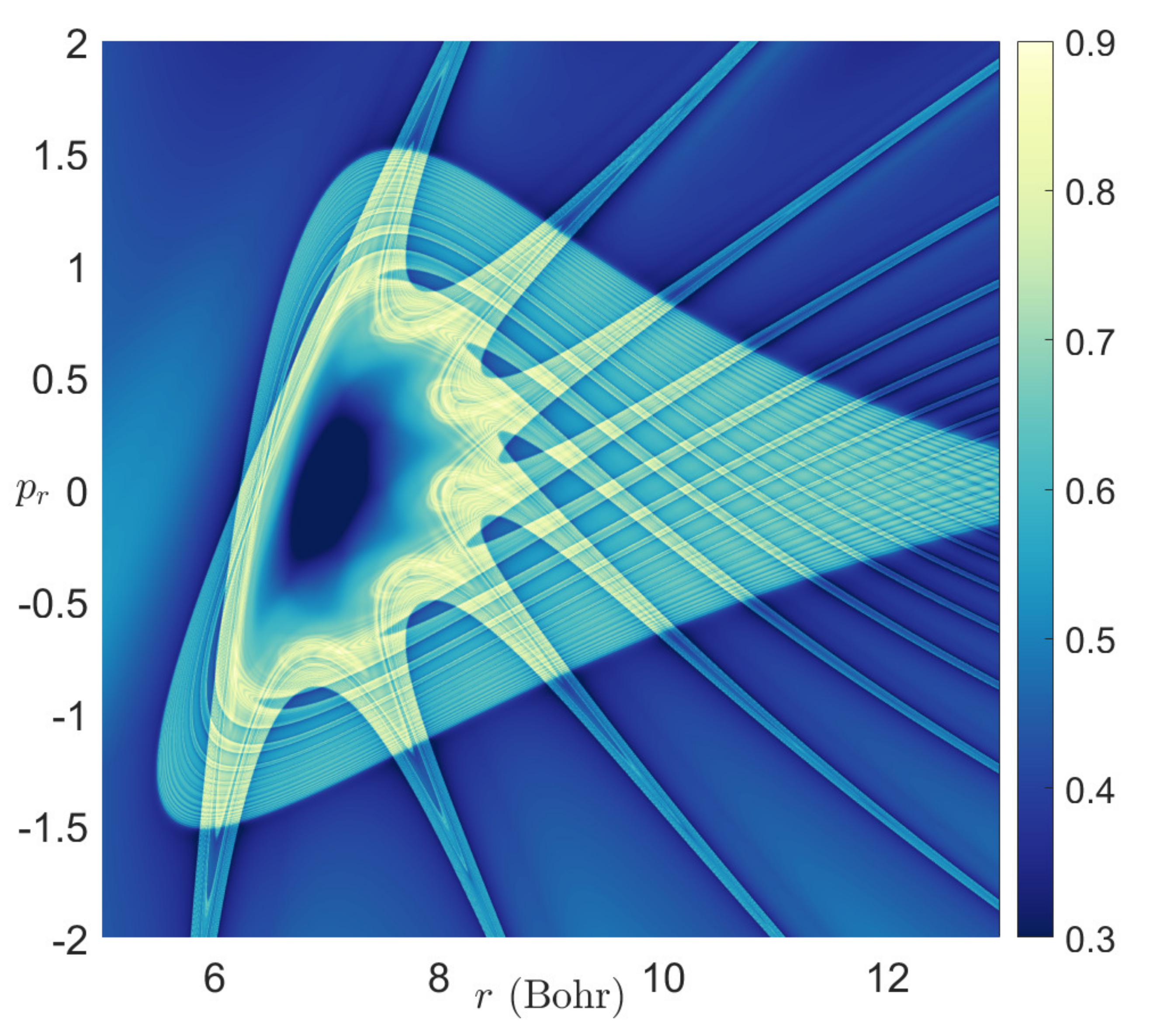}
		F)\includegraphics[scale=0.275]{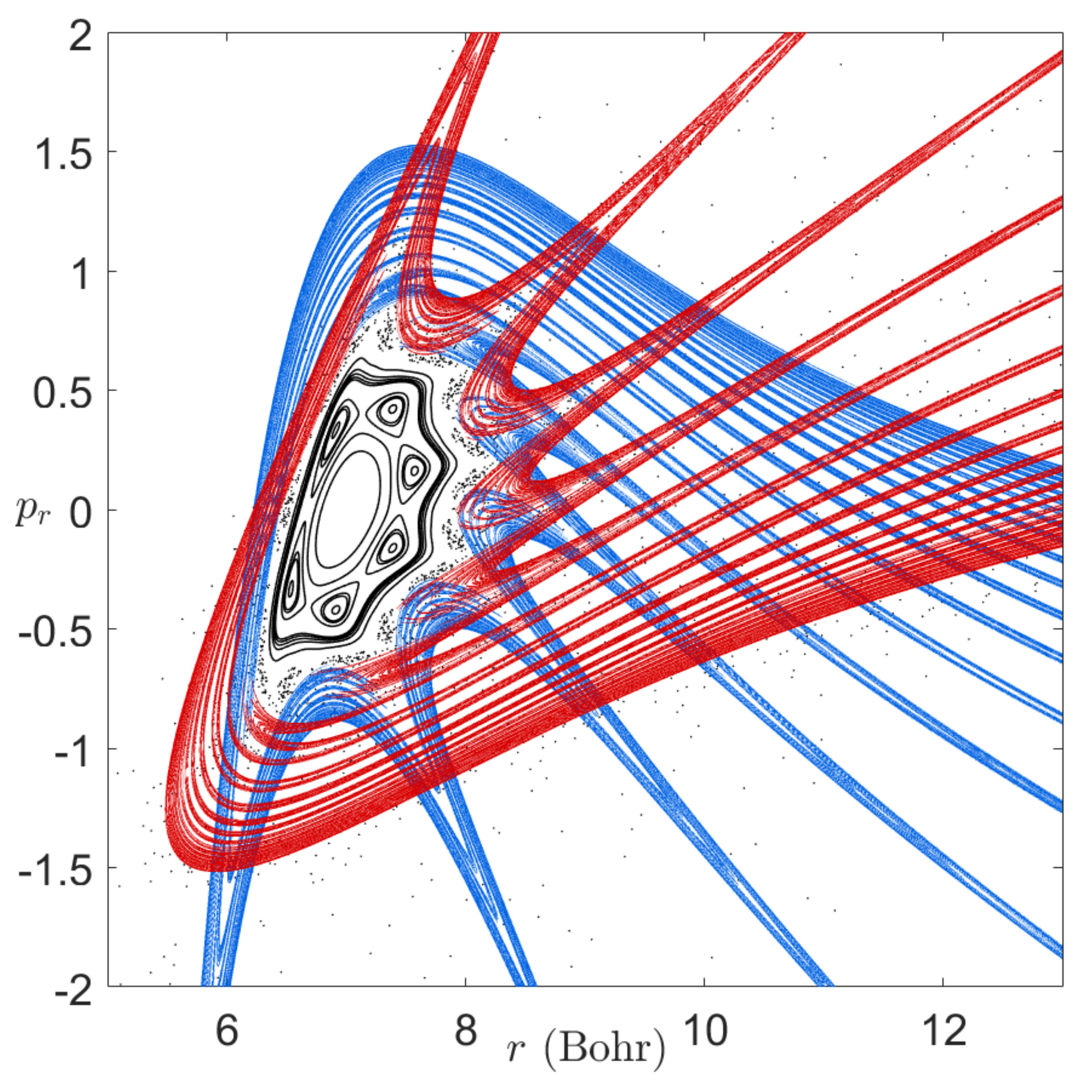}
	\end{center}
	\caption{Phase space revealed by applying DLDs and Poincar\'e maps to the 2D area-preserving map described in Eq. \eqref{2D_map} on the invariant subspace $\gamma = \pi / 2$ and $p_{\gamma} = 0$. On the left column, we display the DLD scalar field, and the right column shows the stable (blue) and unstable (red) manifolds extracted from the gradient of DLDs together with KAM tori obtained from a Poincar\'e map. The first row corresponds to a kicking period $T = 1000$ and $N = 300$ iterations, the second row is for $T = 5000$ and $N = 35$, and the last row uses $T = 8000$ and $N = 20$.}
	\label{fig:phaseSp_diffT}
\end{figure}

We study the phase space structures close to the equilibrium point $(r_{e}+ln(1-\alpha)/\beta,0)$. On the left column of Fig. \ref{fig:phaseSp_diffT} we applied the DLDs for different kicking periods. We notice that as the kicking period is increasing we can decrease the number of iterations in order to reveal the geometric structures. The right column of the same figure shows the Poincar{\'e} map, superimposed with the stable (blue) and unstable (red) manifolds that have been extracted from the gradient of DLDs. In particular in panel B where the kicking period is respectively small we can see the separatrix and we observe that the region around the stable fixed point is regular without any presence of chaos. Around the stable periodic orbits we see invariant curves that represent the KAM (Kolmogorov-Arnold-Moser,  see \cite{kolmogorov1954},\cite{arnold1963}, \cite{moser1962}) invariant tori in the Poincar{\'e} section. These curves are complete barriers to transport in the phase space. That means that any point starting inside the area bounded by a KAM torus will never get out. For higher kicking period we observe again many invariant curves around the stable fixed point (in the panels D and F of Fig. \ref{fig:phaseSp_diffT}). These invariant curves correspond to the intramolecular bonding state of the system because the points that correspond to these invariant curves will stay there for ever. Except  these invariant curves we see  a resonance zone with  many islands that is a sign of chaos. These islands are around a stable periodic orbit  with high order multiplicity. This periodic orbit  is  a stable periodic orbit with period 12  (in panel D)  or a stable periodic orbit  with  period 6 (in panel F). A sign of increasing chaos as we increase the value  $T$ from 1000 to 8000 is  the making of lobes with bigger oscillations (see the panels A, C and E  of Fig. \ref{fig:phaseSp_diffT} ). This is a clear manifestation of the increasing chaos.   

\begin{figure}[htbp]
	\begin{center}
		A)\includegraphics[scale=0.27]{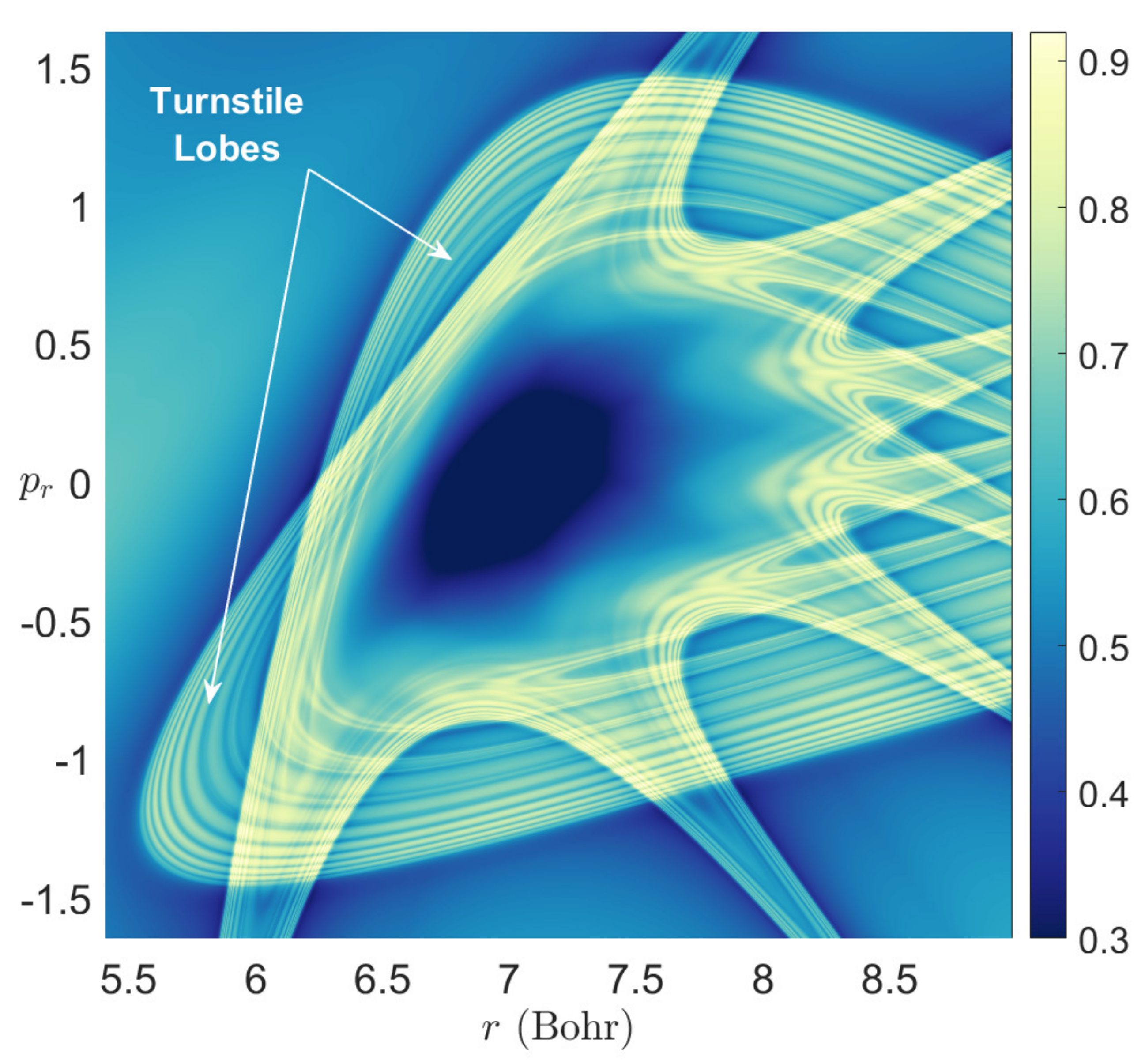} 
		B)\includegraphics[scale=0.27]{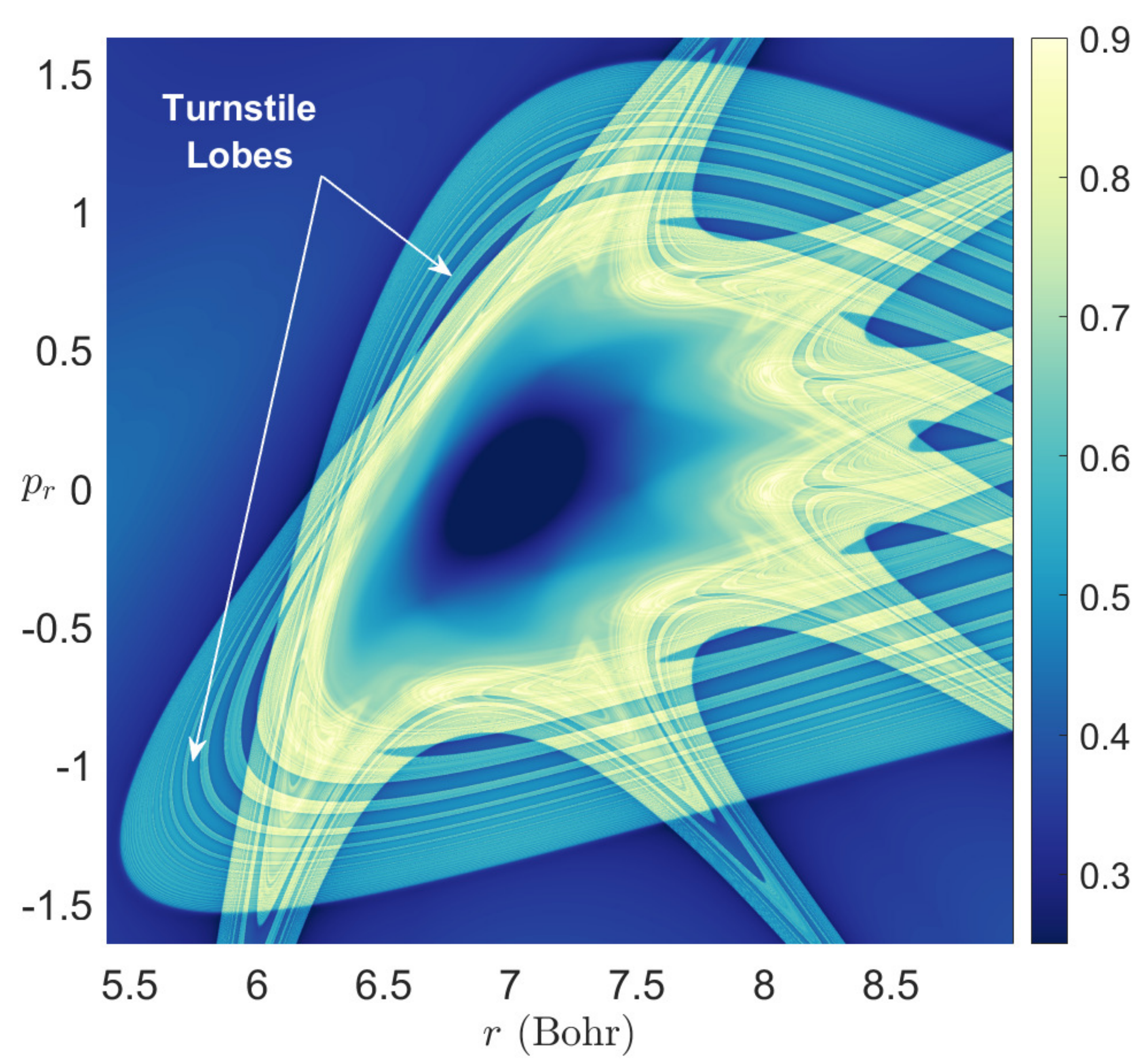} 
		C)\includegraphics[scale=0.27]{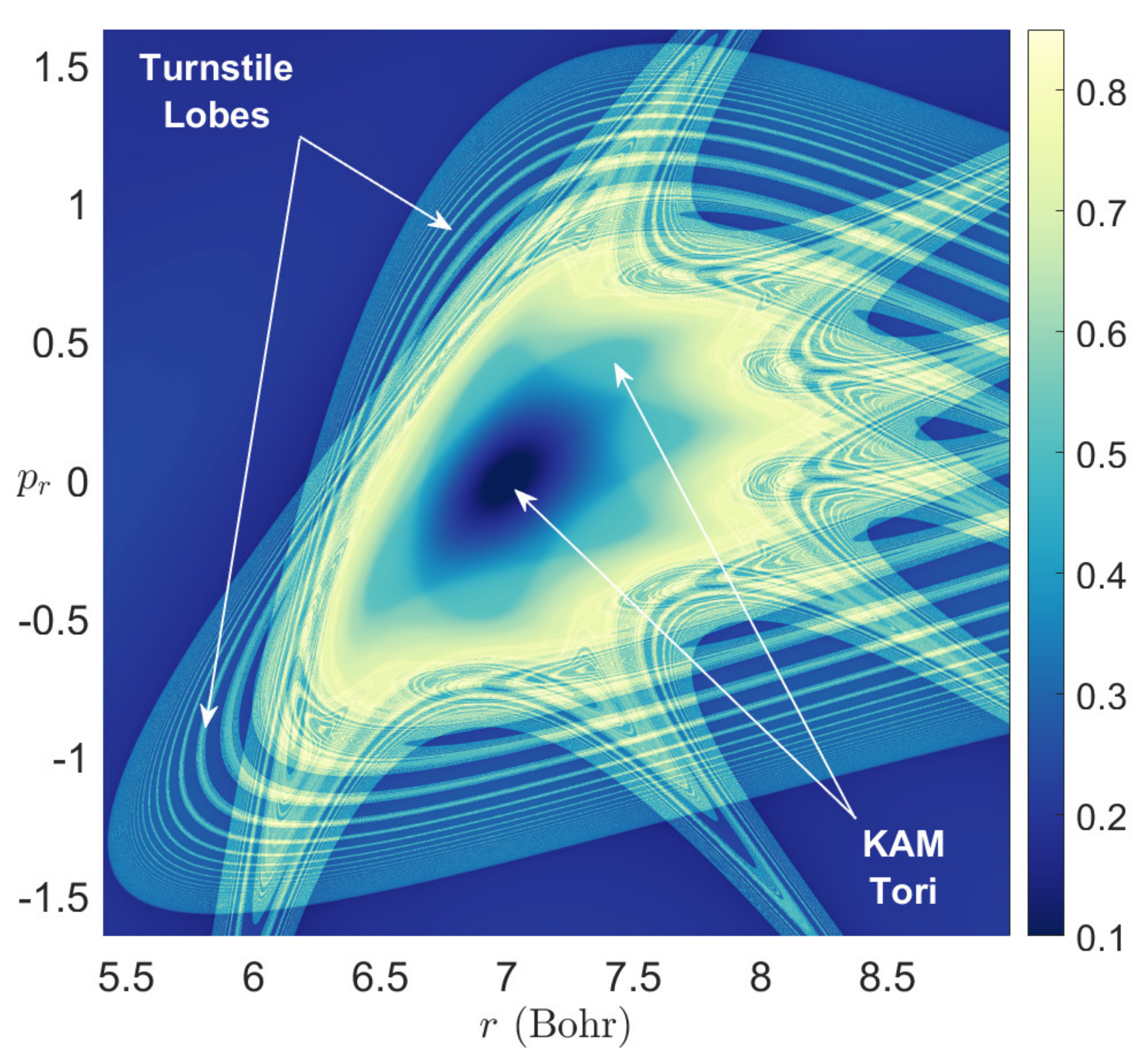} 
		D)\includegraphics[scale=0.27]{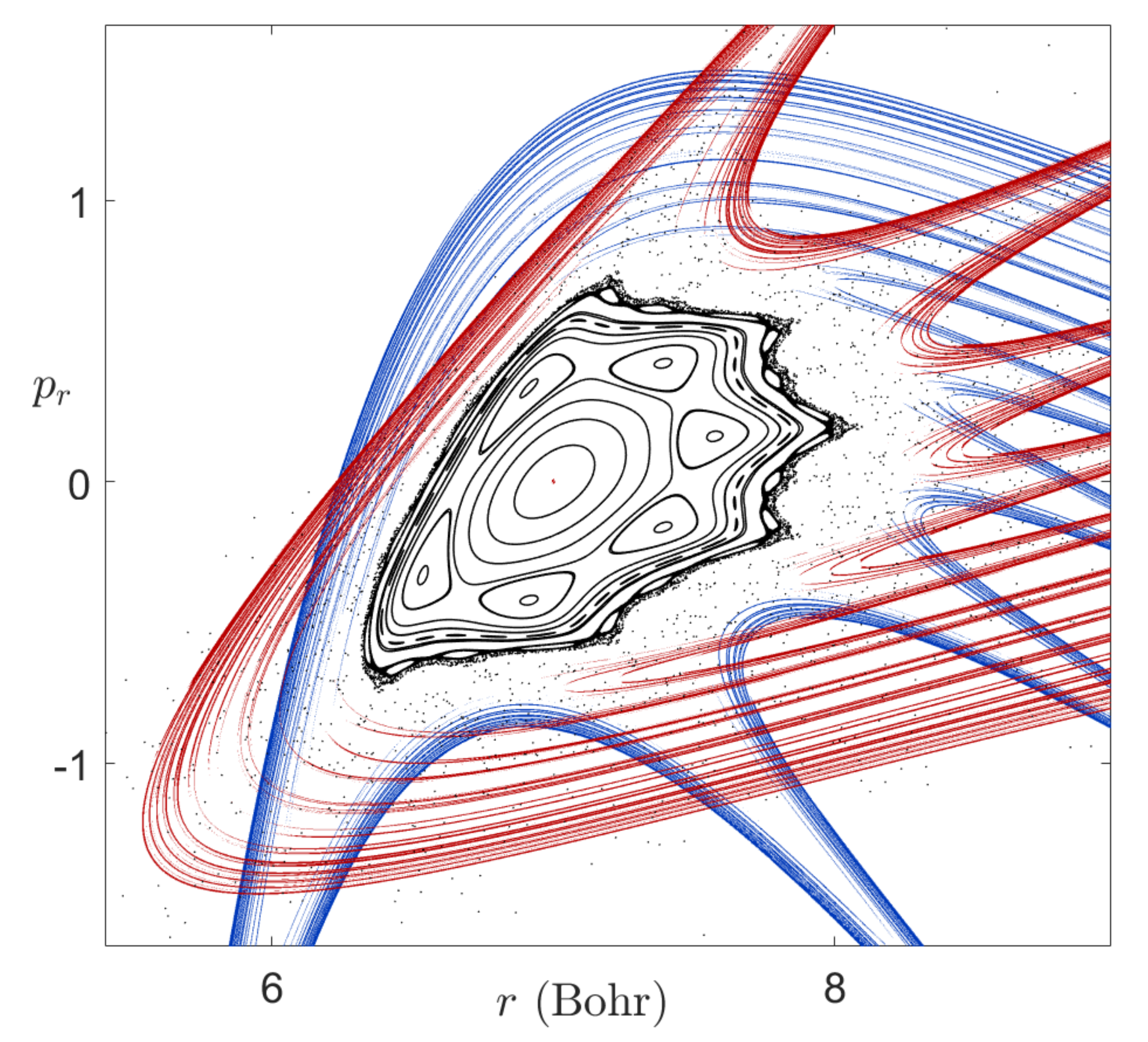}
		E)\includegraphics[scale=0.28]{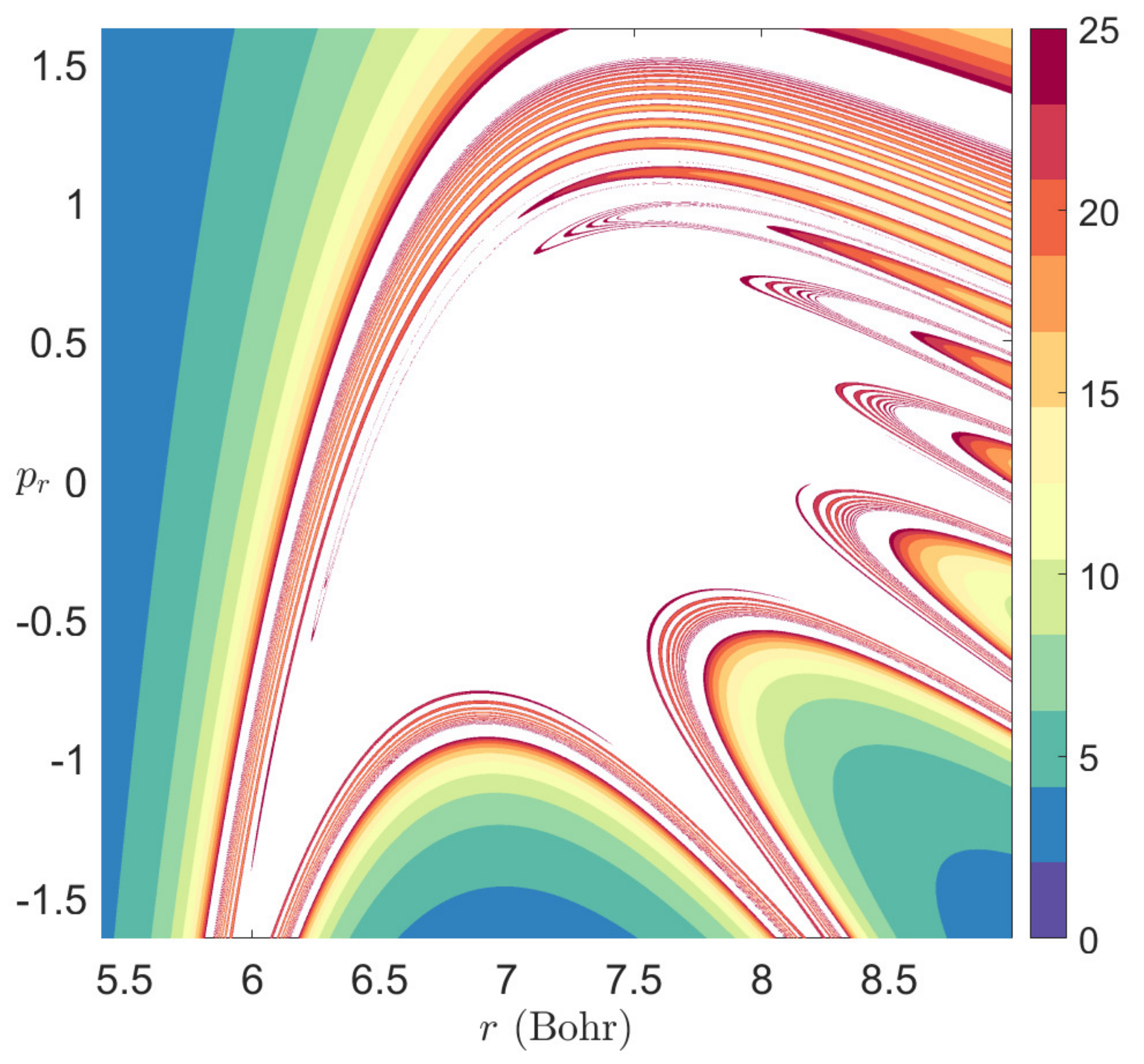}
		F)\includegraphics[scale=0.28]{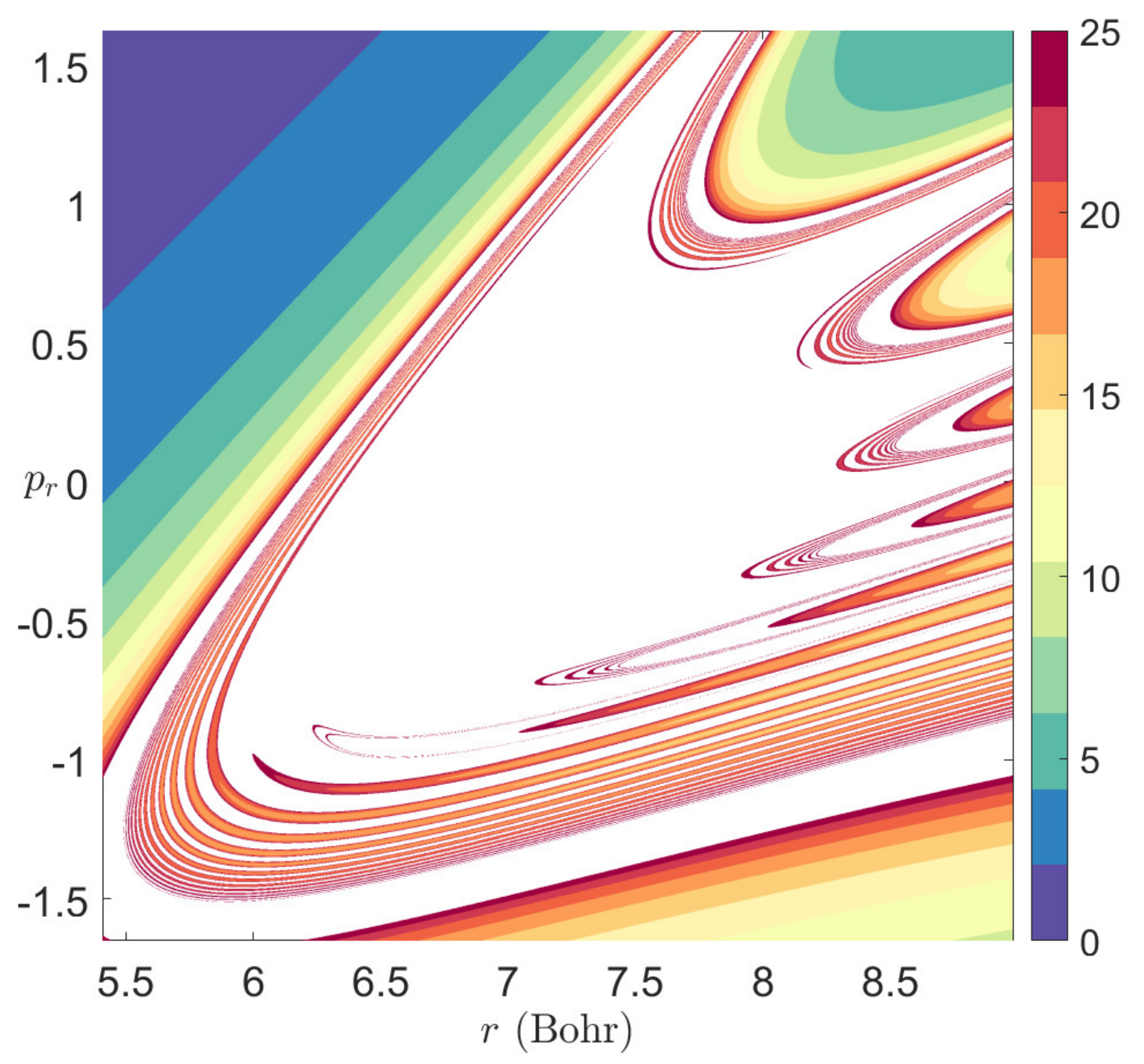}
	\end{center}
	\caption{Phase space structures highlighted by means of applying DLDs ($p = 1$), Poincar\'e maps and escape time plots to the 2D area-preserving map in Eq. \eqref{2D_map} with a kicking period $T = 8000$. This computation has been carried out in a zoom region of the invariant subspace $\gamma = \pi / 2$ and $p_{\gamma} = 0$ to reveal the intricate structure of the turnstile lobes that governs phase space transport. A) DLD scalar field for $N = 15$ iterations; B) corresponds to $N = 25$ iterations; C) uses $N = 50$ iterations; D) stable (blue) and unstable (red) manifolds extracted from the gradient of DLDs together with KAM tori revealed by a Poincar\'e map; E) Forward escape time distribution; F) Backward escape time distribution.}
	\label{fig:lobe_and_KAM_resol}
\end{figure}

\begin{figure}[htbp]
	\begin{center}
		\includegraphics[scale=0.4]{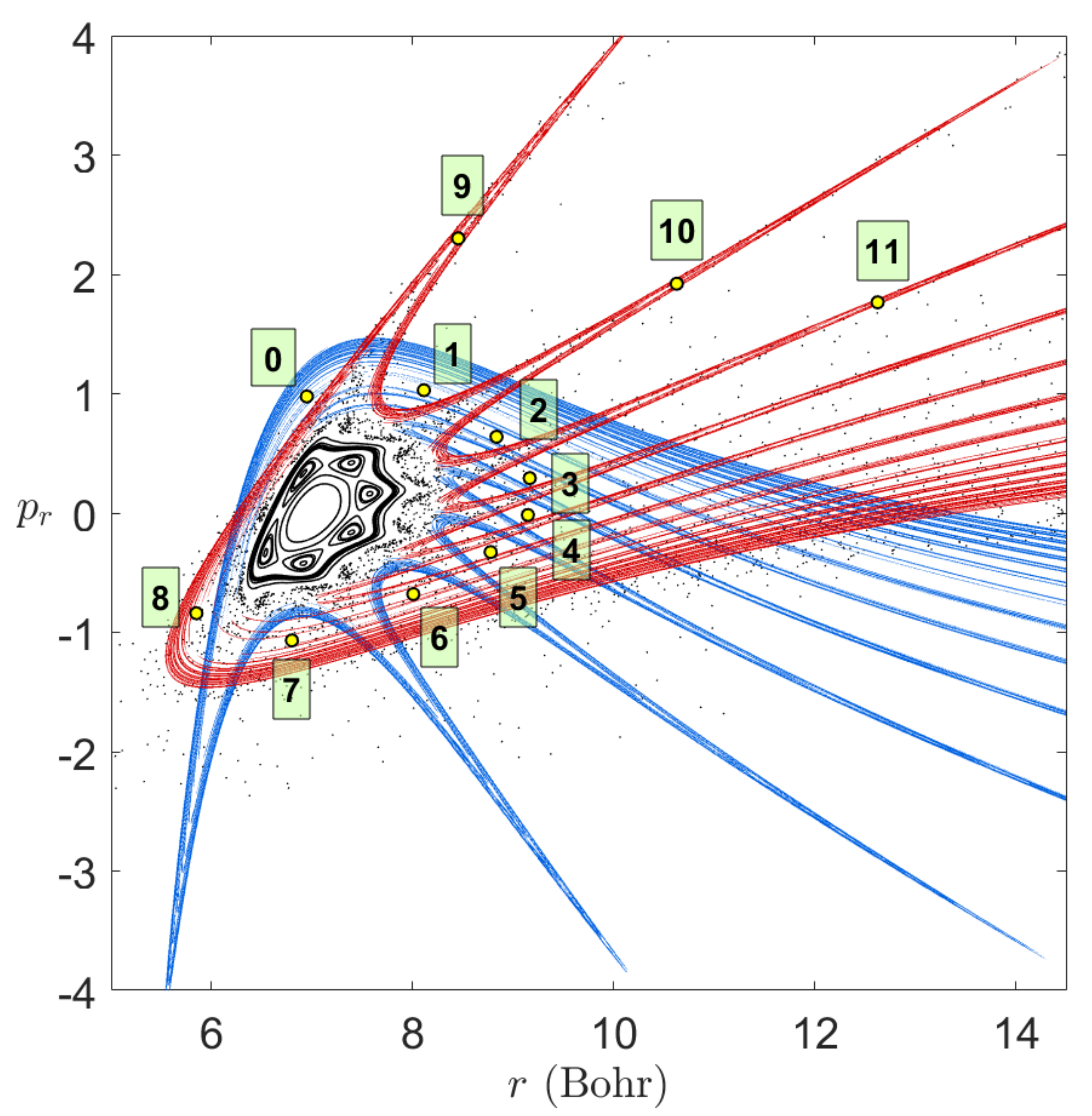} 
	\end{center}
	\caption{Phase space of the 2D area-preserving map in Eq, \eqref{2D_map} for a kicking period $T = 8000$, as revealed by DLDs ($p = 1$) using $N = 15$ iterations and Poincar\'e maps on the invariant subspace $\gamma = \pi / 2$ and $p_{\gamma} = 0$. The stable (blue) and unstable (red) manifolds have been extracted from the gradient of the DLD scalar field. We also depict the forward orbit of an initial condition starting at the location labeled as $0$ in order to illustrate the lobe dynamics governed by the turnstiles formed by the homoclinic intersections of the invariant manifolds of the parabolic point at infinity.}
	\label{fig:lobeDyn}
\end{figure}

The phase space structures  of Fig. \ref{fig:phaseSp_diffT}   correspond to two different cases. The first one corresponds to  intramolecular bonding and second one correspond dissociation in our system respectively: 

\begin{enumerate} 

\item {\bf Case I -  Intramolecular bonding:} In this case, the points lie on the invariant curves around the stable fixed points or  they are inside the bounded region of these invariant curves. These invariant curves correspond to the KAM invariant tori and they are around the stable fixed point or around stable periodic orbits with high order multiplicity. The points  that lie on these invariant curves or they are inside the bounded region of these invariant curves will stay in this region for ever. This is the reason that the points in these regions do not escape (see panels E and F  of Fig. \ref{fig:lobe_and_KAM_resol}). This means that  the first case corresponds to a state of our system that we don't  have dissociation. 

\item {\bf Case II - Dissociation:} In this case, the points are in the region of the irregular distribution of points (see panels D and F of Fig. \ref{fig:phaseSp_diffT}). This region is located  after the region that is occupied by the invariant curves. These points stay at the region  for a time interval before they leave for a larger region of the phase space. This case corresponds to a state of our system that we have dissociation. The time interval, that the points need to escape from the region, decreases as we increase the distance from the stable fixed point until we have escape times close to zero (see panels E and F of Fig. \ref{fig:lobe_and_KAM_resol}). 
The mechanism for this dynamical behavior of these points can be explained through the lobes  of the stable and unstable invariant manifolds of the parabolic fixed point at the infinity (see the lobes in  panels A,B and C of Fig. \ref{fig:lobe_and_KAM_resol}). The points that are in the chaotic region outside the region that is occupied from the invariant curves (see panels D and F of Fig. \ref{fig:phaseSp_diffT}) can be trapped in the lobes of the stable and unstable invariant manifolds of the parabolic fixed point that is at the infinity (see the iterations of the point 0 that is located in one of these lobes in the Fig. \ref{fig:lobeDyn}). The points in these lobes can follow the stable or unstable invariant manifolds in order to be moved close to or far away from  the neighborhood of the parabolic fixed point (that is at the infinity). This means that the points will leave the region around the invariant curves and they will go to the infinity. An example of this behavior is presented in Fig. \ref{fig:lobeDyn}  where the point $0$,that is located to the upper lobe, leaves the region around the stable fixed point and its invariant curves (after 11 successive iterations) and it visits larger regions of the phase space.  
\end{enumerate}

\subsection{Analysis of the 4D Symplectic Map}

\begin{figure}[htbp]
	\begin{center}
		A)\includegraphics[scale=0.25]{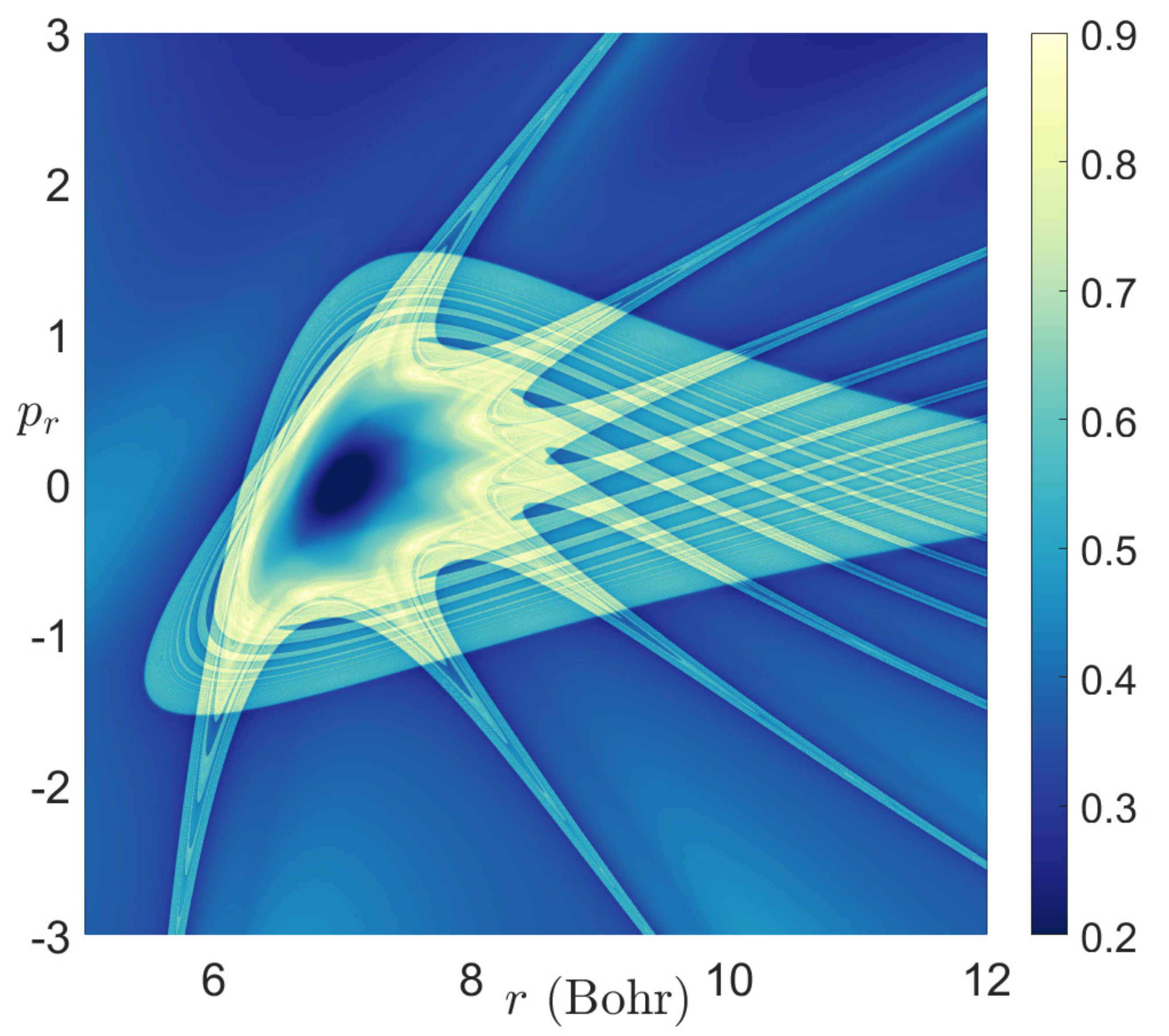} 
		B)\includegraphics[scale=0.25]{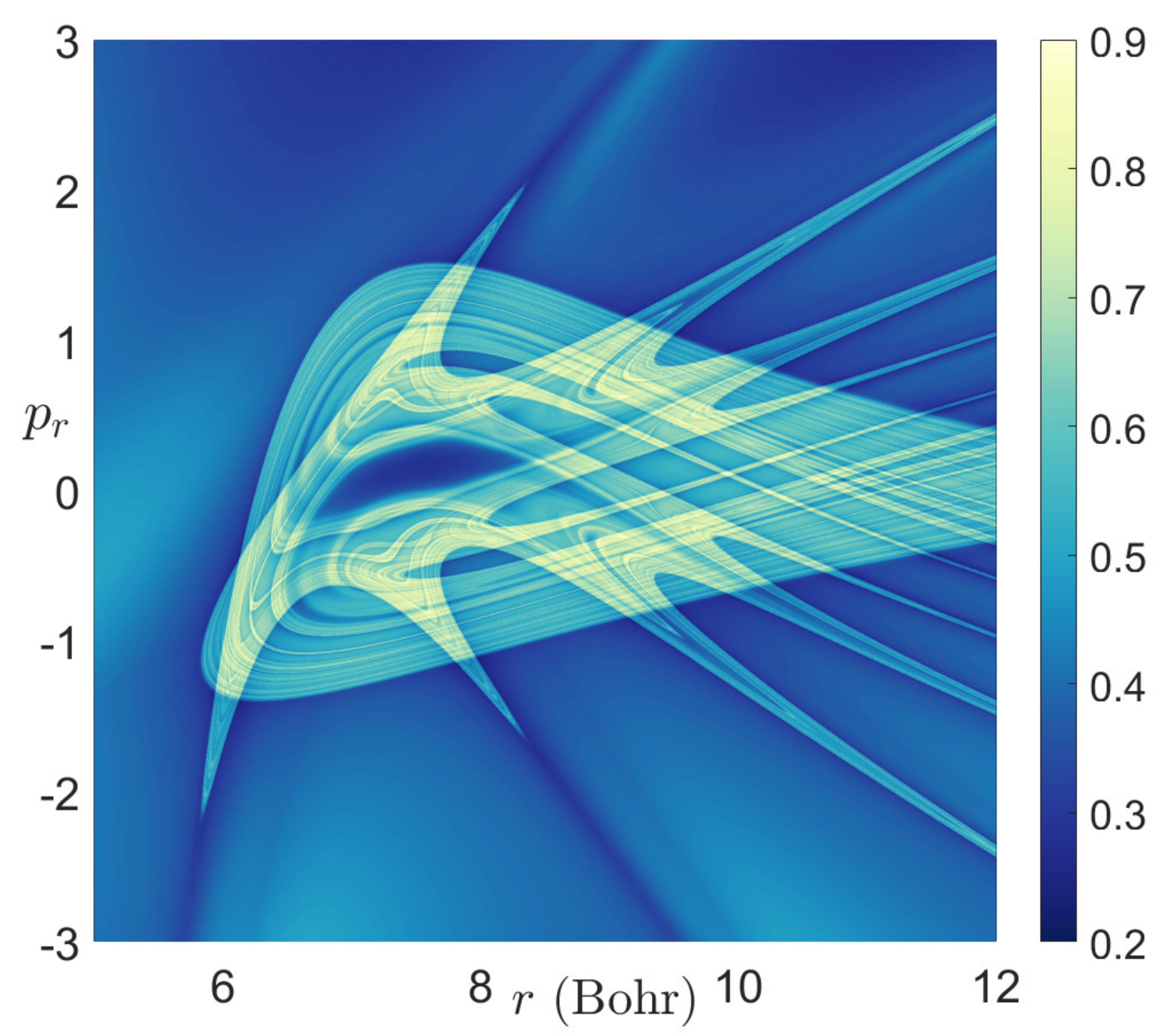} 
		C)\includegraphics[scale=0.25]{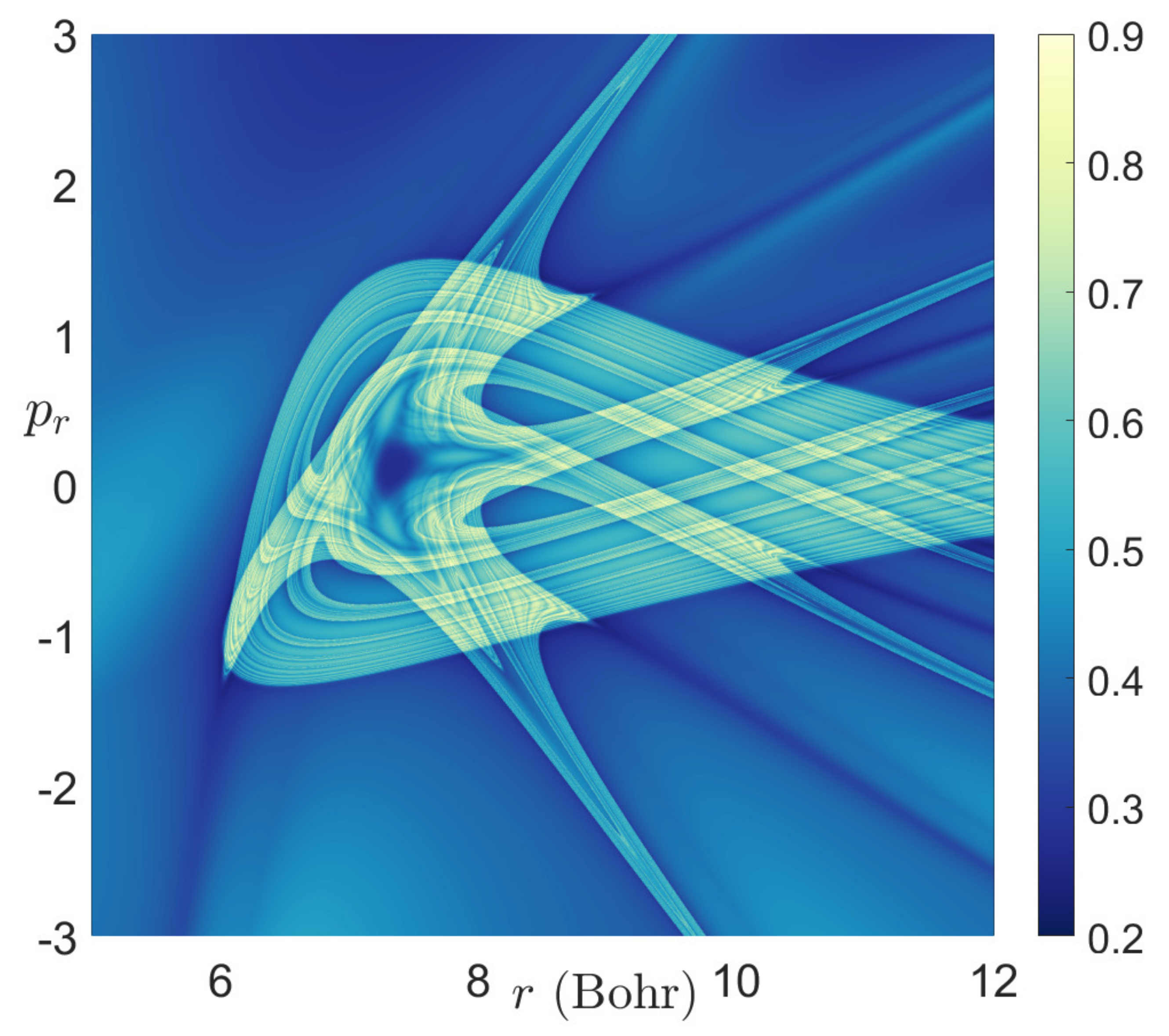} 
		D)\includegraphics[scale=0.25]{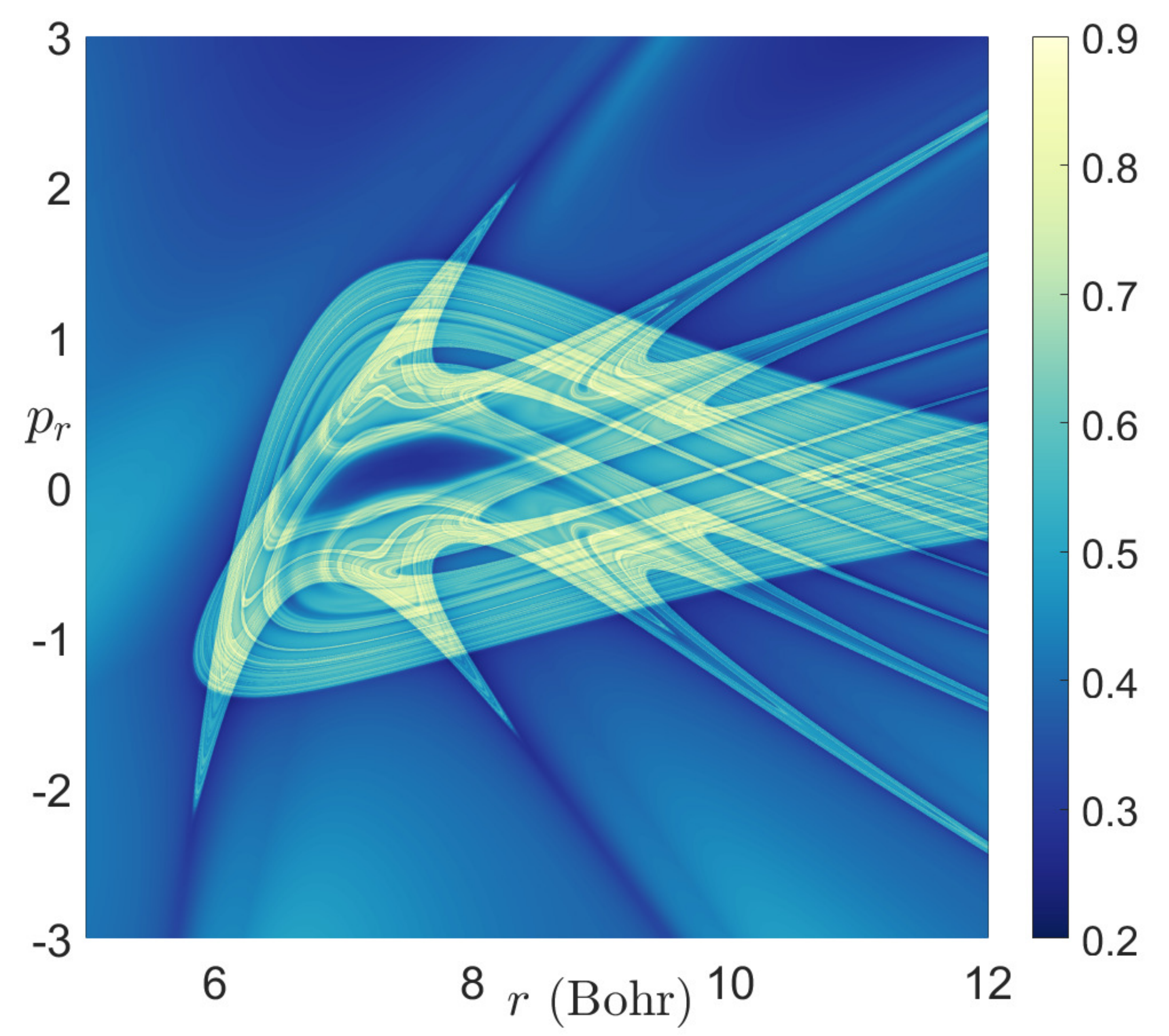} 
	\end{center}
	\caption{Phase space structures obtained by applying Lagrangian descriptors on the two-dimensional slices that result from fixing the angle variable $\gamma$ and its conjugate momentum $p_\gamma$. The number of iterations used in the computation is $N = 25$ and the kicking period of the 4D symplectic map is set to $T = 8000$. In all panels $\gamma = 1/2$, and $p_{\gamma} = 0,\, 1/4, \, 1/2, \, 3/4$ correspond to the panels A, B, C and D respectively. The values given above for the angle and the momentum coordinate have been scaled accordingly by their corresponding periods.}
	\label{fig:fix_gam_pgam}
\end{figure}

\begin{figure}[htbp]
	\begin{center}
	  A)\includegraphics[scale=0.33]{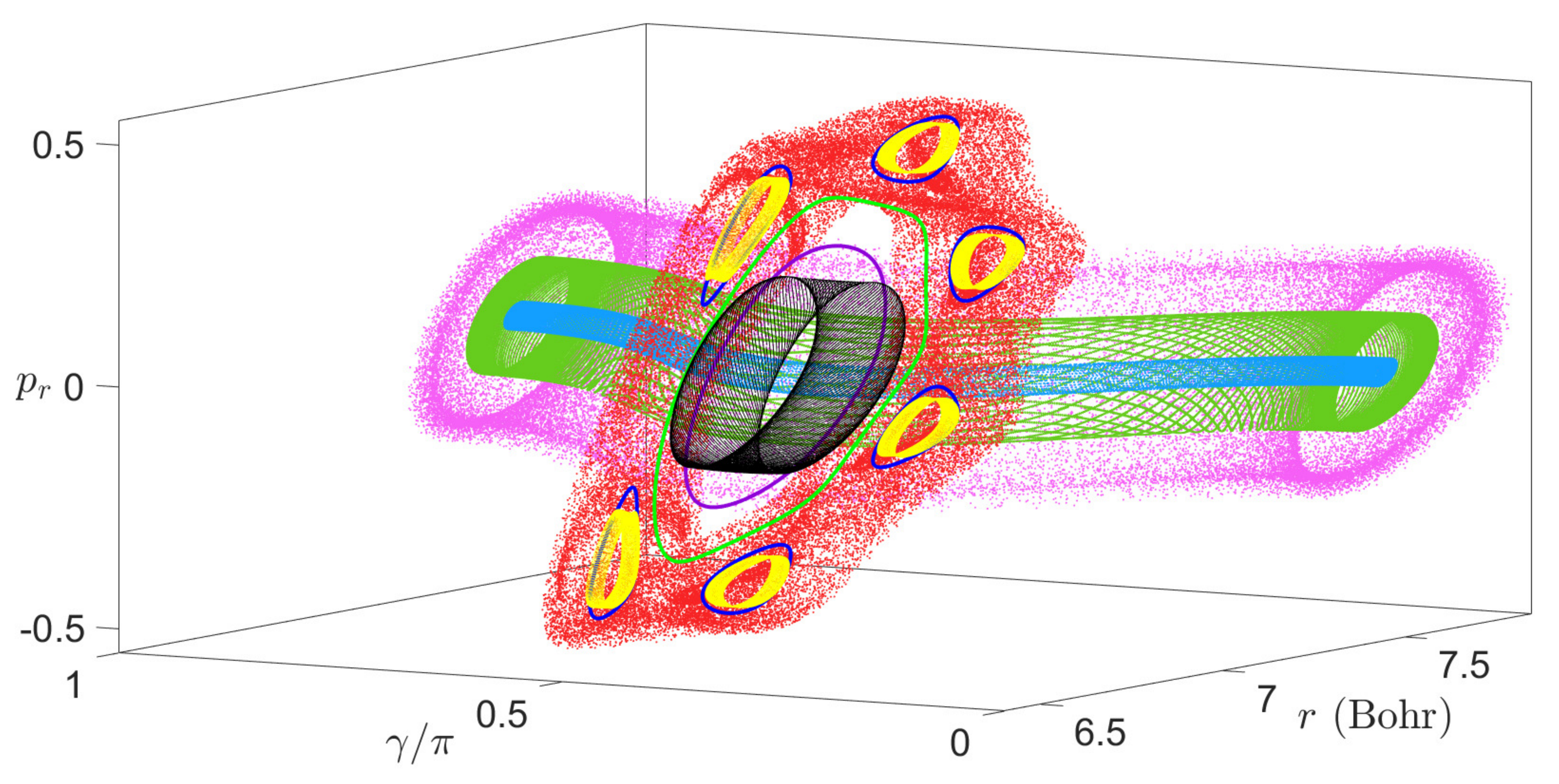}
	  B)\includegraphics[scale=0.21]{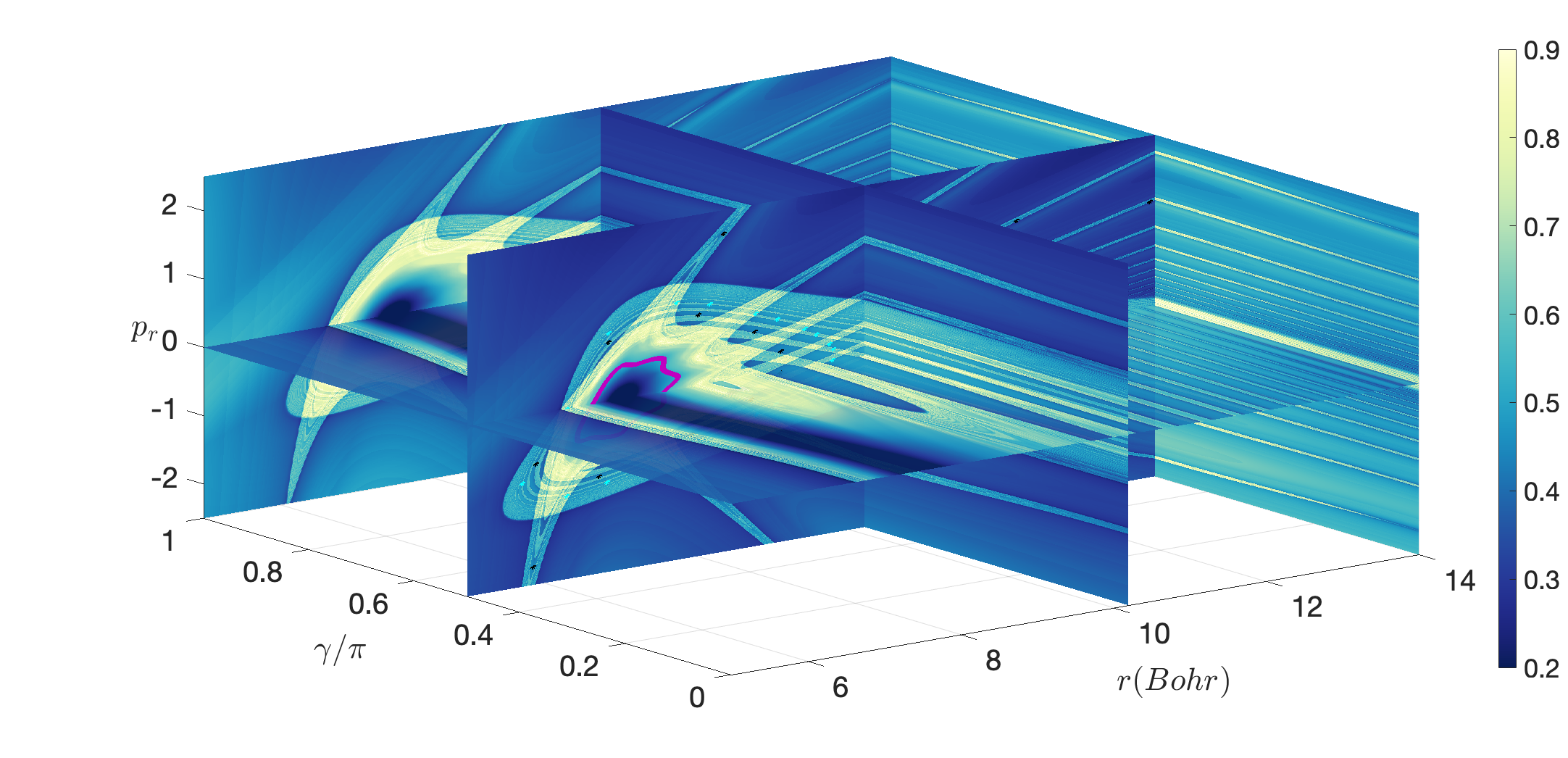}
	\end{center}
	\caption{Three-dimensional representation of the phase space structures obtained in the slice $p_\gamma = 0$ for the 4D symplectic map with kicking period $T = 8000$. A) KAM invariant tori and sticky region (in red) surrounding the yellow and blue tori, as revealed by the method of 3D projections. B) Visualization of the invariant manifolds (KAM tori and stable and unstable manifolds) by means of applying Lagrangian descriptors using $N = 25$ iterations on different 2D slices of the subspace $p_\gamma = 0$. Superimposed with the initial conditions of Fig. \ref{2dplane}.}
	\label{fig:3D_phaseSpace}
\end{figure}

In this section we study the phase space structures of the 4D symplectic map for the van der Waals complex using DLDs \cite{carlos,GG2019b,GG2020a} and 3D projections of the 4D phase space of the 4D symplectic map (representation of the phase space structures in 3D subspaces of the 4D phase space). The  DLD method can help to recover geometrical structures in the phase space and therefore recovers the intersection of the invariant manifolds in the phase space with different slices of the phase space. It is evident that in all of the figures we can visualize manifolds and intersections of high dimensional objects with low dimensional planes. We can compute these invariant manifolds in 2D slices of the 4D phase space of  a  4D symplectic map using DLDs. An example of this computation is presented in Fig. \ref{fig:fix_gam_pgam}. Changing the value of the $p_{\gamma}$ we get the manifolds in panels A, B, C and D  and we see the manifolds to get more distorted as we increase the value of $p_{\gamma}$.  Another interesting finding is that panel D recovers panel B due to the fact that $p_{\gamma}$ is symmetric with respect to the line $p_{\gamma} = 1/2$. In Fig. \ref{fig:3D_phaseSpace} we present the visualization of the invariant manifolds (KAM tori and stable and unstable manifolds) by applying the Lagrangian descriptors on different $2D$ slices. It is evident how the method reveals nicely the phase space structures in 3D. A nice comparison that highlights the importance of the method is the 3D representation of a similar 3 DoF Hamiltonian system presented in the paper \cite{Montoya2020a}. 

\begin{figure}[htbp]
	\begin{center}
	    A)\includegraphics[scale=0.5]{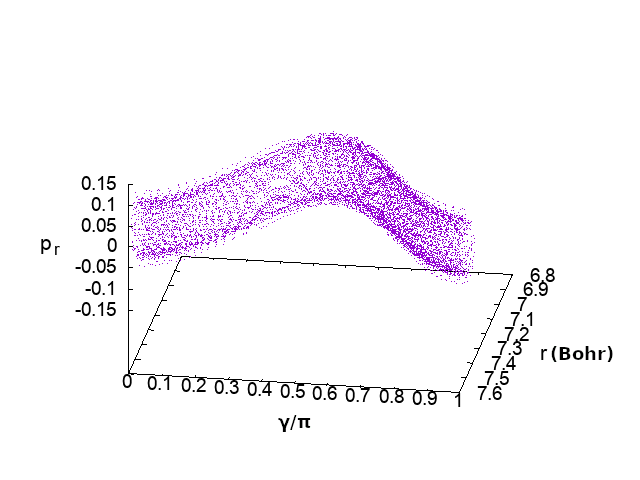} 
		B)\includegraphics[scale=0.5]{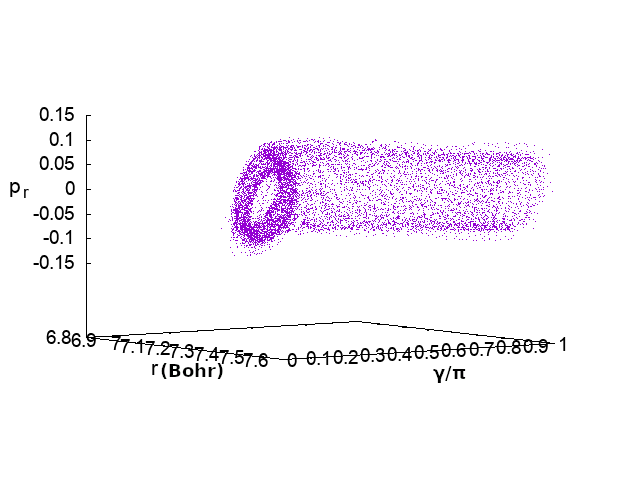} 
	\end{center}
	\caption{The 3D representation $(r,\gamma,p_r)$ of a torus in the neighborhood of a stable fixed point (the initial conditions for this torus is:  $r=7.0998,\gamma=0.501\pi,p_r=p_{\gamma}=0)$.}
	\label{torus}
\end{figure}

\begin{figure}[htbp]
	\begin{center}
	    A)\includegraphics[scale=0.18]{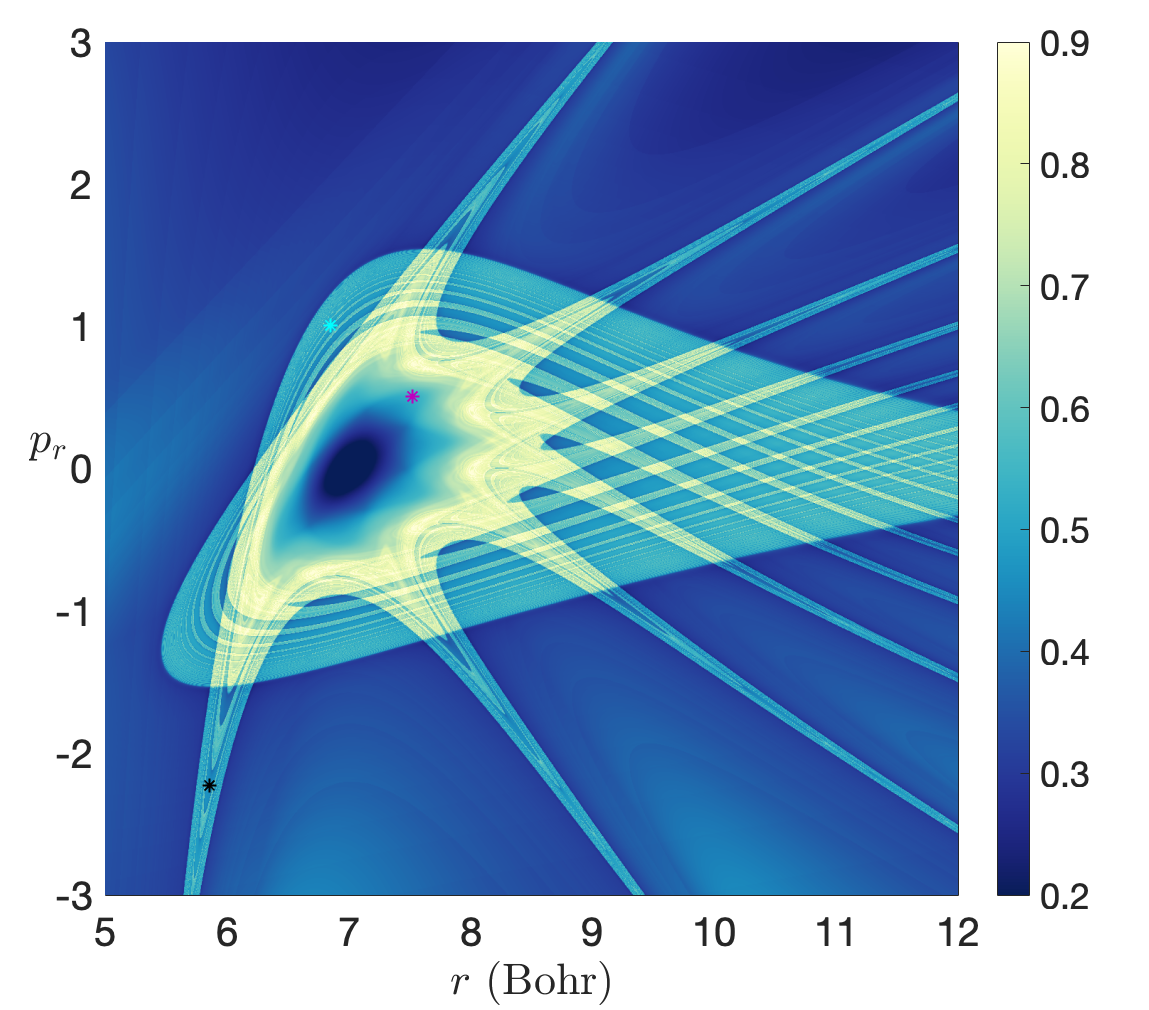} 
		B)\includegraphics[scale=0.18]{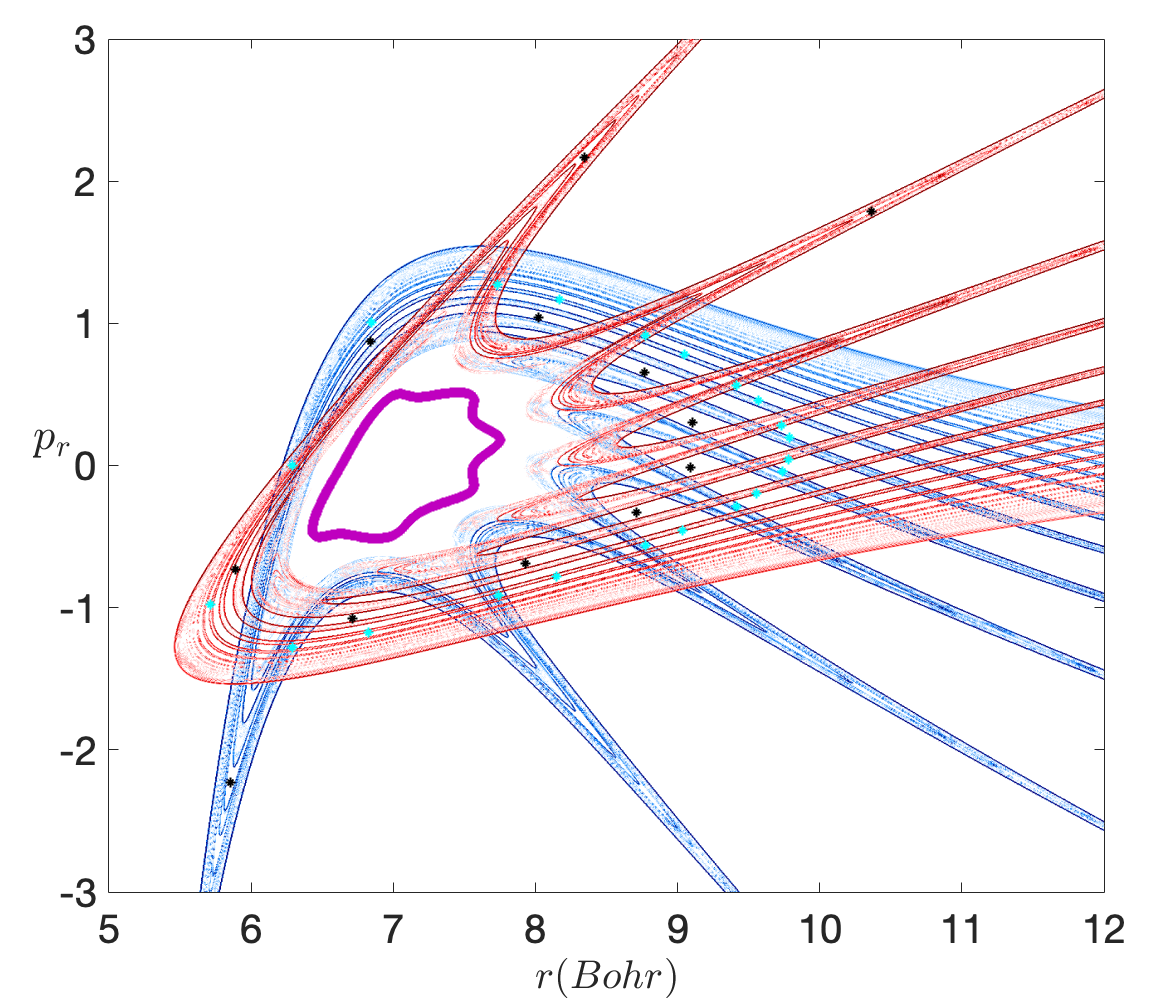} 
	\end{center}
	\caption{A) Phase space structures obtained by applying Lagrangian descriptors on the two-dimensional slices that result from fixing the angle variable $\gamma$ and its conjugate momentum $p_\gamma$. The number of iterations used in the computation is $N = 25$ and the kicking period of the 4D symplectic map is set to $T = 8000$ and $B=4\times 10^{-5}$. In all panels $\gamma = 1/2$, and $p_{\gamma} = 0$. Forward iterations of the initial conditions  $r = 5.8559$, $p_{r}= -2.2287$ (black asterisk of panel A), $r = 7.5227$, $p_{r}= 0.5068$ (magenta asterisk of panel A) $r =6 .8484$, $p_{r}=1.0045$ (cyan asterisk of panel A). B) Stable (blue) and unstable (red) manifolds,}
	\label{2dplane}
\end{figure}

\begin{figure}[htbp]
	\begin{center}
	    \includegraphics[scale=0.15]{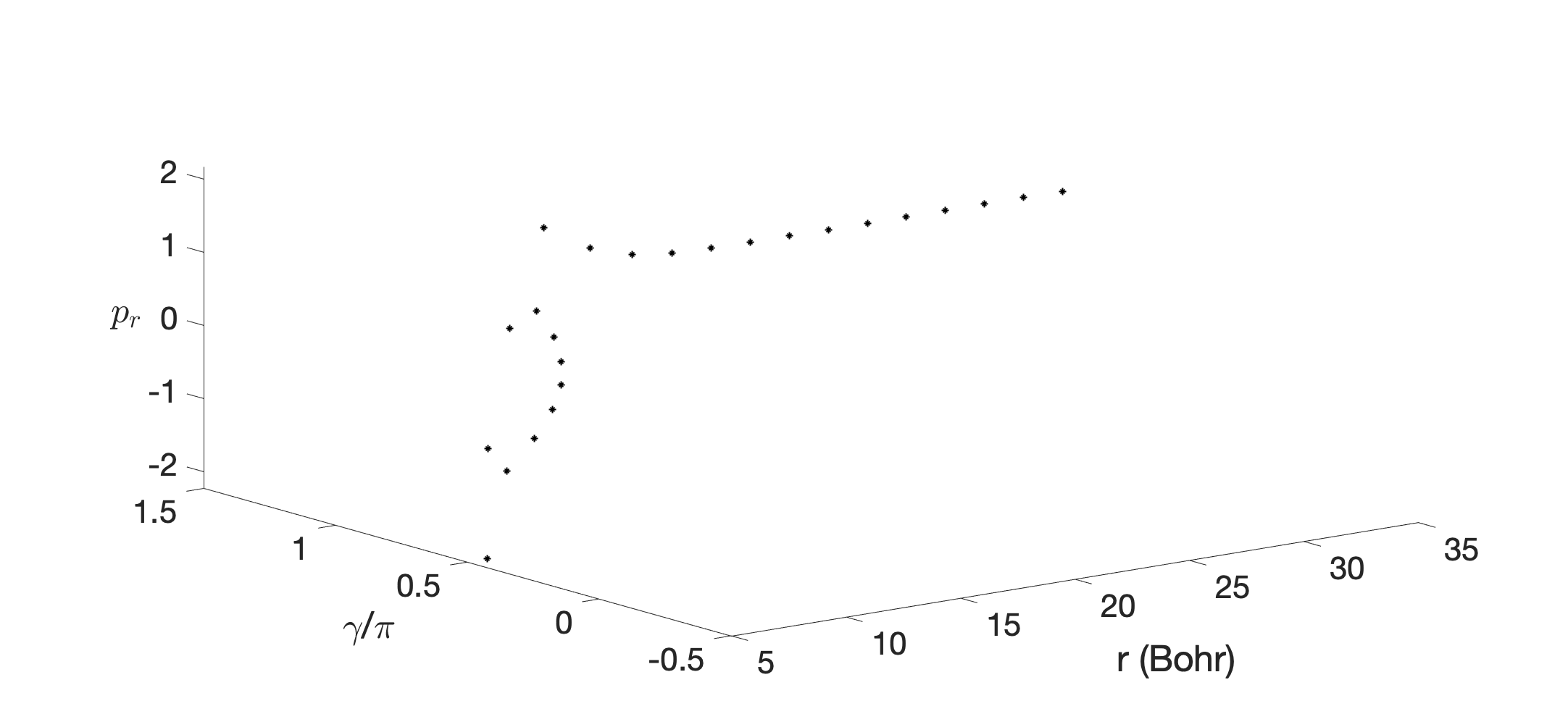} 
	\end{center}
	\caption{The 3D representation $(r,\gamma,p_r)$ of a trajectory that corresponds to the cyan asterisk in Fig.\ref{2dplane}, which is inside a lobe of the stable manifolds of the parabolic fixed point at infinity. Note that $\gamma = 0.5$ and $p_{\gamma} = 0$.} 
	\label{escape2}
\end{figure}

\begin{figure}[htbp]
	\begin{center}
	 A)\includegraphics[scale=0.5]{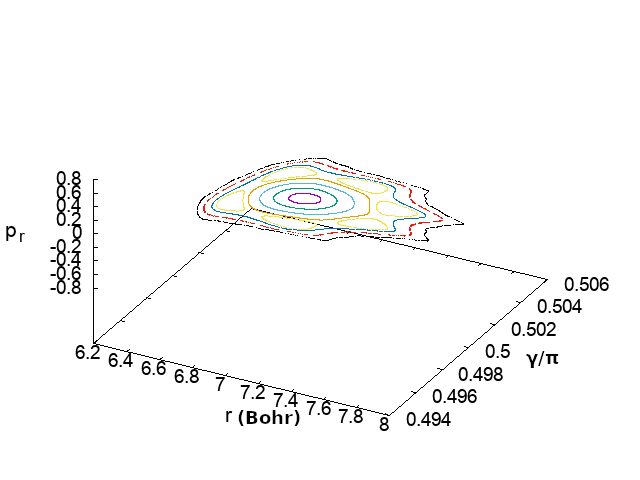} B)\includegraphics[scale=0.5]{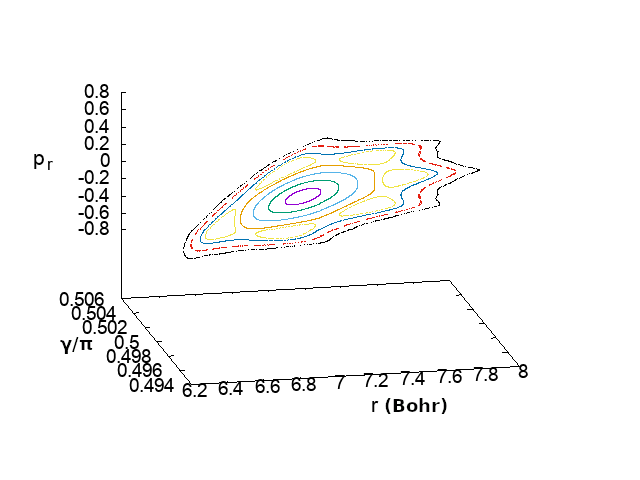} 
	\end{center}
	\caption{The 3D representation $(r,\gamma,p_r)$ of invariant curves (for 1000 iterations) in the neighborhood of a stable fixed point (the initial conditions for these invariant curves are: $r=7.1,7.2,...,7.8, \gamma=0.5\pi,p_r=0.05, p_{\gamma}=0)$.}
	\label{curves1}
\end{figure}

\begin{figure}[htbp]
	\begin{center}
	    \includegraphics[scale=0.55]{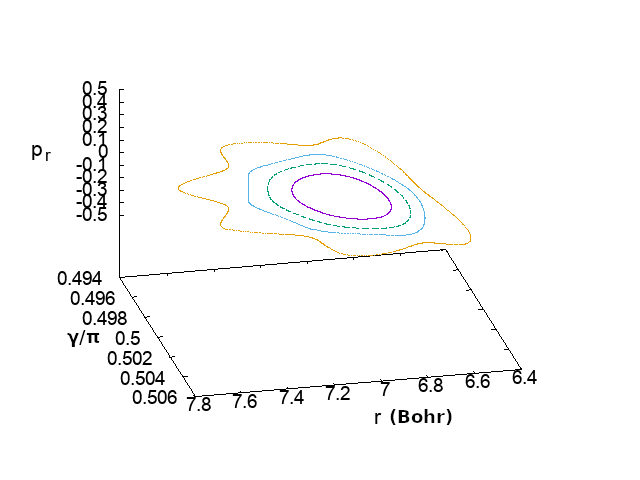} 
	\end{center}
	\caption{The 3D representation $(r,\gamma,p_r)$ of invariant curves (for 1000 iterations) in the neighborhood of a stable fixed point  (the initial conditions for these invariant curves are: $r=7.2,\ldots,7.5, \gamma=0.5\pi,p_r=p_{\gamma}=0)$.}
	\label{curves2}
\end{figure}

\begin{figure}[htbp]
	\begin{center}
	    A)\includegraphics[scale=0.5]{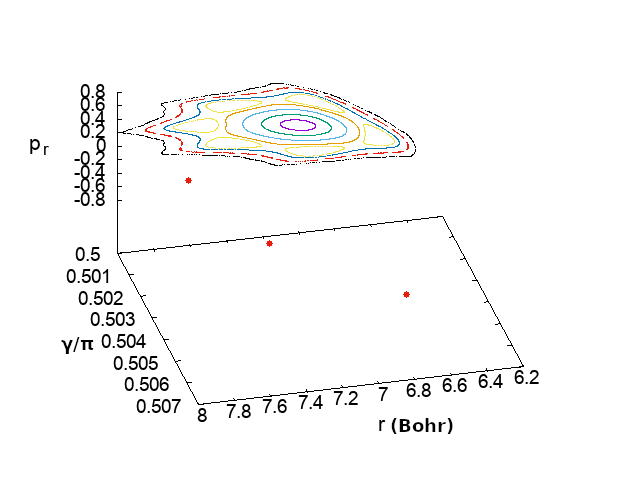} 
		B)\includegraphics[scale=0.5]{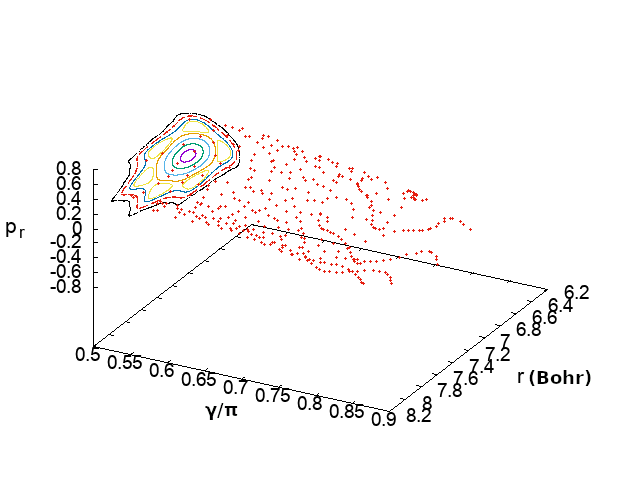} 
	\end{center}
	\caption{The 3D representation $(r,\gamma,p_r)$ of a sticky structure in the neighborhood of a stable fixed point. A) for 1211 iterations; B) for 1600 iterations. The initial condition for this structure is  $r=7.7,\gamma=0.5\pi,p_r=0.05, p_{\gamma}=0)$.}
	\label{sticky1}
\end{figure}

\begin{figure}[htbp]
	\begin{center}
	    A)\includegraphics[scale=0.5]{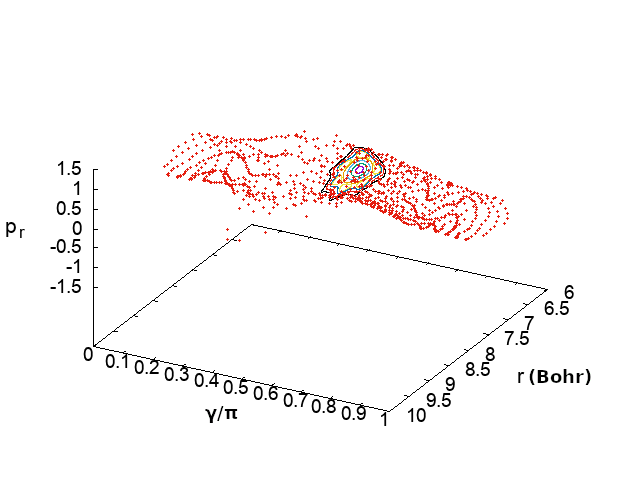} 
		B)\includegraphics[scale=0.5]{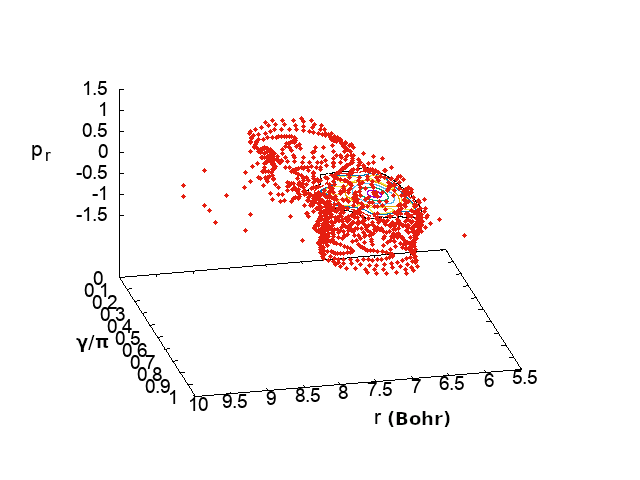} 
	\end{center}
	\caption{The 3D representation $(r,\gamma,p_r)$ of a sticky torus in the neighborhood of a stable fixed point. A) for 2200 iterations; B) for 2246  iterations. The initial condition for this structure is  $r=7.7,\gamma=0.5\pi,p_r=0.05, p_{\gamma}=0)$.}
	\label{sticky2}
\end{figure}

\begin{figure}[htbp]
	\begin{center}
	    \includegraphics[scale=0.55]{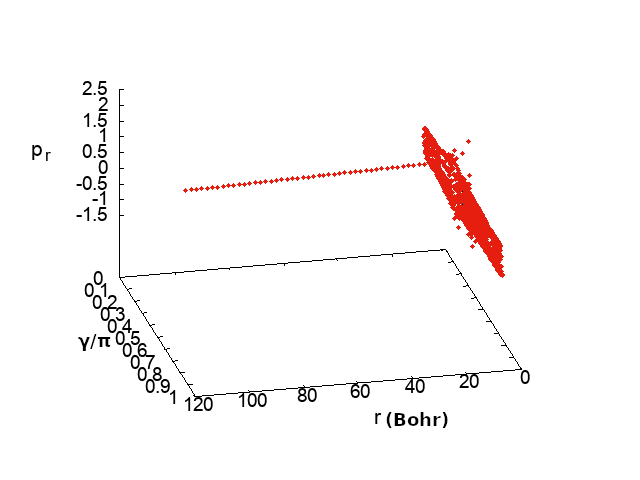} 
	\end{center}
	\caption{The 3D representation $(r,\gamma,p_r)$ of a sticky torus in the neighborhood of a stable fixed point for 2300 iterations. The initial condition for this structure is $r=7.7,\gamma=0.5\pi,p_r=0.05, p_{\gamma}=0)$.}
	\label{sticky3}
\end{figure}

Here, the phase space structures that are responsible in the 4D map model of the van der Waals complex for the intramolecular bonding, dissociation, or intermediate situations between them, can be categorized into three different cases:
\begin{enumerate}
\item {\bf Case I - Intramolecular bonding:} In this case, the points are located on a 2D invariant tori that exist (according to the KAM theorem \cite{kolmogorov1954,moser1962,arnold1963}) around the stable fixed points of the 4D symplectic map. The points  will stay forever on these tori. For example if we perturb the initial conditions of a stable fixed point in the $\gamma$ direction  the successive iterations lie on a invariant tori as we see in Fig. \ref{torus}. These tori correspond to the absence of dissociation for our system.

\item {\bf Case II - Dissociation:} In this case, the points don't lie on a invariant tori but they are located in the lobes of the invariant manifolds of the normally hyperbolic invariant manifold (NHIM) (that is at the infinity). A NHIM is a generalization of unstable periodic orbits to Hamiltonian systems with more than 2DoF and higher dimensional (more than 2) symplectic maps. For the 4D map the NHIM is 2D and the stable and unstable manifolds of the NHIM are 3D. 
 
In Fig. \ref{2dplane}, we present an example of points that are located in the lobes of the invariant   manifolds of the NHIM  in a 2D slide of the 4D phase space (see the cyan asterisk in the upper left lobe in  the panels A and B). The successive iterations of the cyan asterisk shows how it  starts to move away from the initial region (through the lobes of the invariant manifolds) in the 2D slice of the 4D phase space. The points in these lobes can follow the stable or unstable invariant manifolds in order to move close to or far away from  the neighborhood of the NHIM (that is at infinity). Consequently, these points will leave from the initial region where they are located (close to the stable fixed point) and they will go to infinity. This is the reason that  in our example the cyan asterisk will eventually escape to  infinity (see Fig. \ref{escape2}). This means that these points correspond to dissociation for our system.   

\item {\bf Case III - Predissociation:} The phase space structures, in this case, correspond to the states of our system (the complex of $HeI_2$) for which we have  a transition from  intramolecular bonding to  dissociation. In this case  the molecules   are protected, for a long  discrete time (many iterations), from  dissociation. The  phase space structures that correspond to this case are sticky tori and sticky  curves which are around the stable fixed points. The points stay on these structures for many iterations until they leave  these structures. This phenomenon is known as stickiness (\cite{contopoulos2002})  and it has  been studied in 4D Poincar\'{e} maps of Hamiltonian  systems with three degrees of freedom  in  galactic dynamics and general relativity (see \cite{katsanikas2011a,lukes2016dynamics,patsis2014phase2}). Eventually, the points escape to infinity through the manifolds of the NHIM as it was explained in  case II. Some examples are presented in Figs. \ref{curves1}, \ref{curves2},\ref{sticky1},\ref{sticky2} and \ref{sticky3}. If we perturb the stable fixed point, we observe  curves and islands around this point (see  Fig. \ref{curves1}  when we perturb the stable fixed point  in $r$-direction and $p_r$-direction and Fig. \ref{curves2} for a perturbation only in the  $r$-direction). These  curves lie on the 2D plane $(r,p_r)$ of the 4D phase space for $\gamma=pi/2$ and $p_{\gamma}=0$ (see Fig. \ref{curves1},  \ref{curves2}). The points stay on these curves for at least 1000 iterations. Now let's focus on the red curve in Fig. \ref{curves1}. For more than 1208 successive iterations we see that the points start to leave this  curve and the 2D plane $(r,p_r)$  (panel A of Fig. \ref{sticky1})  in the $\gamma$ direction and they start to occupy larger volumes in the phase space (see panel B of Fig. \ref{sticky1}). As we increase the iterations they form a torus (see panel A of Fig.\ref{sticky2})  which contains (inside its a hole) the  curves that are around the stable fixed point. This torus has similar morphology with the torus of Fig. \ref{torus}. The difference is that torus in Fig. \ref{torus} is a KAM invariant torus and the points lie on it forever. On the contrary, the torus in Fig. \ref{sticky2} is sticky, and after 2246 iterations the points start to leave this structure (see panel B of Fig. \ref{sticky2}) before they escape to infinity (see Fig. \ref{sticky3}). Consequently, we have two kinds of sticky structures (or stickiness), sticky curves and sticky tori. The first kind of stickiness corresponds to a state in which we have a transition of the complex HeI$_2$ from intramolecular bonding to intramolecular bonding for larger values for $r$ and $\gamma$. The second kind of stickiness corresponds to a state in which we have a transition from  intramolecular bonding to dissociation. This is the first time that we see that the stickiness is important for the mechanism of the dissociation in chemical reactions. Alternative explanations of predissociation in other models of chemical reactions has been given due to the phenomenon of stable chaos (a known phenomenon from celestial mechanics for the stability of the asteroids \cite{milani1992example}) that involves the interaction of two or more independent anharmonic resonances \cite{karmakar2020stable}. 

\end{enumerate}

\begin{figure}[htbp]
	\begin{center}
		A)\includegraphics[scale=0.28]{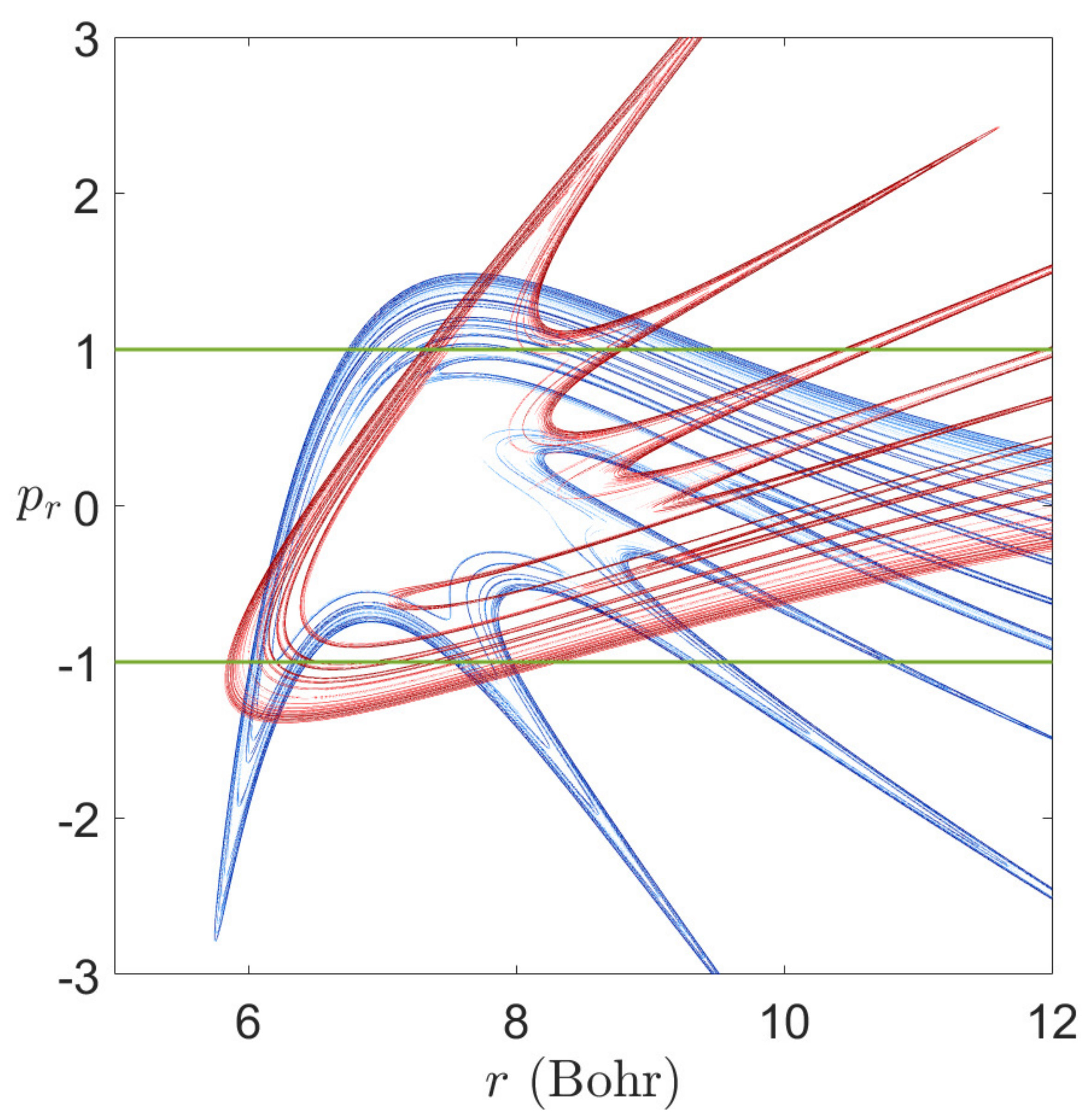} 
		B)\includegraphics[scale=0.28]{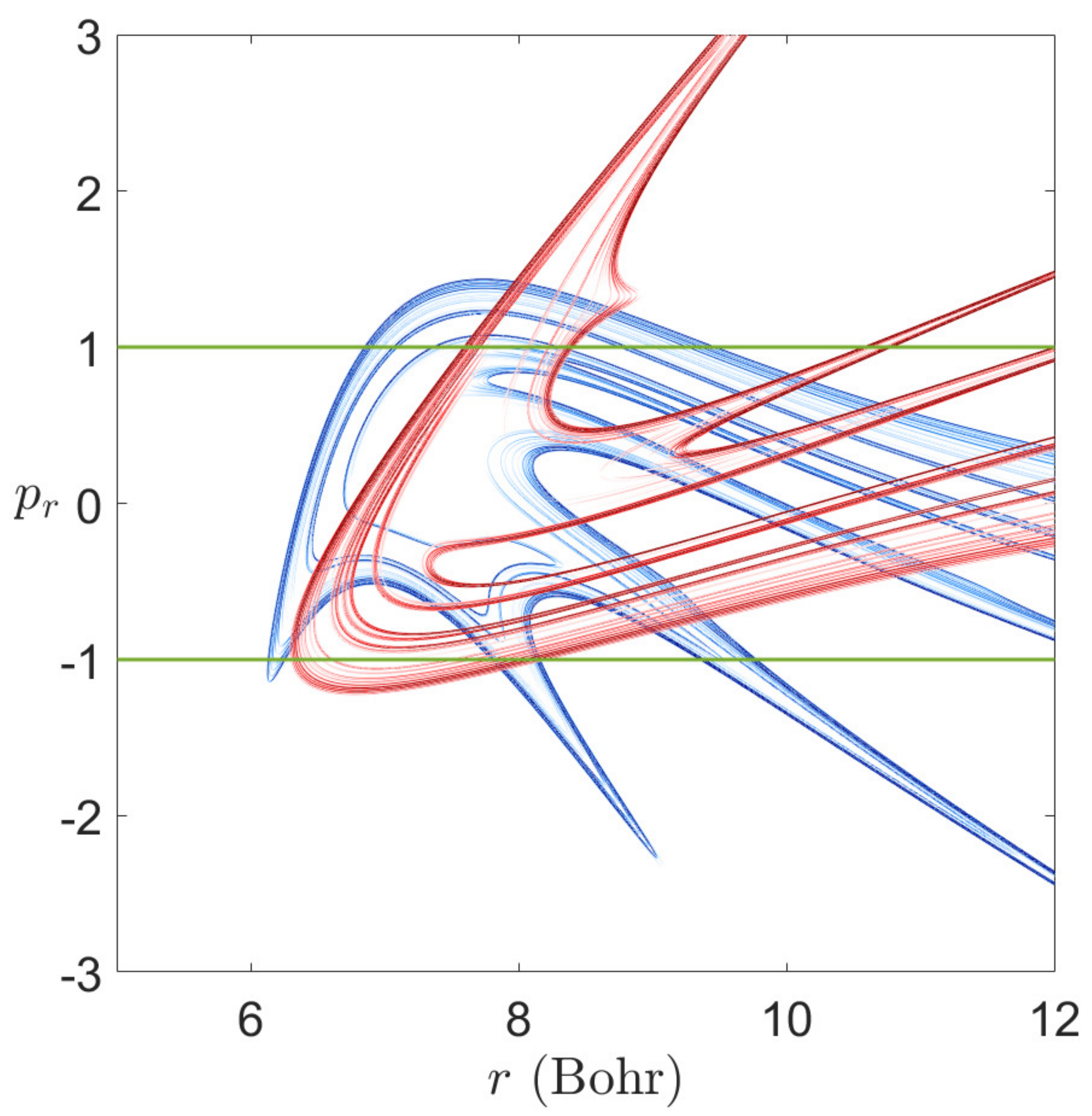} 
		C)\includegraphics[scale=0.28]{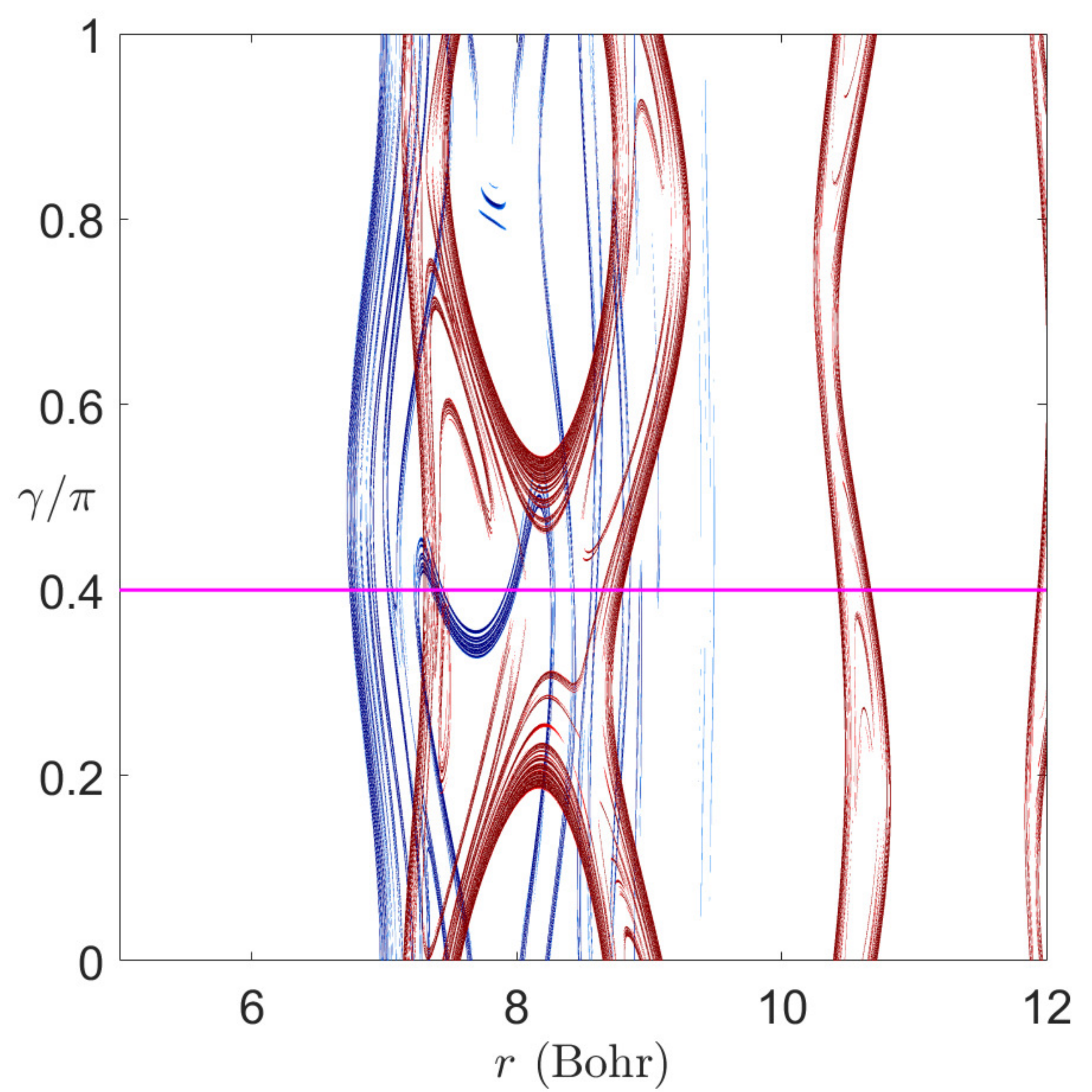} 
		D)\includegraphics[scale=0.28]{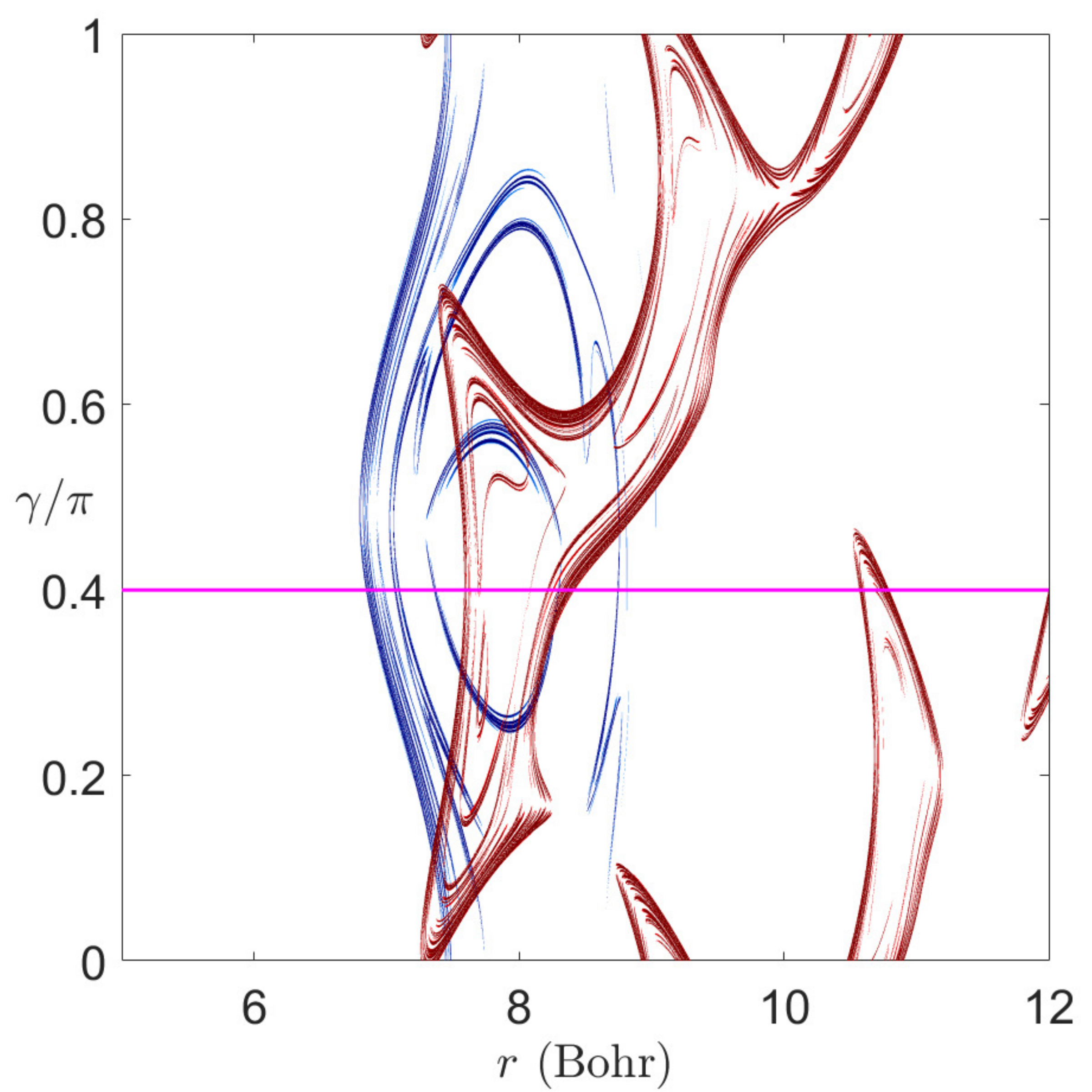} 
		E)\includegraphics[scale=0.28]{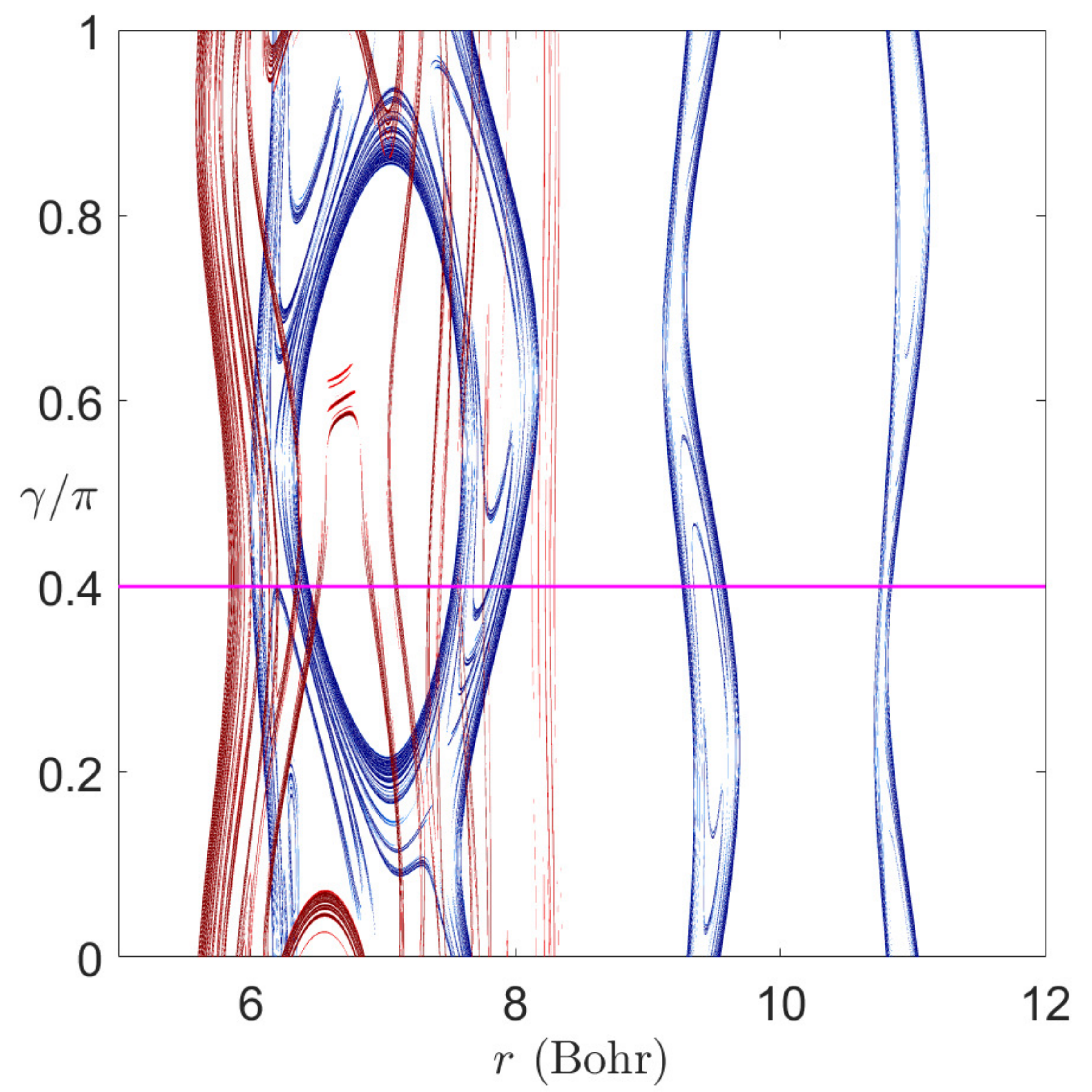} 
		F)\includegraphics[scale=0.28]{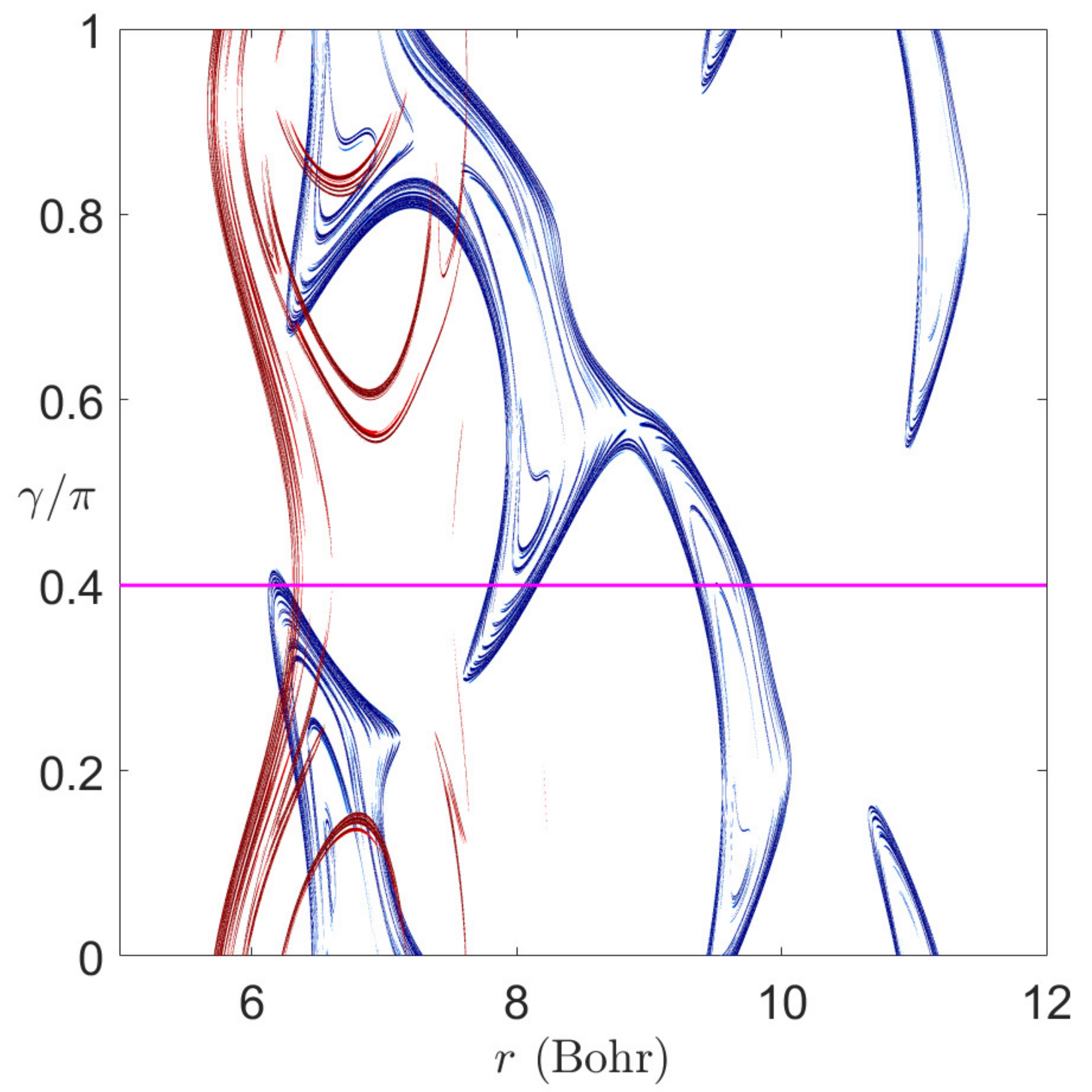} 
	\end{center}
	\caption{Stable (blue) and unstable (red) manifolds in different phase space slices extracted from the application of LDs with $N = 20$ iterations to the 4D symplectic map with kicking period $T = 8000$. The column on the left corresponds to the map with barrier height $B = 2.5 \times 10^{-5}$ (see Table \ref{tab:tab1}), and the right column is for $B = 5.5 \times 10^{-5}$. The top row corresponds to the slice $\gamma = p_{\gamma} = 0.4$, the middle row represents the cut $p_{\gamma} = 0.4$ and $p_r = 1$, and the bottom one is for $p_{\gamma} = 0.4$ and $p_r = -1$. Magenta lines are located at $\gamma = 0.4$ and indicate the positions of the slices depicted in panels A) and B). The green lines in those panels are placed at $p_r = \pm 1$ and correspond to the cuts shown in figures C) to F).}
	\label{fig:broken_lobe}
\end{figure}

We finish the discussion of the results of this work by illustrating how the method of LDs can be used to identify situations where the stable and unstable manifolds of high-dimensional systems intersect transversely without forming lobes. This is a very well known behavior that is common in Hamiltonian systems with 3 or more DOF \cite{wiggins90,ezra,beigie1995,beigie1995b,toda1995}, and it was reported in \cite{ezra} to occur for the van der Waals 4D symplectic map model we are studying here. This happens when the barrier height parameter ($B$) of the model is varied. Our goal is to replicate the results found in \cite{ezra} with Lagrangian descriptors, and show the geometry of the manifolds of the system in different planes. In Fig. \ref{fig:broken_lobe} we depict the stable and unstable manifolds in blue and red color respectively for different 2D phase space slices, where the column on the left corresponds to the barrier height $B = 2.5 \times 10^{-5}$ and the column on the right is calculated for a larger value, $B = 5.5 \times 10^{-5}$. We can clearly see how the lobes that were forming originally for smaller values of $B$ disappear when the barrier height increases, since the stable and unstable manifolds of the NHIM located at infinity do not intersect in the same way. We provide further evidence of this fact by presenting different 2D phase slices, where we show how the turnstile lobe formed by the stable/unstable manifold breaks apart, and therefore there is no volume enclosed in the turnstile anymore in the 4D phase space of the map.


\section{Conclusions}
\label{sec:conc}	
	
In this work we have shown how the method of Lagrangian descriptors can be used to reveal the high-dimensional phase space structures that arise in symplectic maps, by means analyzing their dynamics on low dimensional slices. In order to illustrate the applicability of this technique, we have focused on a 4D symplectic map model that is relevant in chemistry for the study of predissociation mechanisms in van der Waals complexes. This approach allows us to easily and systematically generate three-dimensional representations of the underlying phase space, from which one can gain meaningful insights on the full geometry of the system and the different nonlinear mechanisms that govern its dynamics. Interestingly, our analysis of phase space structure by means of LDs has revealed that we can identify situations where the stable and unstable manifolds of high-dimensional systems (three or more DoF) intersect transversely and do not form turnstile lobes, which can have an important impact on the study of transport for these systems.

Furthermore, in this paper we have studied all the phase space structures that are associated with intramolecular bonding, dissociation and predissociation mechanisms in the 2D and 4D map models for the HeI$_2$ van der Waals complex. The phase space structures are categorized into three cases:

\begin{enumerate}
    \item {\bf Case I - intramolecular bonding}: In this category, the  points lie on KAM invariant curves or invariant tori forever. This means that these solutions of our mapping models  correspond to a  distance  of  the atom of $He$ from the center mass of the two atoms of  $I_2$ that is bounded and we have absence of dissociation. The case of the intramolecular bonding correspond to both the 2D map and to 4D map.
    
    \item {\bf Case II - Dissociation:} In this category, the points are located in the lobes of the invariant manifolds of the parabolic fixed points (for the 2D map) or of the NHIM (for 4D map) and they are guided to the infinity. This means that these solutions of our mapping models  correspond to a  distance of  the atom of $He$ from the center mass of the two atoms of $I_2$   that is not  bounded (goes to the infinity) and we have the presence of the dissociation.
    
    \item {\bf Case III - Predissociation:} In this category, the points are located on sticky phase space structures (sticky curves and sticky tori). The points stay on these structures for many iterations before they leave  and go to the infinity. These solutions of our mapping model  correspond, for a long discrete time interval (many iterations), to a distance of the  atom of $He$ from the center mass of the two atoms of $I_2$   that is  bounded. After this discrete time interval this distance goes to the infinity (dissociation). This means that we have  a transition from a bounded state  to the dissociation, after a long discrete time. The case of the predissociation correspond only to the 4D map.
    
\end{enumerate}

We are perfectly aware that the methodology introduced in this work constitutes just a small step towards the development of a complete understanding regarding the intricate dynamics that take place in the phase space of high-dimensional symplectic maps. Nevertheless, we are confident that the capabilities provided by Lagrangian descriptors to explore and visualize the geometry of the invariant manifolds that characterize phase space transport will be of great value to the nonlinear dynamics community in their efforts to address the outstanding challenges that lie ahead in this discipline. In particular, those aimed at disentangling the dynamical behavior of 3 DoF continuous Hamiltonian systems by means of reducing them to their equivalent 4D symplectic map representations.	
	

\section*{Acknowledgments}

The authors would like to acknowledge the financial support provided by the EPSRC Grant No. EP/P021123/1 and the Office of Naval Research Grant No. N00014-01-1-0769. We are also grateful for discussions of this  work with Prof. Srihari Keshavamurthy during his last research visit to the University of Bristol.

\bibliography{SNreac}	

\begin{thebibliography}{63}%
\makeatletter
\providecommand \@ifxundefined [1]{%
 \@ifx{#1\undefined}
}%
\providecommand \@ifnum [1]{%
 \ifnum #1\expandafter \@firstoftwo
 \else \expandafter \@secondoftwo
 \fi
}%
\providecommand \@ifx [1]{%
 \ifx #1\expandafter \@firstoftwo
 \else \expandafter \@secondoftwo
 \fi
}%
\providecommand \natexlab [1]{#1}%
\providecommand \enquote  [1]{``#1''}%
\providecommand \bibnamefont  [1]{#1}%
\providecommand \bibfnamefont [1]{#1}%
\providecommand \citenamefont [1]{#1}%
\providecommand \href@noop [0]{\@secondoftwo}%
\providecommand \href [0]{\begingroup \@sanitize@url \@href}%
\providecommand \@href[1]{\@@startlink{#1}\@@href}%
\providecommand \@@href[1]{\endgroup#1\@@endlink}%
\providecommand \@sanitize@url [0]{\catcode `\\12\catcode `\$12\catcode
  `\&12\catcode `\#12\catcode `\^12\catcode `\_12\catcode `\%12\relax}%
\providecommand \@@startlink[1]{}%
\providecommand \@@endlink[0]{}%
\providecommand \url  [0]{\begingroup\@sanitize@url \@url }%
\providecommand \@url [1]{\endgroup\@href {#1}{\urlprefix }}%
\providecommand \urlprefix  [0]{URL }%
\providecommand \Eprint [0]{\href }%
\providecommand \doibase [0]{https://doi.org/}%
\providecommand \selectlanguage [0]{\@gobble}%
\providecommand \bibinfo  [0]{\@secondoftwo}%
\providecommand \bibfield  [0]{\@secondoftwo}%
\providecommand \translation [1]{[#1]}%
\providecommand \BibitemOpen [0]{}%
\providecommand \bibitemStop [0]{}%
\providecommand \bibitemNoStop [0]{.\EOS\space}%
\providecommand \EOS [0]{\spacefactor3000\relax}%
\providecommand \BibitemShut  [1]{\csname bibitem#1\endcsname}%
\let\auto@bib@innerbib\@empty
\bibitem [{\citenamefont {Mackay}\ \emph {et~al.}(1984)\citenamefont {Mackay},
  \citenamefont {Meiss},\ and\ \citenamefont {Percival}}]{mackay1984}%
  \BibitemOpen
  \bibfield  {author} {\bibinfo {author} {\bibfnamefont {R.~S.}\ \bibnamefont
  {Mackay}}, \bibinfo {author} {\bibfnamefont {J.~D.}\ \bibnamefont {Meiss}},\
  and\ \bibinfo {author} {\bibfnamefont {I.~C.}\ \bibnamefont {Percival}},\
  }\bibfield  {title} {\bibinfo {title} {Transport in hamiltonian systems},\
  }\href {https://doi.org/https://doi.org/10.1016/0167-2789(84)90270-7}
  {\bibfield  {journal} {\bibinfo  {journal} {Physica D}\ }\textbf {\bibinfo
  {volume} {13}},\ \bibinfo {pages} {55} (\bibinfo {year} {1984})}\BibitemShut
  {NoStop}%
\bibitem [{\citenamefont {Meiss}(1992)}]{meiss1992}%
  \BibitemOpen
  \bibfield  {author} {\bibinfo {author} {\bibfnamefont {J.~D.}\ \bibnamefont
  {Meiss}},\ }\bibfield  {title} {\bibinfo {title} {Symplectic maps,
  variational principles, and transport},\ }\href
  {https://doi.org/10.1103/RevModPhys.64.795} {\bibfield  {journal} {\bibinfo
  {journal} {Rev. Mod. Phys.}\ }\textbf {\bibinfo {volume} {64}},\ \bibinfo
  {pages} {795} (\bibinfo {year} {1992})}\BibitemShut {NoStop}%
\bibitem [{\citenamefont {Meiss}(2015)}]{meiss2015}%
  \BibitemOpen
  \bibfield  {author} {\bibinfo {author} {\bibfnamefont {J.~D.}\ \bibnamefont
  {Meiss}},\ }\bibfield  {title} {\bibinfo {title} {Thirty years of turnstiles
  and transport},\ }\href {https://doi.org/https://doi.org/10.1063/1.4915831}
  {\bibfield  {journal} {\bibinfo  {journal} {Chaos}\ }\textbf {\bibinfo
  {volume} {25}},\ \bibinfo {pages} {097602} (\bibinfo {year}
  {2015})}\BibitemShut {NoStop}%
\bibitem [{\citenamefont {B{\" a}cker}\ and\ \citenamefont
  {Meiss}(2020)}]{backer2020}%
  \BibitemOpen
  \bibfield  {author} {\bibinfo {author} {\bibfnamefont {A.}~\bibnamefont {B{\"
  a}cker}}\ and\ \bibinfo {author} {\bibfnamefont {J.~D.}\ \bibnamefont
  {Meiss}},\ }\bibfield  {title} {\bibinfo {title} {Elliptic bubbles in
  {M}oser's 4{D} quadratic map: {T}he quadfurcation},\ }\href
  {https://doi.org/https://doi.org/10.1137/19M1268665} {\bibfield  {journal}
  {\bibinfo  {journal} {SIAM J. Appl. Dyn. Syst.}\ }\textbf {\bibinfo {volume}
  {19}},\ \bibinfo {pages} {442–479} (\bibinfo {year} {2020})}\BibitemShut
  {NoStop}%
\bibitem [{\citenamefont {Huebner}(2020)}]{huebner}%
  \BibitemOpen
  \bibfield  {author} {\bibinfo {author} {\bibfnamefont {F.}~\bibnamefont
  {Huebner}},\ }\href@noop {} {\emph {\bibinfo {title} {Chaotic transport by a
  turnstile mechanism in 4{D} symplectic maps}}}\ (\bibinfo  {publisher}
  {Technical {U}niversity of {D}resden, {D}resden},\ \bibinfo {year}
  {2020})\BibitemShut {NoStop}%
\bibitem [{\citenamefont {Firmbach}(2021)}]{firmbach}%
  \BibitemOpen
  \bibfield  {author} {\bibinfo {author} {\bibfnamefont {M.}~\bibnamefont
  {Firmbach}},\ }\href@noop {} {\emph {\bibinfo {title} {Chaotic transport and
  partial barriers in 4{D} symplectic maps}}}\ (\bibinfo  {publisher}
  {Technical {U}niversity of {D}resden, {D}resden},\ \bibinfo {year}
  {2021})\BibitemShut {NoStop}%
\bibitem [{\citenamefont {Froeschle}(1970)}]{froeschle1970}%
  \BibitemOpen
  \bibfield  {author} {\bibinfo {author} {\bibfnamefont {C.}~\bibnamefont
  {Froeschle}},\ }\bibfield  {title} {\bibinfo {title} {Numerical study of
  dynamical systems of three degrees of freedom. {I}. {G}raphical displays of
  four-dimensional sections},\ }\href@noop {} {\bibfield  {journal} {\bibinfo
  {journal} {Astron. Astrophys.}\ }\textbf {\bibinfo {volume} {4}},\ \bibinfo
  {pages} {115} (\bibinfo {year} {1970})}\BibitemShut {NoStop}%
\bibitem [{\citenamefont {Udry}\ and\ \citenamefont
  {Pfenniger}(1988)}]{udry1988}%
  \BibitemOpen
  \bibfield  {author} {\bibinfo {author} {\bibfnamefont {S.}~\bibnamefont
  {Udry}}\ and\ \bibinfo {author} {\bibfnamefont {D.}~\bibnamefont
  {Pfenniger}},\ }\bibfield  {title} {\bibinfo {title} {Stochasticity in
  elliptical galaxies},\ }\href@noop {} {\bibfield  {journal} {\bibinfo
  {journal} {Astron. Astrophys.}\ }\textbf {\bibinfo {volume} {198}},\ \bibinfo
  {pages} {135} (\bibinfo {year} {1988})}\BibitemShut {NoStop}%
\bibitem [{\citenamefont {Pfenniger}(1985{\natexlab{a}})}]{pfenniger1985}%
  \BibitemOpen
  \bibfield  {author} {\bibinfo {author} {\bibfnamefont {D.}~\bibnamefont
  {Pfenniger}},\ }\bibfield  {title} {\bibinfo {title} {Numerical study of
  complex instability. {I} - {M}appings.},\ }\href@noop {} {\bibfield
  {journal} {\bibinfo  {journal} {Astron. Astrophys.}\ }\textbf {\bibinfo
  {volume} {150}},\ \bibinfo {pages} {97} (\bibinfo {year}
  {1985}{\natexlab{a}})}\BibitemShut {NoStop}%
\bibitem [{\citenamefont {Pfenniger}(1985{\natexlab{b}})}]{pfenniger1985b}%
  \BibitemOpen
  \bibfield  {author} {\bibinfo {author} {\bibfnamefont {D.}~\bibnamefont
  {Pfenniger}},\ }\bibfield  {title} {\bibinfo {title} {Numerical study of
  complex instability. {II} - {B}arred galaxy bulges},\ }\href@noop {}
  {\bibfield  {journal} {\bibinfo  {journal} {Astron. Astrophys.}\ }\textbf
  {\bibinfo {volume} {150}},\ \bibinfo {pages} {112} (\bibinfo {year}
  {1985}{\natexlab{b}})}\BibitemShut {NoStop}%
\bibitem [{\citenamefont {Skokos}\ \emph {et~al.}(1996)\citenamefont {Skokos},
  \citenamefont {Contopoulos},\ and\ \citenamefont {Polymilis}}]{skokos1997}%
  \BibitemOpen
  \bibfield  {author} {\bibinfo {author} {\bibfnamefont {C.}~\bibnamefont
  {Skokos}}, \bibinfo {author} {\bibfnamefont {G.}~\bibnamefont
  {Contopoulos}},\ and\ \bibinfo {author} {\bibfnamefont {C.}~\bibnamefont
  {Polymilis}},\ }\bibfield  {title} {\bibinfo {title} {Structures in the phase
  space of a four dimensional symplectic map},\ }\href
  {https://doi.org/10.1007/BF00053508} {\bibfield  {journal} {\bibinfo
  {journal} {Mech. Dynam. Astronomy}\ }\textbf {\bibinfo {volume} {65}},\
  \bibinfo {pages} {223} (\bibinfo {year} {1996})}\BibitemShut {NoStop}%
\bibitem [{\citenamefont {Martinet}\ and\ \citenamefont
  {Magnenat}(1981)}]{martinet1981}%
  \BibitemOpen
  \bibfield  {author} {\bibinfo {author} {\bibfnamefont {L.}~\bibnamefont
  {Martinet}}\ and\ \bibinfo {author} {\bibfnamefont {P.}~\bibnamefont
  {Magnenat}},\ }\bibfield  {title} {\bibinfo {title} {Invariant surfaces and
  orbital behaviour in dynamical systems with 3 degrees of freedom},\
  }\href@noop {} {\bibfield  {journal} {\bibinfo  {journal} {Astron.
  Astrophys.}\ }\textbf {\bibinfo {volume} {96}},\ \bibinfo {pages} {68}
  (\bibinfo {year} {1981})}\BibitemShut {NoStop}%
\bibitem [{\citenamefont {Magnenat}(1982)}]{magnenat1982}%
  \BibitemOpen
  \bibfield  {author} {\bibinfo {author} {\bibfnamefont {P.}~\bibnamefont
  {Magnenat}},\ }\bibfield  {title} {\bibinfo {title} {Numerical study of
  periodic orbit properties in a dynamical system with three degrees of
  freedom},\ }\href {https://doi.org/10.1007/BF01243741} {\bibfield  {journal}
  {\bibinfo  {journal} {Celestial Mech.}\ }\textbf {\bibinfo {volume} {28}},\
  \bibinfo {pages} {319} (\bibinfo {year} {1982})}\BibitemShut {NoStop}%
\bibitem [{\citenamefont {Vrahatis}\ \emph {et~al.}(1996)\citenamefont
  {Vrahatis}, \citenamefont {Bountis},\ and\ \citenamefont
  {Kollmann}}]{vrahatis1996}%
  \BibitemOpen
  \bibfield  {author} {\bibinfo {author} {\bibfnamefont {M.~N.}\ \bibnamefont
  {Vrahatis}}, \bibinfo {author} {\bibfnamefont {T.}~\bibnamefont {Bountis}},\
  and\ \bibinfo {author} {\bibfnamefont {M.}~\bibnamefont {Kollmann}},\
  }\bibfield  {title} {\bibinfo {title} {Periodic orbits and invariant surfaces
  of 4{D} nonlinear mappings},\ }\href
  {https://doi.org/10.1142/S0218127496000849} {\bibfield  {journal} {\bibinfo
  {journal} {Int. J. Bifurcation Chaos}\ }\textbf {\bibinfo {volume} {6}},\
  \bibinfo {pages} {1425} (\bibinfo {year} {1996})}\BibitemShut {NoStop}%
\bibitem [{\citenamefont {Pa{\v s}kauskas}\ \emph {et~al.}(2008)\citenamefont
  {Pa{\v s}kauskas}, \citenamefont {Chandre},\ and\ \citenamefont
  {Uzer}}]{paskauskas}%
  \BibitemOpen
  \bibfield  {author} {\bibinfo {author} {\bibfnamefont {R.}~\bibnamefont
  {Pa{\v s}kauskas}}, \bibinfo {author} {\bibfnamefont {C.}~\bibnamefont
  {Chandre}},\ and\ \bibinfo {author} {\bibfnamefont {T.}~\bibnamefont
  {Uzer}},\ }\bibfield  {title} {\bibinfo {title} {Dynamical bottlenecks to
  intramolecular energy flow},\ }\href
  {https://doi.org/https://doi.org/10.1103/PhysRevLett.100.083001} {\bibfield
  {journal} {\bibinfo  {journal} {Phys. Rev. Lett.}\ }\textbf {\bibinfo
  {volume} {100}},\ \bibinfo {pages} {083001} (\bibinfo {year}
  {2008})}\BibitemShut {NoStop}%
\bibitem [{\citenamefont {Pa{\v{s}}kauskas}\ \emph {et~al.}(2009)\citenamefont
  {Pa{\v{s}}kauskas}, \citenamefont {Chandre},\ and\ \citenamefont
  {Uzer}}]{pavskauskas2009bottlenecks}%
  \BibitemOpen
  \bibfield  {author} {\bibinfo {author} {\bibfnamefont {R.}~\bibnamefont
  {Pa{\v{s}}kauskas}}, \bibinfo {author} {\bibfnamefont {C.}~\bibnamefont
  {Chandre}},\ and\ \bibinfo {author} {\bibfnamefont {T.}~\bibnamefont
  {Uzer}},\ }\bibfield  {title} {\bibinfo {title} {Bottlenecks to vibrational
  energy flow in carbonyl sulfide: Structures and mechanisms},\ }\href@noop {}
  {\bibfield  {journal} {\bibinfo  {journal} {The Journal of chemical physics}\
  }\textbf {\bibinfo {volume} {130}},\ \bibinfo {pages} {164105} (\bibinfo
  {year} {2009})}\BibitemShut {NoStop}%
\bibitem [{\citenamefont {Patsis}\ and\ \citenamefont
  {Zachilas}(1994)}]{patsis1994}%
  \BibitemOpen
  \bibfield  {author} {\bibinfo {author} {\bibfnamefont {P.}~\bibnamefont
  {Patsis}}\ and\ \bibinfo {author} {\bibfnamefont {L.}~\bibnamefont
  {Zachilas}},\ }\bibfield  {title} {\bibinfo {title} {Using color and rotation
  for visualizing four-dimensional {P}oincare cross-sections: with applications
  to the orbital behavior of three-dimensional {H}amiltonian system},\ }\href
  {https://doi.org/10.1142/S021812749400112X} {\bibfield  {journal} {\bibinfo
  {journal} {Int. J. Bifurcation Chaos}\ }\textbf {\bibinfo {volume} {4}},\
  \bibinfo {pages} {1399} (\bibinfo {year} {1994})}\BibitemShut {NoStop}%
\bibitem [{\citenamefont {Katsanikas}\ and\ \citenamefont
  {Patsis}(2011)}]{katsanikas2011a}%
  \BibitemOpen
  \bibfield  {author} {\bibinfo {author} {\bibfnamefont {M.}~\bibnamefont
  {Katsanikas}}\ and\ \bibinfo {author} {\bibfnamefont {P.}~\bibnamefont
  {Patsis}},\ }\bibfield  {title} {\bibinfo {title} {The structure of invariant
  tori in a 3{D} galactic potential},\ }\href
  {https://doi.org/10.1142/S0218127411028520} {\bibfield  {journal} {\bibinfo
  {journal} {Int. J. Bifurcation Chaos}\ }\textbf {\bibinfo {volume} {21}},\
  \bibinfo {pages} {467} (\bibinfo {year} {2011})}\BibitemShut {NoStop}%
\bibitem [{\citenamefont {Katsanikas}\ \emph
  {et~al.}(2011{\natexlab{a}})\citenamefont {Katsanikas}, \citenamefont
  {Patsis},\ and\ \citenamefont {Contopoulos}}]{katsanikas2011b}%
  \BibitemOpen
  \bibfield  {author} {\bibinfo {author} {\bibfnamefont {M.}~\bibnamefont
  {Katsanikas}}, \bibinfo {author} {\bibfnamefont {P.}~\bibnamefont {Patsis}},\
  and\ \bibinfo {author} {\bibfnamefont {G.}~\bibnamefont {Contopoulos}},\
  }\bibfield  {title} {\bibinfo {title} {The structure and evolution of
  confined tori near a {H}amiltonian hopf bifurcation},\ }\href
  {https://doi.org/10.1142/S0218127411029811} {\bibfield  {journal} {\bibinfo
  {journal} {Int. Journal Bif. Chaos}\ }\textbf {\bibinfo {volume} {21}},\
  \bibinfo {pages} {2321} (\bibinfo {year} {2011}{\natexlab{a}})}\BibitemShut
  {NoStop}%
\bibitem [{\citenamefont {Katsanikas}\ \emph
  {et~al.}(2011{\natexlab{b}})\citenamefont {Katsanikas}, \citenamefont
  {Patsis},\ and\ \citenamefont {Pinotsis}}]{katsanikas2011c}%
  \BibitemOpen
  \bibfield  {author} {\bibinfo {author} {\bibfnamefont {M.}~\bibnamefont
  {Katsanikas}}, \bibinfo {author} {\bibfnamefont {P.}~\bibnamefont {Patsis}},\
  and\ \bibinfo {author} {\bibfnamefont {A.}~\bibnamefont {Pinotsis}},\
  }\bibfield  {title} {\bibinfo {title} {Chains of rotational tori and
  filamentary structures close to high multiplicity periodic orbits in a 3{D}
  galactic potential},\ }\href {https://doi.org/10.1142/S0218127411029823}
  {\bibfield  {journal} {\bibinfo  {journal} {Int. Journal Bif. Chaos}\
  }\textbf {\bibinfo {volume} {21}},\ \bibinfo {pages} {2331} (\bibinfo {year}
  {2011}{\natexlab{b}})}\BibitemShut {NoStop}%
\bibitem [{\citenamefont {Katsanikas}\ \emph {et~al.}(2013)\citenamefont
  {Katsanikas}, \citenamefont {Patsis},\ and\ \citenamefont
  {Contopoulos}}]{katsanikas2013}%
  \BibitemOpen
  \bibfield  {author} {\bibinfo {author} {\bibfnamefont {M.}~\bibnamefont
  {Katsanikas}}, \bibinfo {author} {\bibfnamefont {P.}~\bibnamefont {Patsis}},\
  and\ \bibinfo {author} {\bibfnamefont {G.}~\bibnamefont {Contopoulos}},\
  }\bibfield  {title} {\bibinfo {title} {Instabilities and stickiness in a 3{D}
  rotating galactic potential},\ }\href
  {https://doi.org/10.1142/S021812741330005X} {\bibfield  {journal} {\bibinfo
  {journal} {Int. Journal Bif. Chaos}\ }\textbf {\bibinfo {volume} {23}},\
  \bibinfo {pages} {1330005} (\bibinfo {year} {2013})}\BibitemShut {NoStop}%
\bibitem [{\citenamefont {Zachilas}\ \emph {et~al.}(2013)\citenamefont
  {Zachilas}, \citenamefont {Katsanikas},\ and\ \citenamefont
  {Patsis}}]{zachilas2013structure}%
  \BibitemOpen
  \bibfield  {author} {\bibinfo {author} {\bibfnamefont {L.}~\bibnamefont
  {Zachilas}}, \bibinfo {author} {\bibfnamefont {M.}~\bibnamefont
  {Katsanikas}},\ and\ \bibinfo {author} {\bibfnamefont {P.}~\bibnamefont
  {Patsis}},\ }\bibfield  {title} {\bibinfo {title} {The structure of phase
  space close to fixed points in a 4{D} symplectic map},\ }\href
  {https://doi.org/10.1142/S0218127413300231} {\bibfield  {journal} {\bibinfo
  {journal} {International Journal of Bifurcation and Chaos}\ }\textbf
  {\bibinfo {volume} {23}},\ \bibinfo {pages} {1330023} (\bibinfo {year}
  {2013})}\BibitemShut {NoStop}%
\bibitem [{\citenamefont {Patsis}\ and\ \citenamefont
  {Katsanikas}(2014{\natexlab{a}})}]{patsis2014phase}%
  \BibitemOpen
  \bibfield  {author} {\bibinfo {author} {\bibfnamefont {P.}~\bibnamefont
  {Patsis}}\ and\ \bibinfo {author} {\bibfnamefont {M.}~\bibnamefont
  {Katsanikas}},\ }\bibfield  {title} {\bibinfo {title} {The phase space of
  boxy--peanut and x-shaped bulges in galaxies--{I}. {P}roperties of
  non-periodic orbits},\ }\href {https://doi.org/10.1093/mnras/stu1988}
  {\bibfield  {journal} {\bibinfo  {journal} {Monthly Notices of the Royal
  Astronomical Society}\ }\textbf {\bibinfo {volume} {445}},\ \bibinfo {pages}
  {3525} (\bibinfo {year} {2014}{\natexlab{a}})}\BibitemShut {NoStop}%
\bibitem [{\citenamefont {Martens}\ \emph {et~al.}(1987)\citenamefont
  {Martens}, \citenamefont {Davis},\ and\ \citenamefont {Ezra}}]{martens1987}%
  \BibitemOpen
  \bibfield  {author} {\bibinfo {author} {\bibfnamefont {C.}~\bibnamefont
  {Martens}}, \bibinfo {author} {\bibfnamefont {M.}~\bibnamefont {Davis}},\
  and\ \bibinfo {author} {\bibfnamefont {G.}~\bibnamefont {Ezra}},\ }\bibfield
  {title} {\bibinfo {title} {Local frequency analysis of chaotic motion in
  multidimensional systems: energy transport and bottlenecks in planar {OCS}},\
  }\href {https://doi.org/10.1016/0009-2614(87)80655-3} {\bibfield  {journal}
  {\bibinfo  {journal} {Chem. Phys. Lett.}\ }\textbf {\bibinfo {volume}
  {142}},\ \bibinfo {pages} {519} (\bibinfo {year} {1987})}\BibitemShut
  {NoStop}%
\bibitem [{\citenamefont {Laskar}(1993)}]{laskar1993}%
  \BibitemOpen
  \bibfield  {author} {\bibinfo {author} {\bibfnamefont {J.}~\bibnamefont
  {Laskar}},\ }\bibfield  {title} {\bibinfo {title} {Frequency analysis for
  multi-dimensional systems. {G}lobal dynamics and diffusion},\ }\href
  {https://doi.org/10.1016/0167-2789(93)90210-R} {\bibfield  {journal}
  {\bibinfo  {journal} {Physica D}\ }\textbf {\bibinfo {volume} {67}},\
  \bibinfo {pages} {257} (\bibinfo {year} {1993})}\BibitemShut {NoStop}%
\bibitem [{\citenamefont {Chandre}\ \emph {et~al.}(2003)\citenamefont
  {Chandre}, \citenamefont {Wiggins},\ and\ \citenamefont
  {Uzer}}]{chandre2003}%
  \BibitemOpen
  \bibfield  {author} {\bibinfo {author} {\bibfnamefont {C.}~\bibnamefont
  {Chandre}}, \bibinfo {author} {\bibfnamefont {S.}~\bibnamefont {Wiggins}},\
  and\ \bibinfo {author} {\bibfnamefont {T.}~\bibnamefont {Uzer}},\ }\bibfield
  {title} {\bibinfo {title} {Time–frequency analysis of chaotic systems},\
  }\href {https://doi.org/10.1016/S0167-2789(03)00117-9} {\bibfield  {journal}
  {\bibinfo  {journal} {Physica D}\ }\textbf {\bibinfo {volume} {181}},\
  \bibinfo {pages} {171} (\bibinfo {year} {2003})}\BibitemShut {NoStop}%
\bibitem [{\citenamefont {Sethi}\ and\ \citenamefont
  {Keshavamurthy}(2012)}]{sethi2012}%
  \BibitemOpen
  \bibfield  {author} {\bibinfo {author} {\bibfnamefont {A.}~\bibnamefont
  {Sethi}}\ and\ \bibinfo {author} {\bibfnamefont {S.}~\bibnamefont
  {Keshavamurthy}},\ }\bibfield  {title} {\bibinfo {title} {Driven coupled
  {M}orse oscillators: visualizing the phase space and characterizing the
  transport},\ }\href {https://doi.org/10.1080/00268976.2012.667166} {\bibfield
   {journal} {\bibinfo  {journal} {Mol. Phys.}\ }\textbf {\bibinfo {volume}
  {110}},\ \bibinfo {pages} {717} (\bibinfo {year} {2012})}\BibitemShut
  {NoStop}%
\bibitem [{\citenamefont {Froeschle}(1972)}]{froeschle1972}%
  \BibitemOpen
  \bibfield  {author} {\bibinfo {author} {\bibfnamefont {C.}~\bibnamefont
  {Froeschle}},\ }\bibfield  {title} {\bibinfo {title} {Numerical study of a
  four-dimensional mapping},\ }\href@noop {} {\bibfield  {journal} {\bibinfo
  {journal} {Astron. Astrophys.}\ }\textbf {\bibinfo {volume} {16}},\ \bibinfo
  {pages} {172} (\bibinfo {year} {1972})}\BibitemShut {NoStop}%
\bibitem [{\citenamefont {Froeschle}\ and\ \citenamefont
  {Lega}(2000)}]{froeschle2000}%
  \BibitemOpen
  \bibfield  {author} {\bibinfo {author} {\bibfnamefont {C.}~\bibnamefont
  {Froeschle}}\ and\ \bibinfo {author} {\bibfnamefont {E.}~\bibnamefont
  {Lega}},\ }\bibfield  {title} {\bibinfo {title} {On the structure of
  symplectic mappings. the fast {L}yapunov indicator: a very sensitive tool},\
  }\href {https://doi.org/10.1023/A:1011141018230} {\bibfield  {journal}
  {\bibinfo  {journal} {Celest. Mech. Dynam. Astronomy}\ }\textbf {\bibinfo
  {volume} {78}},\ \bibinfo {pages} {167} (\bibinfo {year} {2000})}\BibitemShut
  {NoStop}%
\bibitem [{\citenamefont {Bazzani}\ \emph {et~al.}(1998)\citenamefont
  {Bazzani}, \citenamefont {Bongini},\ and\ \citenamefont
  {Turchetti}}]{bazzani1998}%
  \BibitemOpen
  \bibfield  {author} {\bibinfo {author} {\bibfnamefont {A.}~\bibnamefont
  {Bazzani}}, \bibinfo {author} {\bibfnamefont {L.}~\bibnamefont {Bongini}},\
  and\ \bibinfo {author} {\bibfnamefont {G.}~\bibnamefont {Turchetti}},\
  }\bibfield  {title} {\bibinfo {title} {Analysis of resonances in action space
  for symplectic maps},\ }\href {https://doi.org/10.1103/PhysRevE.57.1178}
  {\bibfield  {journal} {\bibinfo  {journal} {Phys. Rev. E}\ }\textbf {\bibinfo
  {volume} {57}},\ \bibinfo {pages} {1178} (\bibinfo {year}
  {1998})}\BibitemShut {NoStop}%
\bibitem [{\citenamefont {Madrid}\ and\ \citenamefont
  {Mancho}(2009)}]{madrid2009}%
  \BibitemOpen
  \bibfield  {author} {\bibinfo {author} {\bibfnamefont {J.~A.~J.}\
  \bibnamefont {Madrid}}\ and\ \bibinfo {author} {\bibfnamefont {A.~M.}\
  \bibnamefont {Mancho}},\ }\bibfield  {title} {\bibinfo {title}
  {{Distinguished trajectories in time dependent vector fields}},\ }\href
  {https://doi.org/10.1063/1.3056050} {\bibfield  {journal} {\bibinfo
  {journal} {Chaos}\ }\textbf {\bibinfo {volume} {19}},\ \bibinfo {pages}
  {013111} (\bibinfo {year} {2009})}\BibitemShut {NoStop}%
\bibitem [{\citenamefont {Mendoza}\ and\ \citenamefont
  {Mancho}(2010)}]{Mancho1}%
  \BibitemOpen
  \bibfield  {author} {\bibinfo {author} {\bibfnamefont {C.}~\bibnamefont
  {Mendoza}}\ and\ \bibinfo {author} {\bibfnamefont {A.~M.}\ \bibnamefont
  {Mancho}},\ }\bibfield  {title} {\bibinfo {title} {The hidden geometry of
  ocean flows},\ }\href
  {https://doi.org/https://journals.aps.org/prl/abstract/10.1103/PhysRevLett.105.038501}
  {\bibfield  {journal} {\bibinfo  {journal} {Phys. Rev. Lett.}\ }\textbf
  {\bibinfo {volume} {105}},\ \bibinfo {pages} {038501} (\bibinfo {year}
  {2010})}\BibitemShut {NoStop}%
\bibitem [{\citenamefont {Mancho}\ \emph {et~al.}(2013)\citenamefont {Mancho},
  \citenamefont {Wiggins}, \citenamefont {Curbelo},\ and\ \citenamefont
  {Mendoza}}]{mancho2013lagrangian}%
  \BibitemOpen
  \bibfield  {author} {\bibinfo {author} {\bibfnamefont {A.~M.}\ \bibnamefont
  {Mancho}}, \bibinfo {author} {\bibfnamefont {S.}~\bibnamefont {Wiggins}},
  \bibinfo {author} {\bibfnamefont {J.}~\bibnamefont {Curbelo}},\ and\ \bibinfo
  {author} {\bibfnamefont {C.}~\bibnamefont {Mendoza}},\ }\bibfield  {title}
  {\bibinfo {title} {Lagrangian descriptors: A method for revealing phase space
  structures of general time dependent dynamical systems},\ }\href
  {https://doi.org/https://doi.org/10.1016/j.cnsns.2013.05.002} {\bibfield
  {journal} {\bibinfo  {journal} {Commun. Nonlinear Sci. Numer. Simul.}\
  }\textbf {\bibinfo {volume} {18}},\ \bibinfo {pages} {3530} (\bibinfo {year}
  {2013})}\BibitemShut {NoStop}%
\bibitem [{\citenamefont {Lopesino}\ \emph {et~al.}(2017)\citenamefont
  {Lopesino}, \citenamefont {Balibrea-Iniesta}, \citenamefont
  {Garc\'ia-Garrido}, \citenamefont {Wiggins},\ and\ \citenamefont
  {Mancho}}]{lopesino2017}%
  \BibitemOpen
  \bibfield  {author} {\bibinfo {author} {\bibfnamefont {C.}~\bibnamefont
  {Lopesino}}, \bibinfo {author} {\bibfnamefont {F.}~\bibnamefont
  {Balibrea-Iniesta}}, \bibinfo {author} {\bibfnamefont {V.~J.}\ \bibnamefont
  {Garc\'ia-Garrido}}, \bibinfo {author} {\bibfnamefont {S.}~\bibnamefont
  {Wiggins}},\ and\ \bibinfo {author} {\bibfnamefont {A.~M.}\ \bibnamefont
  {Mancho}},\ }\bibfield  {title} {\bibinfo {title} {A theoretical {F}ramework
  for {L}agrangian {D}escriptors},\ }\href
  {https://doi.org/10.1142/S0218127417300014} {\bibfield  {journal} {\bibinfo
  {journal} {Int J Bifurc Chaos}\ }\textbf {\bibinfo {volume} {27}},\ \bibinfo
  {pages} {1730001} (\bibinfo {year} {2017})}\BibitemShut {NoStop}%
\bibitem [{\citenamefont {Agaoglou}\ \emph {et~al.}(2020)\citenamefont
  {Agaoglou}, \citenamefont {Aguilar-Sanjuan}, \citenamefont
  {Garc{\'i}a-Garrido}, \citenamefont {Gonz{\'a}lez-Montoya}, \citenamefont
  {Katsanikas}, \citenamefont {Krajňák}, \citenamefont {Naik},\ and\
  \citenamefont {Wiggins}}]{ldbook2020}%
  \BibitemOpen
  \bibfield  {author} {\bibinfo {author} {\bibfnamefont {M.}~\bibnamefont
  {Agaoglou}}, \bibinfo {author} {\bibfnamefont {B.}~\bibnamefont
  {Aguilar-Sanjuan}}, \bibinfo {author} {\bibfnamefont {V.~J.}\ \bibnamefont
  {Garc{\'i}a-Garrido}}, \bibinfo {author} {\bibfnamefont {F.}~\bibnamefont
  {Gonz{\'a}lez-Montoya}}, \bibinfo {author} {\bibfnamefont {M.}~\bibnamefont
  {Katsanikas}}, \bibinfo {author} {\bibfnamefont {V.}~\bibnamefont
  {Krajňák}}, \bibinfo {author} {\bibfnamefont {S.}~\bibnamefont {Naik}},\
  and\ \bibinfo {author} {\bibfnamefont {S.}~\bibnamefont {Wiggins}},\ }\href
  {https://doi.org/10.5281/zenodo.3958985} {\emph {\bibinfo {title} {Lagrangian
  Descriptors: Discovery and Quantification of Phase Space Structure and
  Transport}}}\ (\bibinfo  {publisher} {zenodo: 10.5281/zenodo.3958985},\
  \bibinfo {year} {2020})\BibitemShut {NoStop}%
\bibitem [{\citenamefont {Craven}\ and\ \citenamefont
  {Hernandez}(2015)}]{craven2015lagrangian}%
  \BibitemOpen
  \bibfield  {author} {\bibinfo {author} {\bibfnamefont {G.~T.}\ \bibnamefont
  {Craven}}\ and\ \bibinfo {author} {\bibfnamefont {R.}~\bibnamefont
  {Hernandez}},\ }\bibfield  {title} {\bibinfo {title} {Lagrangian descriptors
  of thermalized transition states on time-varying energy surfaces},\ }\href
  {https://doi.org/10.1103/PhysRevLett.115.148301} {\bibfield  {journal}
  {\bibinfo  {journal} {Phys Rev Lett}\ }\textbf {\bibinfo {volume} {115}},\
  \bibinfo {pages} {148301} (\bibinfo {year} {2015})}\BibitemShut {NoStop}%
\bibitem [{\citenamefont {Naik}\ \emph {et~al.}(2019)\citenamefont {Naik},
  \citenamefont {Garc\'{i}a-Garrido},\ and\ \citenamefont
  {Wiggins}}]{naik2019a}%
  \BibitemOpen
  \bibfield  {author} {\bibinfo {author} {\bibfnamefont {S.}~\bibnamefont
  {Naik}}, \bibinfo {author} {\bibfnamefont {V.~J.}\ \bibnamefont
  {Garc\'{i}a-Garrido}},\ and\ \bibinfo {author} {\bibfnamefont
  {S.}~\bibnamefont {Wiggins}},\ }\bibfield  {title} {\bibinfo {title} {Finding
  {NHIM}: Identifying high dimensional phase space structures in reaction
  dynamics using {L}agrangian descriptors},\ }\href
  {https://doi.org/10.1016/j.cnsns.2019.104907} {\bibfield  {journal} {\bibinfo
   {journal} {Communications in Nonlinear Science and Numerical Simulation}\
  }\textbf {\bibinfo {volume} {79}},\ \bibinfo {pages} {104907} (\bibinfo
  {year} {2019})}\BibitemShut {NoStop}%
\bibitem [{\citenamefont {Garc\'{i}a-Garrido}\ \emph
  {et~al.}(2020{\natexlab{a}})\citenamefont {Garc\'{i}a-Garrido}, \citenamefont
  {Naik},\ and\ \citenamefont {Wiggins}}]{GG2019a}%
  \BibitemOpen
  \bibfield  {author} {\bibinfo {author} {\bibfnamefont {V.~J.}\ \bibnamefont
  {Garc\'{i}a-Garrido}}, \bibinfo {author} {\bibfnamefont {S.}~\bibnamefont
  {Naik}},\ and\ \bibinfo {author} {\bibfnamefont {S.}~\bibnamefont
  {Wiggins}},\ }\bibfield  {title} {\bibinfo {title} {Tilting and squeezing:
  {P}hase space geometry of {H}amiltonian saddle-node bifurcation and its
  influence on chemical reaction dynamics},\ }\href
  {https://doi.org/10.1142/S0218127420300086} {\bibfield  {journal} {\bibinfo
  {journal} {International Journal of Bifurcation and Chaos}\ }\textbf
  {\bibinfo {volume} {30}},\ \bibinfo {pages} {2030008} (\bibinfo {year}
  {2020}{\natexlab{a}})}\BibitemShut {NoStop}%
\bibitem [{\citenamefont {Lopesino}\ \emph {et~al.}(2015)\citenamefont
  {Lopesino}, \citenamefont {Balibrea}, \citenamefont {Wiggins},\ and\
  \citenamefont {Mancho.}}]{carlos}%
  \BibitemOpen
  \bibfield  {author} {\bibinfo {author} {\bibfnamefont {C.}~\bibnamefont
  {Lopesino}}, \bibinfo {author} {\bibfnamefont {F.}~\bibnamefont {Balibrea}},
  \bibinfo {author} {\bibfnamefont {S.}~\bibnamefont {Wiggins}},\ and\ \bibinfo
  {author} {\bibfnamefont {A.~M.}\ \bibnamefont {Mancho.}},\ }\bibfield
  {title} {\bibinfo {title} {{Lagrangian descriptors for two dimensional, area
  preserving autonomous and nonautonomous maps}},\ }\href
  {https://doi.org/10.1016/j.cnsns.2015.02.022} {\bibfield  {journal} {\bibinfo
   {journal} {Commun Nonlinear Sci Numer Simul}\ }\textbf {\bibinfo {volume}
  {27}},\ \bibinfo {pages} {40} (\bibinfo {year} {2015})}\BibitemShut {NoStop}%
\bibitem [{\citenamefont {Garc\'{i}a-Garrido}(2020)}]{GG2019b}%
  \BibitemOpen
  \bibfield  {author} {\bibinfo {author} {\bibfnamefont {V.~J.}\ \bibnamefont
  {Garc\'{i}a-Garrido}},\ }\bibfield  {title} {\bibinfo {title} {An extension
  of discrete {L}agrangian descriptors for unbounded maps},\ }\href
  {https://doi.org/10.1142/S0218127420300128} {\bibfield  {journal} {\bibinfo
  {journal} {International Journal of Bifurcation and Chaos}\ }\textbf
  {\bibinfo {volume} {30}},\ \bibinfo {pages} {2030012} (\bibinfo {year}
  {2020})}\BibitemShut {NoStop}%
\bibitem [{\citenamefont {Wiggins}(1990)}]{wiggins90}%
  \BibitemOpen
  \bibfield  {author} {\bibinfo {author} {\bibfnamefont {S.}~\bibnamefont
  {Wiggins}},\ }\bibfield  {title} {\bibinfo {title} {On the geometry of
  transport in phase space {I}. {T}ransport in $k$ degree-of-freedom
  {H}amiltonian systems, $2 \le k < \infty$},\ }\href
  {https://doi.org/10.1016/0167-2789(90)90159-M} {\bibfield  {journal}
  {\bibinfo  {journal} {Physica D}\ }\textbf {\bibinfo {volume} {44}},\
  \bibinfo {pages} {471} (\bibinfo {year} {1990})}\BibitemShut {NoStop}%
\bibitem [{\citenamefont {Gillilan}\ and\ \citenamefont {Ezra}(1991)}]{ezra}%
  \BibitemOpen
  \bibfield  {author} {\bibinfo {author} {\bibfnamefont {E.}~\bibnamefont
  {Gillilan}}\ and\ \bibinfo {author} {\bibfnamefont {G.}~\bibnamefont
  {Ezra}},\ }\bibfield  {title} {\bibinfo {title} {Transport and turnstiles in
  multidimensional {H}amiltonian mappings for unimolecular fragmentation:
  Application to van der {W}aals predissociation},\ }\href
  {https://doi.org/doi.org/10.1063/1.459840} {\bibfield  {journal} {\bibinfo
  {journal} {J. Chem. Phys.}\ }\textbf {\bibinfo {volume} {94}},\ \bibinfo
  {pages} {2648} (\bibinfo {year} {1991})}\BibitemShut {NoStop}%
\bibitem [{\citenamefont {Beigie}(1995{\natexlab{a}})}]{beigie1995}%
  \BibitemOpen
  \bibfield  {author} {\bibinfo {author} {\bibfnamefont {D.}~\bibnamefont
  {Beigie}},\ }\bibfield  {title} {\bibinfo {title} {Codimension-one
  partitioning and phase space transport in multi-degree-of-freedom
  {H}amiltonian systems with non-toroidal invariant manifold intersections},\
  }\href {https://doi.org/10.1016/0960-0779(94)E0133-A} {\bibfield  {journal}
  {\bibinfo  {journal} {Chaos, Solitons \& Fractals}\ }\textbf {\bibinfo
  {volume} {5}},\ \bibinfo {pages} {177} (\bibinfo {year}
  {1995}{\natexlab{a}})}\BibitemShut {NoStop}%
\bibitem [{\citenamefont {Beigie}(1995{\natexlab{b}})}]{beigie1995b}%
  \BibitemOpen
  \bibfield  {author} {\bibinfo {author} {\bibfnamefont {D.}~\bibnamefont
  {Beigie}},\ }\bibfield  {title} {\bibinfo {title} {Multiple separatrix
  crossing in multi-degree-of-freedom {H}amiltonian flows},\ }\bibfield
  {journal} {\bibinfo  {journal} {Journal of Nonlinear Science}\ }\textbf
  {\bibinfo {volume} {5}},\ \href {https://doi.org/10.1007/BF01869100}
  {10.1007/BF01869100} (\bibinfo {year} {1995}{\natexlab{b}})\BibitemShut
  {NoStop}%
\bibitem [{\citenamefont {Toda}(1995)}]{toda1995}%
  \BibitemOpen
  \bibfield  {author} {\bibinfo {author} {\bibfnamefont {M.}~\bibnamefont
  {Toda}},\ }\bibfield  {title} {\bibinfo {title} {Crisis in chaotic scattering
  of a highly excited van der {W}aals complex},\ }\href
  {https://doi.org/10.1103/PhysRevLett.74.2670} {\bibfield  {journal} {\bibinfo
   {journal} {Physical Review Letters}\ }\textbf {\bibinfo {volume} {74}},\
  \bibinfo {pages} {2670} (\bibinfo {year} {1995})}\BibitemShut {NoStop}%
\bibitem [{\citenamefont {Richter}\ \emph {et~al.}(2014)\citenamefont
  {Richter}, \citenamefont {Lange}, \citenamefont {B\"acker},\ and\
  \citenamefont {Ketzmerick}}]{richter2014}%
  \BibitemOpen
  \bibfield  {author} {\bibinfo {author} {\bibfnamefont {M.}~\bibnamefont
  {Richter}}, \bibinfo {author} {\bibfnamefont {S.}~\bibnamefont {Lange}},
  \bibinfo {author} {\bibfnamefont {A.}~\bibnamefont {B\"acker}},\ and\
  \bibinfo {author} {\bibfnamefont {R.}~\bibnamefont {Ketzmerick}},\ }\bibfield
   {title} {\bibinfo {title} {Visualization and comparison of classical
  structures and quantum states of four-dimensional maps},\ }\href
  {https://doi.org/10.1103/PhysRevE.89.022902} {\bibfield  {journal} {\bibinfo
  {journal} {Phys. Rev. E}\ }\textbf {\bibinfo {volume} {89}},\ \bibinfo
  {pages} {022902} (\bibinfo {year} {2014})}\BibitemShut {NoStop}%
\bibitem [{\citenamefont {Gaspard}\ and\ \citenamefont
  {Rice}(1989)}]{gaspard1989}%
  \BibitemOpen
  \bibfield  {author} {\bibinfo {author} {\bibfnamefont {P.}~\bibnamefont
  {Gaspard}}\ and\ \bibinfo {author} {\bibfnamefont {S.~A.}\ \bibnamefont
  {Rice}},\ }\bibfield  {title} {\bibinfo {title} {{H}amiltonian mapping models
  of molecular fragmentation},\ }\href {https://doi.org/10.1021/j100356a014}
  {\bibfield  {journal} {\bibinfo  {journal} {The Journal of Physical
  Chemistry}\ }\textbf {\bibinfo {volume} {93}},\ \bibinfo {pages} {6947}
  (\bibinfo {year} {1989})}\BibitemShut {NoStop}%
\bibitem [{\citenamefont {Gonzalez~Montoya}\ and\ \citenamefont
  {Wiggins}(2020)}]{montoya2020b}%
  \BibitemOpen
  \bibfield  {author} {\bibinfo {author} {\bibfnamefont {F.}~\bibnamefont
  {Gonzalez~Montoya}}\ and\ \bibinfo {author} {\bibfnamefont {S.}~\bibnamefont
  {Wiggins}},\ }\bibfield  {title} {\bibinfo {title} {Phase space structure and
  escape time dynamics in a van der {W}aals model for exothermic reactions},\
  }\href {https://doi.org/10.1103/PhysRevE.102.062203} {\bibfield  {journal}
  {\bibinfo  {journal} {Phys. Rev. E}\ }\textbf {\bibinfo {volume} {102}},\
  \bibinfo {pages} {062203} (\bibinfo {year} {2020})}\BibitemShut {NoStop}%
\bibitem [{\citenamefont {Gentry}(1984)}]{gentry1984vibrationally}%
  \BibitemOpen
  \bibfield  {author} {\bibinfo {author} {\bibfnamefont {W.~R.}\ \bibnamefont
  {Gentry}},\ }\href@noop {} {\emph {\bibinfo {title} {Vibrationally Excited
  States of Polyatomic van der Waals Molecules: Lifetimes and Decay
  Mechanisms}}}\ (\bibinfo  {publisher} {ACS Publications},\ \bibinfo {year}
  {1984})\BibitemShut {NoStop}%
\bibitem [{\citenamefont {Janda}(1985)}]{janda1985predissociation}%
  \BibitemOpen
  \bibfield  {author} {\bibinfo {author} {\bibfnamefont {K.~C.}\ \bibnamefont
  {Janda}},\ }\bibfield  {title} {\bibinfo {title} {Predissociation of
  polyatomic van der {W}aals molecules},\ }\href@noop {} {\bibfield  {journal}
  {\bibinfo  {journal} {Advances in chemical physics}\ }\textbf {\bibinfo
  {volume} {60}},\ \bibinfo {pages} {201} (\bibinfo {year} {1985})}\BibitemShut
  {NoStop}%
\bibitem [{\citenamefont {García-Garrido}(2020)}]{GG2020a}%
  \BibitemOpen
  \bibfield  {author} {\bibinfo {author} {\bibfnamefont {V.~J.}\ \bibnamefont
  {García-Garrido}},\ }\bibfield  {title} {\bibinfo {title} {Unveiling the
  fractal structure of julia sets with {L}agrangian descriptors},\ }\href
  {https://doi.org/https://doi.org/10.1016/j.cnsns.2020.105417} {\bibfield
  {journal} {\bibinfo  {journal} {Communications in Nonlinear Science and
  Numerical Simulation}\ }\textbf {\bibinfo {volume} {91}},\ \bibinfo {pages}
  {105417} (\bibinfo {year} {2020})}\BibitemShut {NoStop}%
\bibitem [{\citenamefont {Kolmogorov}(1954)}]{kolmogorov1954}%
  \BibitemOpen
  \bibfield  {author} {\bibinfo {author} {\bibfnamefont {A.}~\bibnamefont
  {Kolmogorov}},\ }\bibfield  {title} {\bibinfo {title} {On the conservation of
  conditionally periodic motions under small perturbation of the
  {H}amiltonian},\ }\href@noop {} {\bibfield  {journal} {\bibinfo  {journal}
  {Dokl. Akad. Nauk SSSR}\ }\textbf {\bibinfo {volume} {98}},\ \bibinfo {pages}
  {527} (\bibinfo {year} {1954})}\BibitemShut {NoStop}%
\bibitem [{\citenamefont {Arnold}(1963)}]{arnold1963}%
  \BibitemOpen
  \bibfield  {author} {\bibinfo {author} {\bibfnamefont {V.}~\bibnamefont
  {Arnold}},\ }\bibfield  {title} {\bibinfo {title} {Proof of an {K}olmogorov's
  theorem on the conservation of conditionally periodic motions with a small
  variation in the {H}amiltonian},\ }\href@noop {} {\bibfield  {journal}
  {\bibinfo  {journal} {Russ. Math. Surv.}\ }\textbf {\bibinfo {volume} {18}},\
  \bibinfo {pages} {9} (\bibinfo {year} {1963})}\BibitemShut {NoStop}%
\bibitem [{\citenamefont {Moser}(1962)}]{moser1962}%
  \BibitemOpen
  \bibfield  {author} {\bibinfo {author} {\bibfnamefont {J.}~\bibnamefont
  {Moser}},\ }\bibfield  {title} {\bibinfo {title} {On invariant curves of
  area-preserving mappings of an annulus},\ }\href@noop {} {\bibfield
  {journal} {\bibinfo  {journal} {Nachr. Akad. Wiss. G{\"o}ttingen II
  (Vandenhoeck \& Ruprecht)}\ ,\ \bibinfo {pages} {1–20}} (\bibinfo {year}
  {1962})}\BibitemShut {NoStop}%
\bibitem [{\citenamefont {Gonzalez~Montoya}\ \emph {et~al.}(2020)\citenamefont
  {Gonzalez~Montoya}, \citenamefont {Borondo},\ and\ \citenamefont
  {Jung}}]{Montoya2020a}%
  \BibitemOpen
  \bibfield  {author} {\bibinfo {author} {\bibfnamefont {F.}~\bibnamefont
  {Gonzalez~Montoya}}, \bibinfo {author} {\bibfnamefont {F.}~\bibnamefont
  {Borondo}},\ and\ \bibinfo {author} {\bibfnamefont {C.}~\bibnamefont
  {Jung}},\ }\bibfield  {title} {\bibinfo {title} {Atom scattering off a
  vibrating surface: An example of chaotic scattering with three degrees of
  freedom},\ }\href {https://doi.org/10.1016/j.cnsns.2020.105282} {\bibfield
  {journal} {\bibinfo  {journal} {Commun. Nonlinear Sci. Numer. Simulat.}\
  }\textbf {\bibinfo {volume} {90}},\ \bibinfo {pages} {105282} (\bibinfo
  {year} {2020})}\BibitemShut {NoStop}%
\bibitem [{\citenamefont {Contopoulos}(2002)}]{contopoulos2002}%
  \BibitemOpen
  \bibfield  {author} {\bibinfo {author} {\bibfnamefont {G.}~\bibnamefont
  {Contopoulos}},\ }\bibfield  {title} {\bibinfo {title} {Order and chaos in
  dynamical astronomy},\ }\href@noop {} {\bibfield  {journal} {\bibinfo
  {journal} {Springer-Verlag}\ } (\bibinfo {year} {2002})}\BibitemShut
  {NoStop}%
\bibitem [{\citenamefont {Lukes-Gerakopoulos}\ \emph
  {et~al.}(2016)\citenamefont {Lukes-Gerakopoulos}, \citenamefont {Katsanikas},
  \citenamefont {Patsis},\ and\ \citenamefont {Seyrich}}]{lukes2016dynamics}%
  \BibitemOpen
  \bibfield  {author} {\bibinfo {author} {\bibfnamefont {G.}~\bibnamefont
  {Lukes-Gerakopoulos}}, \bibinfo {author} {\bibfnamefont {M.}~\bibnamefont
  {Katsanikas}}, \bibinfo {author} {\bibfnamefont {P.~A.}\ \bibnamefont
  {Patsis}},\ and\ \bibinfo {author} {\bibfnamefont {J.}~\bibnamefont
  {Seyrich}},\ }\bibfield  {title} {\bibinfo {title} {Dynamics of a spinning
  particle in a linear in spin {H}amiltonian approximation},\ }\href
  {https://doi.org/10.1103/PhysRevD.94.024024} {\bibfield  {journal} {\bibinfo
  {journal} {Physical Review D}\ }\textbf {\bibinfo {volume} {94}},\ \bibinfo
  {pages} {024024} (\bibinfo {year} {2016})}\BibitemShut {NoStop}%
\bibitem [{\citenamefont {Patsis}\ and\ \citenamefont
  {Katsanikas}(2014{\natexlab{b}})}]{patsis2014phase2}%
  \BibitemOpen
  \bibfield  {author} {\bibinfo {author} {\bibfnamefont {P.}~\bibnamefont
  {Patsis}}\ and\ \bibinfo {author} {\bibfnamefont {M.}~\bibnamefont
  {Katsanikas}},\ }\bibfield  {title} {\bibinfo {title} {The phase space of
  boxy--peanut and x-shaped bulges in galaxies--{II}. {T}he relation between
  face-on and edge-on boxiness},\ }\href
  {https://doi.org/10.1093/mnras/stu1970} {\bibfield  {journal} {\bibinfo
  {journal} {Monthly Notices of the Royal Astronomical Society}\ }\textbf
  {\bibinfo {volume} {445}},\ \bibinfo {pages} {3546} (\bibinfo {year}
  {2014}{\natexlab{b}})}\BibitemShut {NoStop}%
\bibitem [{\citenamefont {Milani}\ and\ \citenamefont
  {Nobili}(1992)}]{milani1992example}%
  \BibitemOpen
  \bibfield  {author} {\bibinfo {author} {\bibfnamefont {A.}~\bibnamefont
  {Milani}}\ and\ \bibinfo {author} {\bibfnamefont {A.~M.}\ \bibnamefont
  {Nobili}},\ }\bibfield  {title} {\bibinfo {title} {An example of stable chaos
  in the solar system},\ }\href {https://doi.org/10.1038/357569a0} {\bibfield
  {journal} {\bibinfo  {journal} {Nature}\ }\textbf {\bibinfo {volume} {357}},\
  \bibinfo {pages} {569} (\bibinfo {year} {1992})}\BibitemShut {NoStop}%
\bibitem [{\citenamefont {Karmakar}\ \emph {et~al.}(2020)\citenamefont
  {Karmakar}, \citenamefont {Yadav},\ and\ \citenamefont
  {Keshavamurthy}}]{karmakar2020stable}%
  \BibitemOpen
  \bibfield  {author} {\bibinfo {author} {\bibfnamefont {S.}~\bibnamefont
  {Karmakar}}, \bibinfo {author} {\bibfnamefont {P.~K.}\ \bibnamefont
  {Yadav}},\ and\ \bibinfo {author} {\bibfnamefont {S.}~\bibnamefont
  {Keshavamurthy}},\ }\bibfield  {title} {\bibinfo {title} {Stable chaos and
  delayed onset of statisticality in unimolecular dissociation reactions},\
  }\href {https://doi.org/10.1038/s42004-019-0252-y} {\bibfield  {journal}
  {\bibinfo  {journal} {Communications Chemistry}\ }\textbf {\bibinfo {volume}
  {3}},\ \bibinfo {pages} {1} (\bibinfo {year} {2020})}\BibitemShut {NoStop}%
\bibitem [{\citenamefont {Agaoglou}\ \emph {et~al.}(2019)\citenamefont
  {Agaoglou}, \citenamefont {Aguilar-Sanjuan}, \citenamefont
  {Garc{\'i}a-Garrido}, \citenamefont {Garc{\'i}a-Meseguer}, \citenamefont
  {Gonz{\'a}lez-Montoya}, \citenamefont {Katsanikas}, \citenamefont
  {Krajňák}, \citenamefont {Naik},\ and\ \citenamefont
  {Wiggins}}]{Agaoglou2019}%
  \BibitemOpen
  \bibfield  {author} {\bibinfo {author} {\bibfnamefont {M.}~\bibnamefont
  {Agaoglou}}, \bibinfo {author} {\bibfnamefont {B.}~\bibnamefont
  {Aguilar-Sanjuan}}, \bibinfo {author} {\bibfnamefont {V.~J.}\ \bibnamefont
  {Garc{\'i}a-Garrido}}, \bibinfo {author} {\bibfnamefont {R.}~\bibnamefont
  {Garc{\'i}a-Meseguer}}, \bibinfo {author} {\bibfnamefont {F.}~\bibnamefont
  {Gonz{\'a}lez-Montoya}}, \bibinfo {author} {\bibfnamefont {M.}~\bibnamefont
  {Katsanikas}}, \bibinfo {author} {\bibfnamefont {V.}~\bibnamefont
  {Krajňák}}, \bibinfo {author} {\bibfnamefont {S.}~\bibnamefont {Naik}},\
  and\ \bibinfo {author} {\bibfnamefont {S.}~\bibnamefont {Wiggins}},\ }\href
  {https://doi.org/10.5281/zenodo.3568210} {\emph {\bibinfo {title} {Chemical
  Reactions: A Journey into Phase Space}}}\ (\bibinfo  {publisher}
  {zenodo:10.5281/zenodo.3568210},\ \bibinfo {year} {2019})\BibitemShut
  {NoStop}%
\bibitem [{\citenamefont {Balibrea-Iniesta}\ \emph {et~al.}(2016)\citenamefont
  {Balibrea-Iniesta}, \citenamefont {Lopesino}, \citenamefont {Wiggins},\ and\
  \citenamefont {Mancho}}]{balibrea2016}%
  \BibitemOpen
  \bibfield  {author} {\bibinfo {author} {\bibfnamefont {F.}~\bibnamefont
  {Balibrea-Iniesta}}, \bibinfo {author} {\bibfnamefont {C.}~\bibnamefont
  {Lopesino}}, \bibinfo {author} {\bibfnamefont {S.}~\bibnamefont {Wiggins}},\
  and\ \bibinfo {author} {\bibfnamefont {A.~M.}\ \bibnamefont {Mancho}},\
  }\bibfield  {title} {\bibinfo {title} {Lagrangian descriptors for stochastic
  differential equations: A tool for revealing the phase portrait of stochastic
  dynamical systems},\ }\href {https://doi.org/10.1142/S0218127416300366}
  {\bibfield  {journal} {\bibinfo  {journal} {International Journal of
  Bifurcation and Chaos}\ }\textbf {\bibinfo {volume} {26}},\ \bibinfo {pages}
  {1630036} (\bibinfo {year} {2016})}\BibitemShut {NoStop}%
\bibitem [{\citenamefont {Garc\'{i}a-Garrido}\ \emph
  {et~al.}(2020{\natexlab{b}})\citenamefont {Garc\'{i}a-Garrido}, \citenamefont
  {Agaoglou},\ and\ \citenamefont {Wiggins}}]{GG2020b}%
  \BibitemOpen
  \bibfield  {author} {\bibinfo {author} {\bibfnamefont {V.~J.}\ \bibnamefont
  {Garc\'{i}a-Garrido}}, \bibinfo {author} {\bibfnamefont {M.}~\bibnamefont
  {Agaoglou}},\ and\ \bibinfo {author} {\bibfnamefont {S.}~\bibnamefont
  {Wiggins}},\ }\bibfield  {title} {\bibinfo {title} {Exploring isomerization
  dynamics on a potential energy surface with an index-2 saddle using
  lagrangian descriptors},\ }\href
  {https://doi.org/10.1146/annurev.pc.32.100181.001111} {\bibfield  {journal}
  {\bibinfo  {journal} {Commun Nonlinear Sci Numer Simulat}\ }\textbf {\bibinfo
  {volume} {89}},\ \bibinfo {pages} {105331} (\bibinfo {year}
  {2020}{\natexlab{b}})}\BibitemShut {NoStop}%
\end{thebibliography}%

\appendix


\section{Equilibrium points of the Hamiltonian Model}
In this appendix we will describe the location and stability of the equilibrium points of the Hamiltonian model studied in this work. Firstly, we will find the equilibrium points $(r^{\ast},\gamma^{\ast},p_{r}^{\ast},p_{\gamma}^{\ast})$ of Hamilton's equations in Eq. \eqref{eq:hameq_2dof}, which is a straightforward task, since they correspond to critical points of the PES, that is, $\nabla V(r^{\ast},\gamma^{\ast}) = 0$, and are located in configuration space, since $p_{r}^{\ast} = p_{\gamma}^{\ast} = 0$. They are given by:
\begin{equation}
\gamma^{\ast} = \dfrac{k\pi}{2} \;,\; k \in \mathbb{Z} \quad,\quad  r^{\ast} = r_e + \dfrac{\ln(1 + \alpha \cos (2 \gamma^{\ast}))}{\beta}
\label{eq:eq_pts_w2DoF}
\end{equation}
Interestingly, there is also another equilibrium point of the system located at $r^{\ast} = \infty$. Moreover, since the PES is periodic in the variable $\gamma$ with period $\pi$, we will only need to consider the values $\gamma^{\ast} = 0 \, , \, \pi/2$. In order to determine the linear stability of the equilibrium points, we need the second order partial derivatives:
\begin{equation}
\begin{cases}
V_{rr} \equiv \dfrac{\partial^{2} V}{\partial r^2} = 2 \beta^2 D \left[-1 +2\left(1 + \alpha \cos (2\gamma )\right) e^{-\beta (r-r_e)}\right] e^{-\beta (r-r_e)} \\[.4cm]
V_{\gamma\gamma} \equiv \dfrac{\partial^2 V}{\partial \gamma^2} = -4 \alpha D \cos (2\gamma) \, e^{-2\beta (r-r_e)} \\[.4cm]
V_{r\gamma} \equiv \dfrac{\partial^{2} V}{\partial r \partial \gamma} = \dfrac{\partial^{2} V}{\partial \gamma \partial r} = 4 \alpha \beta D \sin (2\gamma) \, e^{-2\beta (r-r_e)}
\end{cases}
\label{eq:pder2}
\end{equation}
for the construction of the Hessian matrix, which is a submatrix of the Jacobian. Observe that, for the equilibrium point at infinity all the partial derivatives vanish, so that the Jacobian matrix has all zero eigenvalues and consequently this equilibrium point is said to be parabolic. On the other hand, when $r^{\ast}$ is a finite equilibrium point we get:
\begin{equation}
\begin{cases}
\dfrac{\partial^{2} V}{\partial r^2}\left(r^{\ast},\gamma^{\ast}\right) = \dfrac{2\beta^2 D}{1 + \alpha \cos(2\gamma^{\ast})} \\[.4cm]
\dfrac{\partial^2 V}{\partial \gamma^2}\left(r^{\ast},\gamma^{\ast}\right) = - \dfrac{4 \alpha D \cos(2\gamma^{\ast})}{\left(1 + \alpha \cos(2\gamma^{\ast})\right)^2} \\[.5cm]
\dfrac{\partial^{2} V}{\partial r \partial \gamma}\left(r^{\ast},\gamma^{\ast}\right) = \dfrac{\partial^{2} V}{\partial \gamma \partial r}\left(r^{\ast},\gamma^{\ast}\right) = 0
\end{cases}
\end{equation}
and the Jacobian matrix has the form:
\begin{equation}
J(r^{\ast},\gamma^{\ast},p_{r}^{\ast},p_{\gamma}^{\ast}) = \begin{pmatrix}
0 & 0 & \dfrac{1}{m} & \hspace{.2cm} 0 \hspace{.1cm}  \\[.2cm]
0 & 0 & 0 & \hspace{.2cm} \dfrac{1}{I} \hspace{.1cm} \\[.2cm]
-V_{rr} & -V_{r\gamma} & 0 & \hspace{.2cm} 0 \hspace{.1cm} \\[.2cm]
-V_{r\gamma} & -V_{\gamma\gamma} & 0 & \hspace{.2cm} 0 \hspace{.1cm} 
\end{pmatrix} =
\begin{pmatrix}
0 & 0 & \hspace{.2cm} \dfrac{1}{m} & \hspace{.4cm} 0 \hspace{.2cm} \\[.2cm]
0 & 0 & \hspace{.2cm} 0 & \hspace{.4cm} \dfrac{1}{I} \hspace{.2cm} \\[.2cm]
-\dfrac{2\beta^2 D}{1 + \alpha \cos(2\gamma^{\ast})} & 0 & \hspace{.2cm} 0 & \hspace{.4cm} 0 \hspace{.2cm} \\[.2cm]
0 & \dfrac{4 \alpha D \cos(2\gamma^{\ast})}{\left(1 + \alpha \cos(2\gamma^{\ast})\right)^2} & \hspace{.2cm} 0 & \hspace{.4cm} 0 \hspace{.2cm}
\end{pmatrix}
\label{eq:jacob_2dof}
\end{equation}

The eigenvalues for the equilibrium point $\left(r_e + \frac{\ln(1 + \alpha)}{\beta},0,0,0\right)$ are given by:
\begin{equation}	
\lambda_{1,2} = \pm \dfrac{2}{1 + \alpha} \sqrt{\dfrac{\alpha D}{I}} \quad,\quad \lambda	_{3,4} = \pm \beta \sqrt{\dfrac{2D}{m (1 + \alpha)}} \, i
\end{equation}
so that the equilibrium point is an index-1 saddle with saddle$\times$center stability. On the other hand, for the equilibrium point $\left(r_e + \frac{\ln(1 - \alpha)}{\beta},\pi/2,0,0\right)$ we get the eigenvalues:
\begin{equation}
\omega_{1,2} = \pm \dfrac{2}{1 - \alpha} \sqrt{\dfrac{\alpha D}{I}} \, i \quad,\quad \omega_{3,4} = \pm \beta \sqrt{\dfrac{2D}{m (1 - \alpha)}} \, i
\end{equation}
which determines that this point is a center equilibrium. Observe that the value of the radial coordinate at this equilibrium point is $r_e + \frac{\ln(1 - \alpha)}{\beta}$, which coincides with the value $r_{min}$ we introduced in Eq. \eqref{model_params}. It is also important to remark that the values attained by the potential energy at these equilibria are:
\begin{equation}
V_1 = V\left(r_e + \frac{\ln(1 + \alpha)}{\beta},0\right) = -\dfrac{D}{1+\alpha} \quad,\quad V_2 = V\left(r_e + \frac{\ln(1 - \alpha)}{\beta},\pi/2\right) = -\dfrac{D}{1-\alpha} 
\end{equation}
If we define $B = \Delta V = V_1 - V_2$ as the potential barrier height that the HeI$_2$ complex has to overcome for internal rotation, and $W = -V_2$ is the well depth of the Morse-like oscillator in the plane $\gamma = \pi/2$, we can write the coupling parameter $\alpha$ in terms of $B$ and $W$ to yield:
\begin{equation}
\alpha = \dfrac{B/W}{2-(B/W)}
\end{equation}
and we recover the expression we already introduced in Eq. \eqref{model_params} when defining the vdW PES.

Now we are ready to find the fixed points $(r^{\ast},\gamma^{\ast},p_{r}^{\ast},p_{\gamma}^{\ast})$ of the 4D symplectic map in Eq. (\ref{eq:map4D}). The fixed points are located at:
\begin{equation}
p_{r}^{\ast} = p_{\gamma}^{\ast} = 0 \quad,\quad \gamma^{\ast} = \dfrac{k\pi}{2} \;,\; k \in \mathbb{Z} \quad,\quad  r^{\ast} = r_e + \dfrac{\ln(1 + \alpha \cos (2 \gamma^{\ast}))}{\beta}
\label{eq:fix_pts_4D}
\end{equation}
and there is also a fixed point at infinity, $r^{\ast} = \infty$. In the next step, we compute their linear stability by analyzing the eigenvalues of the Jacobian matrix obtained by linearizing the map about the fixed points:
\begin{equation}
J(r^{\ast},\gamma^{\ast},p_{r}^{\ast},p_{\gamma}^{\ast}) = \begin{pmatrix}
1 - \dfrac{T^2}{m} \, V_{rr} & - \dfrac{T^2}{m} \, V_{r\gamma} & \dfrac{T}{m} & \hspace{.2cm} 0 \hspace{.1cm}  \\[.4cm]
- \dfrac{T^2}{I} \, V_{r\gamma} & 1 - \dfrac{T^2}{I} \, V_{\gamma\gamma} & 0 & \hspace{.2cm} \dfrac{T}{I} \hspace{.1cm} \\[.4cm]
-T V_{rr} & - T V_{r\gamma} & 1 & \hspace{.2cm} 0 \hspace{.1cm} \\[.4cm]
- T V_{r\gamma} & - T V_{\gamma\gamma} & 0 & \hspace{.2cm} 1 \hspace{.1cm} 
\end{pmatrix} = 
\begin{pmatrix}
1 - \dfrac{T^2}{m} \, V_{rr} & 0 & \dfrac{T}{m} & \hspace{.2cm} 0 \hspace{.1cm}  \\[.3cm]
0 & 1 - \dfrac{T^2}{I} \, V_{\gamma\gamma} & 0 & \hspace{.2cm} \dfrac{T}{I} \hspace{.1cm} \\[.3cm]
-T V_{rr} & 0 & 1 & \hspace{.2cm} 0 \hspace{.1cm} \\[.3cm]
0 & - T V_{\gamma\gamma} & 0 & \hspace{.2cm} 1 \hspace{.1cm} 
\end{pmatrix}
\label{eq:jacob_symp4d}
\end{equation}
For the fixed point at infinity, the Jacobian matrix has only one eigenvalue, $\lambda = 1$, with multiplicity 4. Hence, this point has parabolic stability. On the other hand, we discuss next the stability of the remaining fixed points of the 4D map.

The eigenvalues of the equilibrium point $\left(r_e+\dfrac{\ln(1+\alpha)}{\beta},0,0,0\right)$ are given by:
\begin{equation}	
\lambda_{1,2} = 1-\dfrac{T^{2}V^{\ast}_{rr}}{2m}\pm \dfrac{1}{2}\left(\dfrac{T^{4}}{m^{2}}V_{rr}^{\ast 2}-4\dfrac{T^{2}}{m}V^{\ast}_{rr}\right)^{1/2} \quad,\quad \lambda	_{3,4} = 1-\dfrac{T^{2}V^{\ast}_{\gamma \gamma}}{2m}\pm \dfrac{1}{2}\left(\dfrac{T^{4}}{m^{2}}V_{\gamma \gamma}^{\ast2}-4\dfrac{T^{2}}{m}V^{\ast}_{\gamma \gamma}\right)^{1/2}
\end{equation}
where $V^{\ast}_{rr}= \dfrac{2\beta ^{2} D}{(1+\alpha)}$ and $V^{\ast}_{\gamma \gamma}=-\dfrac{4\alpha D}{(1+\alpha)^{2}}$. The eigenvalues of the equilibrium point $\left(r_e+\dfrac{\ln(1-\alpha)}{\beta},\dfrac{\pi}{2},0,0\right)$ are given by:
\begin{equation}	
\lambda_{1,2} =  1-\dfrac{T^{2}V^{\star}_{rr}}{2m}\pm \dfrac{1}{2}\left(\dfrac{T^{4}}{m^{2}}V_{rr}^{\star 2}-4\dfrac{T^{2}}{m}V^{\star}_{rr}\right)^{1/2} \quad,\quad \lambda_{3,4} = 1- \dfrac{T^{2}V^{\star}_{\gamma \gamma}}{2m}\pm \dfrac{1}{2}\left(\dfrac{T^{4}}{m^{2}}V_{\gamma \gamma}^{\star2}-4\frac{T^{2}}{m}V^{\star}_{\gamma \gamma}\right)^{1/2}
\end{equation}
where $V^{\star}_{rr}=\dfrac{2\beta ^{2} D}{(1-\alpha)}$ and $V^{\star}_{\gamma \gamma}=\dfrac{4\alpha D}{(1-\alpha)^{2}}$.


\section{The Method of Lagrangian Descriptors}

This appendix gives a brief description on the diagnostic tool that we have used in this work to analyze the phase space structures of the 4D symplectic map modelling the dynamics of van der Waals complexes. The method of Lagrangian descriptors is a technique that was originally introduced in the context of fluid mechanics to study transport and mixing in geophysical flows \cite{madrid2009,Mancho1}, but in the past years it has been found to provide useful information for the analysis of the high dimensional phase space of chemical systems, see e.g. \cite{Agaoglou2019,ldbook2020} and references therein. Although the method was initially developed to analyze continuous-time dynamical systems \cite{mancho2013lagrangian,lopesino2017} with general time dependence, it has also been applied to maps \cite{carlos}, stochastic dynamical systems \cite{balibrea2016} and complex maps \cite{GG2020a}.

The idea behind the method of Lagrangian descriptors is very simple. Take any initial condition and accumulate along its trajectory the values attained by a positive scalar function that depends on the phase space variables. This calculation is carried out both forward and backward in time as the system evolves. Once this computation is done for a grid of initial conditions chosen on a given phase space slice where one would like to reveal structure, the scalar output obtained from the method will highlight the location of the invariant stable and unstable manifolds intersecting this slice. These manifolds will appear as 'singular features' of the scalar field, that is, they are detected at points where the values of LDs display an abrupt change, which indicates distinct dynamical behavior. Forward integration of trajectories detects stable manifolds while backward evolution does the same for unstable manifolds. It is important to remark that this technique also reveals the structure of KAM tori, but in this case the tori regions correspond to smooth values of the LD function. In fact, tori can be visualized by computing long-term time averages of LDs as discussed in \cite{lopesino2017,GG2019b,naik2019a}.

We explain here the basic setup for the method of Discrete Lagrangian Descriptors (DLDs) to analyze maps. This tool was first introduced in \citet{carlos} for the study of 2D area-preserving maps. Consider a domain $D \subset \mathbb{R}^{k}$ and a function $f$ that defines the discrete-time dynamical system,
\begin{equation}
\mathbf{z}_{i+1} = f(\mathbf{z}_{i}) \quad,\quad \text{with} \quad \mathbf{z}_i = (x^i_1,\ldots,x^i_k) \in D \;,\quad \forall i \in \mathbb{N} \cup \lbrace0\rbrace
\;.
\label{discrete_DS}
\end{equation}
We assume that this mapping is invertible so that,
\begin{equation}
\mathbf{z}_{i} = f^{-1}(\mathbf{z}_{i+1}) \quad,\quad \text{with} \quad \textbf{z}_i = (x^i_1,\ldots,x^i_k) \in D \;,\quad \forall i \in \mathbb{Z}^{-}
\label{discrete_DS_back}
\end{equation}
Given a fixed number of iterations $N > 0$ of the maps $f$ and $f^{-1}$, and take $p \in (0,1]$ to specify the $l^p$ norm that we will use for constructing the method, then the DLD applied to systems (\ref{discrete_DS}) and (\ref{discrete_DS_back}) is given by the following function:
\begin{equation}
MD_{p}\left(\mathbf{z}_0,N\right) = \sum_{i=-N}^{N-1} ||\mathbf{z}_{i+1}-\mathbf{z}_{i}||_{p} = \sum_{i=-N}^{N-1} \sum_{j = 1}^{k} |x^j_{i+1}-x^j_{i}|^{p}
\label{DLD}
\end{equation}
where $\mathbf{z}_0 \in D$ is any initial condition. Observe that Eq. (\ref{DLD}) can be split into two quantities:
\begin{equation}
MD_{p}\left(\mathbf{z}_0,N\right) = MD_{p}^{+}\left(\mathbf{z}_0,N\right) + MD_{p}^{-}\left(\mathbf{z}_0,N\right) \;, 
\end{equation}
where:
\begin{equation}
MD_{p}^{+} = \sum_{i=0}^{N-1} ||\textbf{z}_{i+1}-\mathbf{z}_{i}||_{p} = \sum_{i=0}^{N-1} \sum_{j = 1}^{k} |x^j_{i+1}-x^j_{i}|^{p} \;,
\label{DLD+}
\end{equation}
measures the forward evolution of the orbit of $\mathbf{z}_0$ and,
\begin{equation}
MD_{p}^{-} = \sum_{i=-N}^{-1} ||\textbf{z}_{i+1}-\mathbf{z}_{i}||_{p} = \sum_{i=-N}^{-1} \sum_{j = 1}^{k} |x^j_{i+1}-x^j_{i}|^{p}  \;,
\label{DLD-}
\end{equation}
accounts for the backward evolution of the orbit of $\mathbf{z}_0$. In \cite{carlos} it is shown that DLDs highlight stable and unstable manifolds at points where the $MD_p$ field becomes non-differentiable (discontinuities or unboundedness of its gradient). These non-differentiable points are displayed in a simple way when depicting the $MD_p$ scalar field, and are known in the literature as singular features. Moreover, \cite{carlos} demonstrates that $MD_{p}^{+}$ detects the stable manifolds of the fixed points and $MD_{p}^{-}$ does the same for the unstable manifolds of the fixed points. Therefore, the fixed points with respect to the map (\ref{discrete_DS}) will be located at the intersections of these structures.

In the definition of DLDs, the number of iterations $N$ considered for the computation of the orbit of any initial condition plays a crucial role, since it controls the level of detail with which the method reveals the geometrical structures in the phase space of the system. The underlying reason is that for a large number of iterations, we are incorporating into the analysis more information about the past and future history of the orbits of the map. However, issues might arise when applying this technique to analyze unbounded maps where orbits can escape to infinity very fast or even in finite time. This problem was circumvented in \cite{GG2019b} by means of fixing a phase space region before the analysis is carried out, and computing the value attained by the DLD along the orbit of an initial condition until the fixed number of iterations is reached or the orbit leaves the predefined domain, what happens first. This approach works nicely for many open systems such as the H\'enon map that was studied in \cite{GG2019b} with this adapted version of DLDs.

However, in the context of the 4D symplectic Morse map that we study in this work, we have observed that this strategy to avoid escaping trajectories is still posing some difficulties for the detection of phase space structure. For this reason, we have decided to adopt a different strategy that was developed in \cite{GG2020a} for the study of complex maps and Julia sets. Since the unbounded dynamics of the Morse map takes place in the radial plane $r-p_r$, and we know that the map has a fixed point at infinity, we compactify this plane by wrapping it around a unit sphere, and identifying all the points at infinity with the north pole of the sphere. This is done by means of using a stereographic projection ($\mathcal{S}$) that maps the points on the surface of the sphere, known as the Riemann sphere, onto its equatorial plane that is identified with the $r-p_r$ plane. Given an initial condition $\mathbf{z}_0 = (r_0,\gamma_0,p_{r,0},p_{\gamma,0})$, where $\mathbf{w}_0 = (r_0,p_{r,0})$ represents its coordinates in the radial plane, then we define the DLD scalar function as:
\begin{equation}
\mathcal{D}_{p}\left(\mathbf{z}_0,N\right) = 
= \sum_{i=-N}^{N-1} \big|\big|\mathcal{S}^{-1}(\mathbf{w}_{i+1}) - \mathcal{S}^{-1}(\mathbf{w}_{i})\big|\big|_p \;.
\label{DLD_comp}
\end{equation} 
In this work we have chosen $p = 1$ as the exponent of the $l^p$ norm used to define the DLD function. Notice that, although we can compute the DLD scalar function on any set of initial conditions, this is typically done by selecting two dimensional planes of the 4D phase space of the map. This is of great help for the visualization of structures of the high-dimensional space because we can use these slices to reconstruct the high-dimensional system. It is important to remark here that the implementation of DLDs that we have used relies only on the projections of the iterates of the radial components of the map onto the Riemann sphere. The reason behind this choice is that the dynamics in the angular degree of freedom $\gamma$ is periodic, and thus, bounded, and in our analysis we are interested in the analysis of the dissociation mechanism in phase space, which is characterized by the situation where the radial coordinates of the map go to infinity. What the method is doing is to accumulate the $p$-norm of the vectors constructed between sequential points of the map in the radial plane, but this operation is carried out using the corresponding projections of the iterates onto the Riemann sphere to avoid the issues caused by orbits escaping to infinity at an increasing rate. For more details on how to construct this version of DLDs, refer to \cite{GG2020b}.

\end{document}